\documentclass[a4paper,fleqn,10pt]{article}
\usepackage{amsmath}
\usepackage{amssymb}
\usepackage[a4paper,pdfborder={0 0 0}]{hyperref}
\usepackage[all]{hypcap}
\usepackage{array}
\usepackage{calc}
\usepackage{longtable}
\usepackage{multirow}
\usepackage{axodraw4j}
\usepackage{pstricks}
\usepackage{graphicx}
\usepackage{xspace}
\usepackage{cite}
\usepackage{mcite}
\usepackage[format=hang,labelfont=bf]{caption}
\usepackage{subfig}
\usepackage{sectsty}
\allsectionsfont{\sffamily}
\subsubsectionfont{\mdseries\itshape\large}
\let\spreprint\empty
\newcommand{\preprint}[1]{\def\spreprint{\protect#1}}
\let\sinstitute\empty
\newcommand{\institute}[1]{\def\sinstitute{\protect#1}}
\makeatletter
\renewcommand{\maketitle}{\begingroup
  \null\thispagestyle{empty}
    \ifx\spreprint\empty
      \vskip 5ex
    \else
      \flushright\large\spreprint\vskip 2ex
    \fi
    \vskip 5ex
    \flushleft
      {\sffamily\bfseries\huge\@title}\vskip 2ex
      \@author\vskip 2ex
      \ifx\sinstitute\empty
      \else
        {\small\sinstitute}
      \fi
    \vskip 5ex
  \endgroup
}
\makeatother
\renewenvironment{abstract}{\begin{center}
  {\large\sffamily\bfseries Abstract: }
  \begin{minipage}[t]{0.75\textwidth}
}{\end{minipage}\end{center}\vskip 10ex}

\newcommand{\myfigure}[3]{
  \begin{figure}[#1]
    \begin{center}
      #2\\
      \parbox[t]{\widthof{#2}}{\caption{#3}}
    \end{center}
  \end{figure}
}

\newcommand{\Sherpa}{S\protect\scalebox{0.8}{HERPA}\xspace}
\newcommand{\Comix}{C\protect\scalebox{0.8}{OMIX}\xspace}

\newcommand{\Pythia}{P\protect\scalebox{0.8}{YTHIA}\xspace}

\newcommand{\Professor}{P\protect\scalebox{0.8}{ROFESSOR}\xspace}
\long\def\symbolfootnote[#1]#2{\begingroup%
\def\thefootnote{\fnsymbol{footnote}}\footnote[#1]{#2}\endgroup}
\newcommand{\abs}[1]{\left| #1\right|}
\newcommand{\rbr}[1]{\left( #1\right)}
\newcommand{\abr}[1]{\langle #1\rangle}
\newcommand{\cbr}[1]{\left\{ #1\right\}}
\newcommand{\sbr}[1]{\left[ #1\right]}

\newcommand{\order}{\mathcal{O}}

\hypersetup{
  pdfauthor={Tancredi Carli, Thomas Gehrmann, Stefan Hoeche},
  pdftitle={Hadronic final states in deep-inelastic scattering with Sherpa},
  pdfkeywords={QCD, DIS, HERA, Matrix Element, Parton Shower, Sherpa}
}
\preprint{ZU-TH 20/09\\MCNET/09/19}
\author{Tancredi Carli$^1$, Thomas Gehrmann$^2$,
  Stefan H{\"o}che$^2$}
\title{Hadronic final states in\\[2mm] deep-inelastic scattering with \Sherpa}
\institute{$^1$ CERN, Department of Physics,
  CH-1211 Geneva 23, Switzerland\\
  $^2$ Institut f{\"u}r Theoretische Physik, 
  Universit{\"a}t Z{\"u}rich, CH-8057 Z{\"u}rich, Switzerland}
\begin{document}
\maketitle
\begin{abstract}
  We extend the multi-purpose Monte-Carlo event generator \Sherpa to include processes 
in deeply inelastic lepton-nucleon scattering. Hadronic 
final states in this kinematical setting are characterised by the presence 
of multiple kinematical scales, which were up to now accounted for only 
by specific resummations in individual kinematical regions. Using an 
extension of the recently introduced method for
merging truncated parton showers with higher-order tree-level
matrix elements, it is possible to obtain predictions which are reliable 
in all kinematical limits. 
Different hadronic final states, defined by jets or individual hadrons, 
in deep-inelastic scattering
are analysed and the corresponding results are compared to HERA data.
The various sources of 
theoretical uncertainties of the approach are discussed 
and quantified. The extension 
to deeply inelastic processes provides the opportunity to 
validate the merging of matrix elements and parton showers in 
multi-scale kinematics inaccessible in other collider environments. It also
allows to use HERA data on hadronic final states in the tuning of 
hadronisation models.

\end{abstract}
\section{Introduction} 
\label{sec:intro}
Deep-inelastic lepton-nucleon scattering (DIS) allows to analyse the structure 
of the nucleon by means of a pointlike probe, and provides an experimental 
framework for a multitude of studies of strong interaction dynamics. 
The kinematical situation of deeply inelastic events offers access to 
configurations which cannot be probed at other colliders. The (space-like) 
virtuality of the exchanged photon sets a hard scale for the scattering 
process, which is also the unique hard scale in inclusive deep-inelastic structure functions. 
Studying more exclusive properties of the hadronic final state allows access 
to multiple other scales, given for example by the transverse momenta of final-state jets. 
Such multiple-scale configurations are impossible to realise in $e^+e^-$ annihilation,
where the centre-of-mass energy is the only hard scale of the process, and difficult 
to access in purely hadronic collisions (vector-boson plus multi-jet production being an example 
of such a multi-scale configuration). Hadronic final states in DIS
thus offer the unique opportunity to study multi-scale processes 
in QCD and provide the advantage of a relatively clean experimental setting.
A wealth of corresponding experimental data is available 
from the HERA collider experiments H1 and ZEUS. 

The kinematical situation of a deeply inelastic scattering process with incoming proton momentum $p$
and incoming and outgoing electron momenta $k$ and $k'$ is characterised by the virtuality $Q^2$ of the 
exchanged boson, carrying momentum $q=k-k'$, and by the Bj{\o}rken variable $x$, which can be inferred 
purely from the outgoing electron momentum as
\begin{align}
  Q^2 =& -q^2 = (k-k')^2 &&\text{and} &x=&\frac{Q^2}{2\, q\cdot p}
\end{align}
Measurements are usually either performed in the Breit frame, or in the centre-of-mass frame of proton 
and virtual photon, called the hadronic centre-of-mass frame.\footnote{
  The hadronic centre-of-mass frame is defined by $\vec{q}+\vec{p}=0$, while in the Breit frame
  $2 x \vec{p} + \vec{q} = 0$. The two frames are related by a longitudinal boost.}
 The centre-of-mass energy squared is then given 
by $W^2 = {Q^2(1-x)}/{x}$. In lepton-hadron collisions, one generally distinguishes photoproduction processes, 
where the exchange photon is quasi-real, $Q^2\to 0$ with $W^2$ fixed, and deeply inelastic processes. 
This distinction is made experimentally by imposing a minimum cut on $Q^2$, 
typically of the order a few GeV$^2$.

Inclusive structure functions as the basic quantities in deep-inelastic processes depend on $x$ and $Q^2$ only, 
and the description of the proton structure in terms of parton distributions is formulated in 
the space of these variables. The evolution of the parton distributions with increasing $Q^2$ 
is determined by the Dokshitzer-Gribov-Lipatov-Altarelli-Parisi (DGLAP) 
equations~\cite{Gribov:1972ri,*Lipatov:1974qm,*Dokshitzer:1977sg,*Altarelli:1977zs},
which are known to next-to-next-to-leading order~\cite{Curci:1980uw,*Furmanski:1980cm,*Moch:2004pa,*Vogt:2004mw} in QCD. 
These equations form the basis of the QCD-improved parton model, and allow for a determination of the process-independent
parton distribution functions from global fits~\cite{Martin:2009iq,Ball:2008by,Nadolsky:2008zw,JimenezDelgado:2008hf} 
to data from lepton-hadron and hadron-hadron collisions. Higher-order corrections to the DGLAP 
equations~\cite{Curci:1980uw,*Furmanski:1980cm,*Moch:2004pa,*Vogt:2004mw}
contain powers of logarithms in $x$ or $(1-x)$, which potentially spoil the convergence of the perturbative 
expansion at large and small $x$. In both limits, resummation formalisms for the large logarithmic corrections 
are available: threshold resummation at large $x$ and BFKL resummation~\cite{Kuraev:1977fs,*Balitsky:1978ic} at small $x$. 
A unified DGLAP/BFKL resummation is provided by the CCFM 
equation~\cite{Ciafaloni:1987ur,*Catani:1989yc,*Catani:1989sg,*Marchesini:1994wr}. The currently 
available inclusive structure function data can however be described entirely by the DGLAP framework. Considerable 
experimental and theoretical effort was made especially in order to establish observables sensitive to BFKL effects. 
In this context, specific hadronic final states, such as forward jets~\cite{Aktas:2005up,*Chekanov:2007pa} 
are usually investigated and jet rapidity correlations appear to be promising observables.  

Among hadronic final states, jet cross sections offer the most direct probes of parton-level dynamics. 
The H1 and ZEUS experiments have performed many different measurements of jet production processes, ranging 
from single-jet-inclusive and di-jet (which is often called $(2+1)$-jet because of the extra 
proton remnant jet) cross sections to multi-jet cross sections and jet correlations. While 
the former are used for precision determinations of the strong coupling and of parton distributions, 
the latter offer detailed insight into the production dynamics, and can highlight the kinematical limitations of 
the DGLAP framework. In this framework, next-to-leading order (NLO) QCD predictions are available for single-jet-inclusive 
and di-jet cross sections~\cite{Mirkes:1995ks,*Catani:1996jh} and for three-jet production~\cite{Nagy:2001xb,*Nagy:2005gn}. 
For central jet production, and provided that $Q^2$ is not much smaller than the transverse energies of the jets in the Breit frame, $E_{T,B}$, these 
calculations yield a very good description of the experimental data~\cite{Aktas:2007pb,*Chekanov:2006xr}. In the situation 
of $Q^2\ll E_{T,B}^2$, large logarithmic corrections of the form $\ln(Q^2/E_{T,B}^2)$ appear to all orders in perturbation theory.
By attributing a parton content to the virtual photon entering the hard process~\cite{Schuler:1996en}, these can be resummed. 
Including the contributions from the virtual photon structure into the NLO QCD calculations~\cite{Potter:1998jt} extends
the kinematical range where those are applicable (including the description of forward jet production~\cite{Kramer:1999jr}), and 
allows for a smooth transition from deep-inelastic to photoproduction processes.

Within the framework of Monte-Carlo event generation, multi-jet production is usually described 
through parton showers, starting from a leading-order process. Various parton-shower models exist, 
which are either based on DGLAP evolution~\cite{Bengtsson:1986et,*Sjostrand:1985xi,
 Marchesini:1983bm,*Marchesini:1987cf,Sjostrand:2004ef,Kuhn:2000dk,*Krauss:2005re},
or CCFM evolution~\cite{Marchesini:1990zy,*Marchesini:1992jw,*Jung:2001hx,*Golec-Biernat:2007pu}.
Other methods use the colour dipole model~\cite{Lonnblad:1992tz,*Winter:2007ye} or an approach based
on the Catani-Seymour subtraction technique~\cite{Schumann:2007mg,Dinsdale:2007mf}. All of those models
have in common, that they are capable of describing parton-level final states in certain regions of 
the phase space only, which are defined by the respective resummation prescription. By construction,
none of them allows to correctly account for multi-parton correlations, and therefore their
predictions should be corrected using higher-order matrix elements. 
Studies of electroweak gauge boson production~\cite{Gleisberg:2005qq,*Alwall:2007fs}
and the production of coloured heavy states~\cite{Mangano:2006rw,*Alwall:2008qv} indicate
that an improved description of high transverse momentum jets in the DGLAP framework can be 
obtained by an appropriate combination of tree-level matrix elements and parton 
showers~\cite{Catani:2001cc,*Krauss:2002up,Mangano:2001xp}. 
Such merged calculations are reliable in most kinematical limits. They can be thought of
as unifying the leading logarithmic expressions of the different resummation prescriptions 
(DGLAP, BFKL, virtual photon structure) in a single calculation. 
Recently, new and powerful techniques have been developed, which extend those 
methods~\cite{Hoeche:2009rj,Hamilton:2009ne}. Their advantage is that parton
shower radiation patterns are recovered to the accuracy provided by the shower model
and therefore any statements about the logarithmic accuracy of the parton shower
also holds for a merged calculation in this scheme. These novel techniques have so far been used 
in two relevant cases, namely $e^+e^-$ annihilation into hadrons and Drell-Yan like production
of electroweak gauge bosons. The technical prerequisites for realising the approach 
in~\cite{Hoeche:2009rj} were implemented into the multi-purpose Monte-Carlo event generator 
\Sherpa{}~\cite{Gleisberg:2003xi,*Gleisberg:2008ta} in full generality.
Since hadronic final states in deep-inelastic scattering depend 
on multiple kinematical scales, related observables provide an independent and particularly 
sensitive test of the quality of the approach and its implementation. It is the aim of this work to present 
a first study of this class of processes with \Sherpa, and to confront results with data from the HERA experiments.
Moreover we propose an extension of the merging algorithm in~\cite{Hoeche:2009rj}, which accounts
for the proper simulation of low-$Q^2$, high-$E_{T,B}^2$ events.

Hadronic final states also offer insight into strong interaction dynamics at lower scales. Using kinematical spectra 
of identified hadrons, it is possible to probe the parton-to-hadron transition (hadronisation), which can not be computed 
from first principles, but is usually described using semi-empirical models~\cite{Andersson:1983ia,*Andersson:1998tv,%
Field:1982dg,*Gottschalk:1983fm,*Webber:1983if,Krauss:2010xy,*Winter:2003tt}. 
These models are typically tuned to data from $e^+e^-$ collider experiments and the obtained fits are assumed 
to be universal. Including data from deep-inelastic processes instead allows to probe different flavour combinations
and beam remnant fragmentation, and to resolve parameter degeneracies. It is therefore vital to have means for 
simulating partonic final states in deep-inelastic processes reliably, not only to describe jet spectra, 
but also to reduce uncertainties in fragmentation models, which can then be tuned to experimental data in a combined fit.
Therefore we also provide some examples for the influence of the parton-level inputs on different fragmentation models.

The outline of this paper is as follows. Section~\ref{sec:mcdetails} introduces
the technical details of the Monte Carlo simulation, including the proposal of
an improved merging technique for low-$Q^2$ events. Section~\ref{sec:results} 
presents the results of our analysis and discusses theoretical uncertainties. 
Finally Sec.~\ref{sec:conclusions} contains some concluding remarks.

\section{Event generation techniques}
\label{sec:mcdetails}
\myfigure{t}{\hspace{1cm}
  \subfloat[][]{\scalebox{0.33}{\begin{picture}(360,58) (271,-247)
    \SetWidth{1.0}
    \SetColor{Black}
    \Photon(464,-166)(368,-166){5}{6}
    \Line[arrow,arrowpos=0.5,arrowlength=5,arrowwidth=2,arrowinset=0.2](320,-228)(368,-166)
    \Line[arrow,arrowpos=0.5,arrowlength=5,arrowwidth=2,arrowinset=0.2](368,-166)(320,-104)
    \Text(405,-200)[lb]{\huge$Q^2$}
    \Line(629,-191)(581,-191)
    \Line(629,-173)(581,-173)
    \Line[arrow,arrowpos=0.5,arrowlength=5,arrowwidth=2,arrowinset=0.2](630,-182)(582,-182)
    \GOval(582,-182)(22,6)(0){0.882}
    \Line[arrow,arrowpos=0.5,arrowlength=5,arrowwidth=2,arrowinset=0.2](576,-182)(480,-182)
    \Line[arrow,arrowpos=0.5,arrowlength=5,arrowwidth=2,arrowinset=0.2](480,-150)(576,-150)
    \Arc[clock](480,-166)(16,-90,-270)
  \end{picture}}\label{fig:diskin_one}}\hspace{2cm}
  \subfloat[][]{\scalebox{0.33}{\begin{picture}(392,162) (271,-143)
    \SetWidth{1.0}
    \SetColor{Black}
    \Line[arrow,arrowpos=0.5,arrowlength=5,arrowwidth=2,arrowinset=0.2](320,-126)(368,-62)
    \Line[arrow,arrowpos=0.5,arrowlength=5,arrowwidth=2,arrowinset=0.2](368,-62)(320,2)
    \Line[arrow,arrowpos=0.5,arrowlength=5,arrowwidth=2,arrowinset=0.2](480,-62)(470,18)
    \Line[arrow,arrowpos=0.5,arrowlength=5,arrowwidth=2,arrowinset=0.2](450,-142)(480,-62)
    \Line(661,-71)(613,-71)
    \Line[arrow,arrowpos=0.5,arrowlength=5,arrowwidth=2,arrowinset=0.2](662,-62)(614,-62)
    \Line(661,-53)(613,-53)
    \Gluon(608,-62)(496,-62){7.5}{9}
    \GOval(614,-62)(22,6)(0){0.882}
    \GOval(480,-62)(16,16)(0){0.6}
    \Photon(464,-62)(368,-62){5}{6}
    \Text(405,-96)[lb]{\huge$Q^2$}
    \Text(485,2)[lb]{\huge$E_{T,B}$}
  \end{picture}}\hspace{1cm}\label{fig:diskin_two}}
}{Schematic view of the scattering kinematics in the Breit frame for leading-order 
  $e^\pm q\to e^\pm q$ scattering and 2-jet production processes
  in DIS. The lightly shaded blob denotes the incoming proton. For 2-jet events with
  large jet transverse energy, $E_{T,B}^2\gtrsim Q^2$, the $2\to 2$ process depicted by 
  the dark shaded blob in Fig.~\protect\subref{fig:diskin_two} sets the hard scale.
  \label{fig:diskin}}

The most striking difference between deep-inelastic scattering and processes 
like Drell-Yan lepton pair production is the nearly arbitrary hard scale $Q^2$, 
at which the proton structure can be probed by the virtual photon.
While this presents an excellent opportunity for {\em measuring} the QCD dynamics
of the process, it also constitutes the main obstacle for {\em simulating} it 
with Monte Carlo techniques. The reason for these problems and the solution
adopted in the context of this work are outlined below.

\subsection{Parton shower evolution}
\label{sec:shower}
Existing parton shower simulations are often based on virtuality ordering~\cite{
  Bengtsson:1986et,*Sjostrand:1985xi,Kuhn:2000dk,*Krauss:2005re} or transverse momentum 
ordering~\cite{Sjostrand:2004ef,Dinsdale:2007mf,Schumann:2007mg}. 
The hard scale, i.e.\ the maximum evolution parameter 
for a set of colour connected partons, is then usually taken to be the maximum virtuality
involved in the production of these partons. In Drell-Yan like events, for example,
the hard scale is taken as the invariant mass of the final state lepton pair. 
In the case of DIS it is taken as the (negative) photon virtuality $Q^2$. 
Using this choice it is a priori impossible to fill the complete available phase space 
with parton-shower emissions, since $Q^2$ tends to be close to zero.
Next-to-leading order calculations indicate, however, that the emission probability 
for additional partons is large, even if $Q^2$ is low, due to the possibly large hadronic 
centre-of-mass energy. 

The leading contribution in this context stems from the interaction,
$e^\pm g\to e^\pm q\bar{q}$, where the ``sub-process'' $\gamma^*g\to q\bar{q}$ plays 
the role of the hardest interaction, if the transverse energy squared of the final state 
quarks in the Breit frame, $E_{T,B}^2$, is larger than $Q^2$
(cf.\ Fig.~\ref{fig:diskin}). A similar problem was noted in~\cite{Plehn:2005cq} 
in the context of supersymmetric particle production at hadron colliders. 
To circumvent it, and allow a wider range of the phase space to be accessible for parton
shower radiation, so-called ``power shower'' schemes were employed to artificially 
increase the starting scale of the initial-state shower. Although apparently in conflict
with factorisation assumptions, this approach has recently received some theoretical support
from the proposition of modified DGLAP evolution equations~\cite{Dokshitzer:2005bf,
  *Dokshitzer:2009wd}. 

Another approach to overcome the restriction of the shower phase space
by a low factorisation scale would be to employ an ordering parameter different 
from virtuality or transverse momentum; one which is more suited for the description 
of radiation off colour dipoles connected to initial-state hadrons. In fact, the proper 
framework for treating initial-state radiation is given by the CCFM equations~\cite{
  Ciafaloni:1987ur,*Catani:1989yc,*Catani:1989sg,*Marchesini:1994wr}, 
which order branchings in terms of emission angles, cf.\ also~\cite{Marchesini:1983bm,*Marchesini:1987cf}.
The CCFM scheme allows the transverse momentum of emitted partons to become 
larger than the factorisation scale and can therefore provide a generic solution of the 
above problem. It has successfully been used in several Monte-Carlo event 
generators~\cite{Marchesini:1990zy,*Marchesini:1992jw,*Jung:2001hx,*Golec-Biernat:2007pu}. We do, however, not resort 
to the CCFM technique here. The fact that a transverse  momentum ordered parton shower 
can only sensibly describe parton spectra below the factorisation scale will rather be 
compensated by a special technique for merging matrix elements and truncated showers,
which is introduced in Sec.~\ref{sec:mets}.

In the context of this work we employ the parton-shower algorithm initially presented 
in~\cite{Schumann:2007mg}, which is based on the Catani-Seymour (CS) subtraction method, 
cf.~\cite{Catani:1996vz,*Catani:2002hc}. Modifications of the original approach to account 
for recoil effects into the final state from splitting initial state partons with final 
state spectator were recently proposed~\cite{Platzer:2009jq} and are refined in~\cite{Hoeche:2009xy}.
It is interesting to investigate the corresponding effect on the parton shower predictions.
Figure~\ref{fig:recoil} shows differential $n$-jet rates, i.e.\ the scale where an $n$-jet
event is clustered into an $n-1$-jet event, using the exclusive $k_T$-jet 
algorithm~\cite{Catani:1992zp}. The difference between the predictions are sizeable,
when switching between the original and the modified recoil scheme, especially in the 
low-$k_T$ domain, 1~GeV$\,\lesssim k_T\lesssim\,$10~GeV.
This implies that the choice of the recoil scheme in a given parton-shower simulation 
should be part of an uncertainty analysis, much like the variation of renormalisation 
and factorisation scales. We comment on this subject in Secs.~\ref{sec:mets} and ~\ref{sec:results}.
To improve the parton-shower prediction in the domain of hard emissions and therefore 
alleviate the merging with NLO real emission matrix elements, the shower splitting kernels 
can be modified to include matrix element corrections. The corrected splitting kernels 
amount to antenna functions~\cite{GehrmannDeRidder:2005cm,*Daleo:2006xa}, which were used for parton showers only in 
$e^+e^-$ annihilation up to now~\cite{Giele:2007di}. The corresponding 
procedure is outlined in Appendix~\ref{sec:mecorrection}. Figures~\ref{fig:mecorr_new} 
and~\ref{fig:mecorr_old} show the influence of these corrections on the $k_T$-jet rates 
in the Breit frame. We observe a substantial change in the total rate of emissions.
In the following, matrix element corrected splitting kernels are therefore employed.

\myfigure{t}{
  \subfloat[][]{\includegraphics[width=5.5cm]{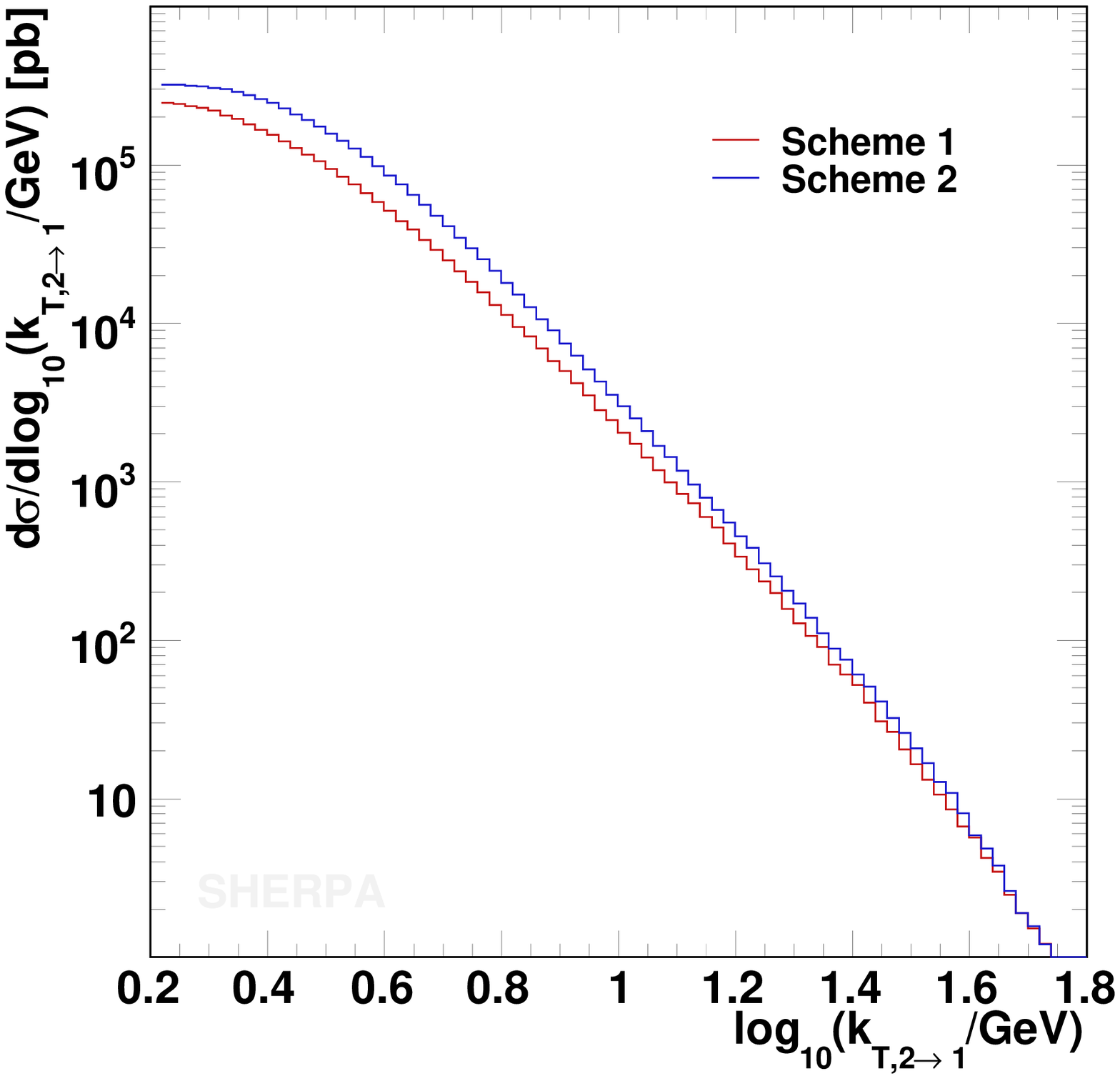}
  \label{fig:recoil}}\hspace*{-5mm}
  \subfloat[][]{\includegraphics[width=5.5cm]{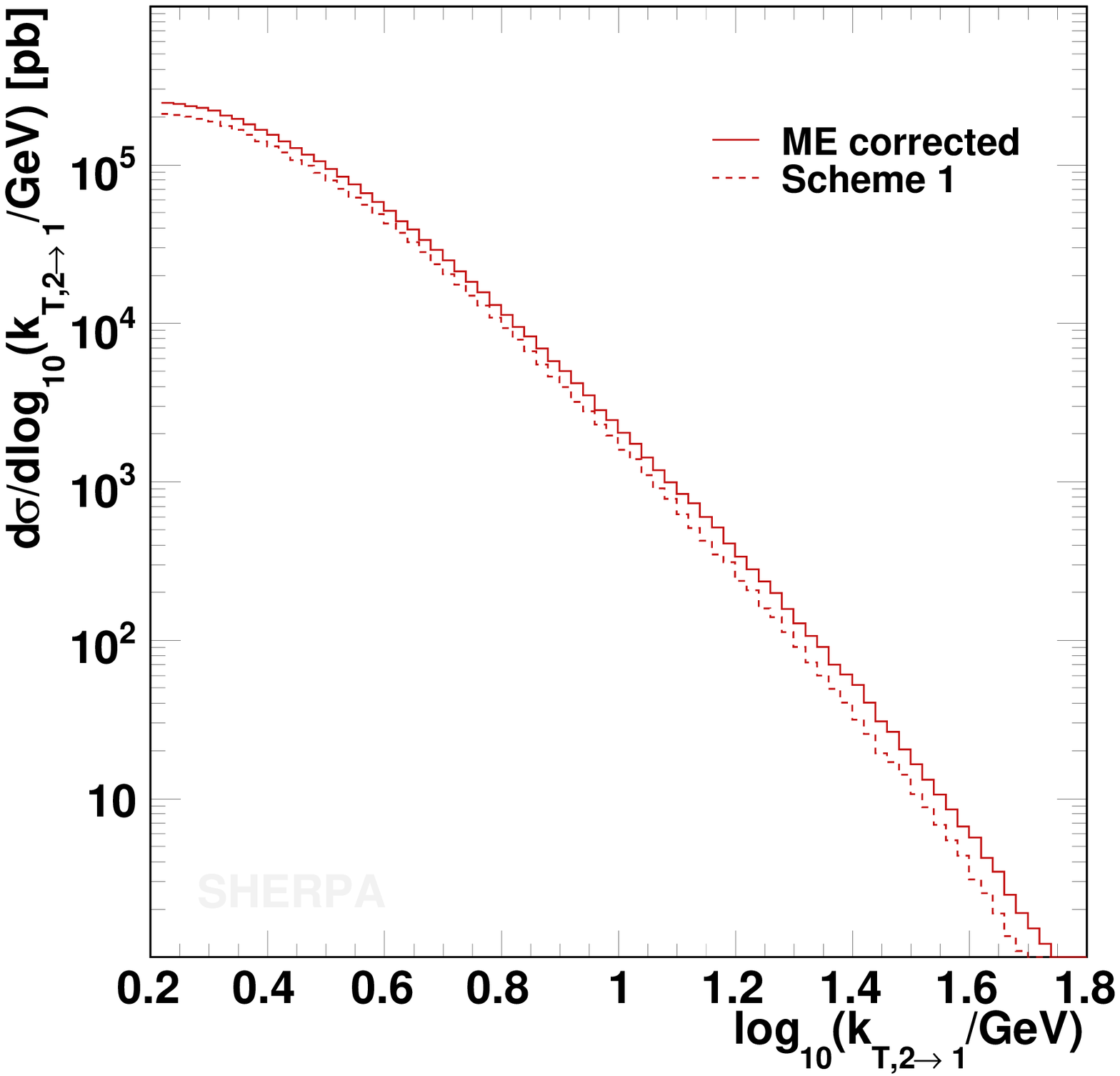}
  \label{fig:mecorr_new}}\hspace*{-5mm}
  \subfloat[][]{\includegraphics[width=5.5cm]{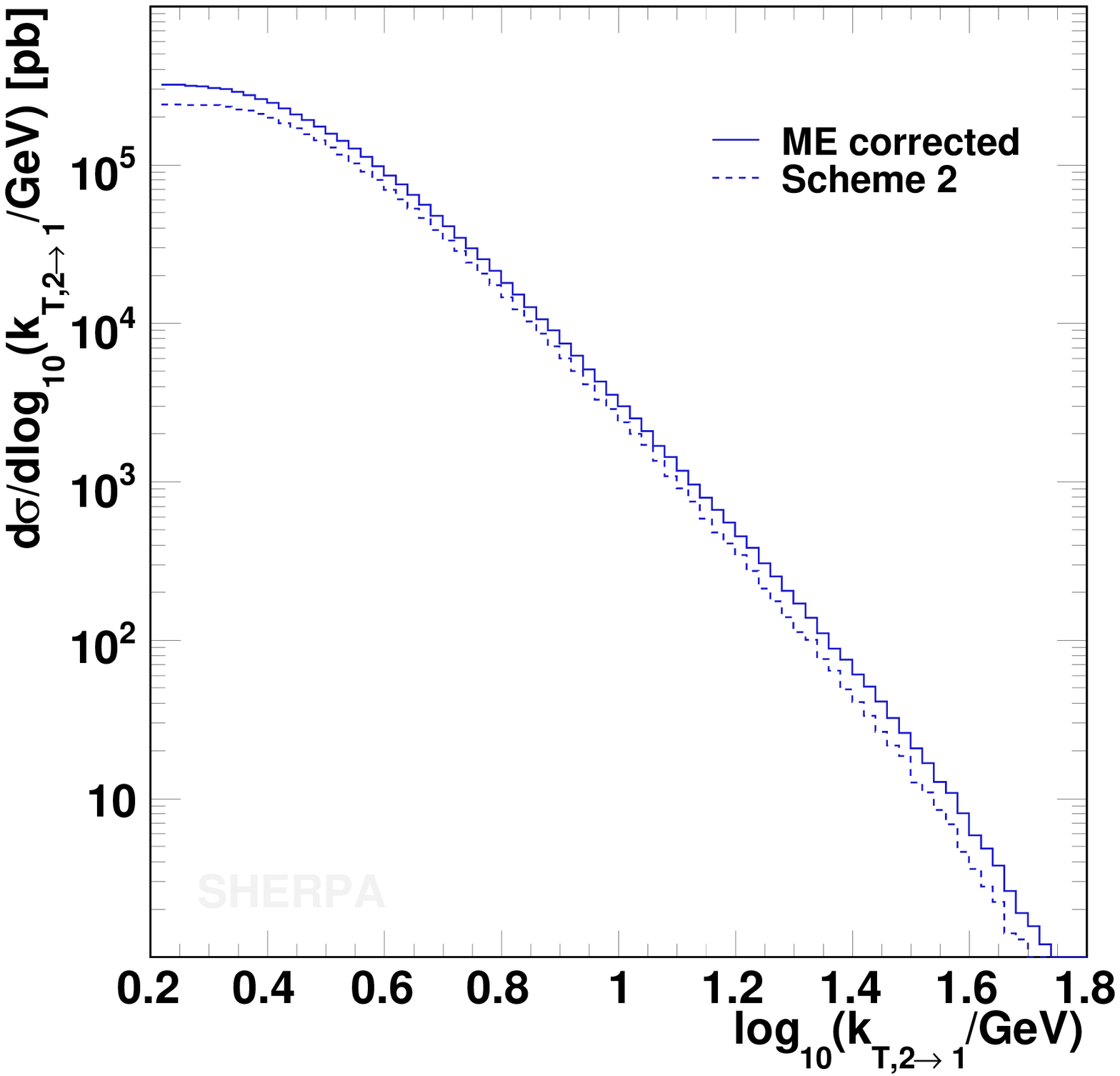}
  \label{fig:mecorr_old}}
}{Differential 2-jet rates defined by the exclusive $k_T$-jet algorithm in the Breit frame
  for deep-inelastic scattering events with $Q^2>4\,{\rm GeV}^2$.
  Part~\protect\subref{fig:recoil} compares the influence of different recoil strategies,
  while parts~\protect\subref{fig:mecorr_new} and~\protect\subref{fig:mecorr_old} show the effect 
  of matrix element corrections. 
  Monte Carlo samples were generated using the parton shower model of~\protect\cite{Schumann:2007mg}.
  Scheme 1 stands for the recoil strategy in~\protect\cite{Platzer:2009jq,Hoeche:2009xy}, while
  scheme 2 labels the original strategy employed in~\protect\cite{Schumann:2007mg}.
  \label{fig:shower}}

\subsection{Merging matrix elements and parton showers}
\label{sec:mets}

Next-to-leading order calculations and parton-shower simulation present two essentially 
different approaches to perturbative QCD. Fixed-order calculations seek to determine
all finite corrections to the leading-order process and are usually most important when
measuring inclusive quantities. Parton-shower simulation aims at a proper resummation
of large logarithmic corrections to the leading-order result, while preserving the overall 
cross section of the initial event sample. For the first emission this presents an 
approximation to the real NLO correction, whose quality largely depends on the underlying
assumptions about the splitting kinematics and the recoil scheme, as outlined in 
Sec.~\ref{sec:shower}. Thus, parton-shower simulations
are inherently incapable to describe the precise correlations between more than a few
final-state QCD particles properly. While the number of particles of interest is still low
($\order(1-6)$), the corresponding problems can easily be corrected by employing full
tree-level matrix elements instead of splitting kernels. Their computation has been automated 
in various approaches and poses no conceptional problem. The task is then reduced to 
finding an efficient and versatile algorithm for implementing the parton-shower correction 
in a generic way. Several methods attempted to solve this problem in the past~\cite{
  Catani:2001cc,*Krauss:2002up,Lonnblad:2001iq,*Lavesson:2008ah,Mangano:2001xp}, 
while two especially suitable approaches were suggested only recently~\cite{Hoeche:2009rj,Hamilton:2009ne}.
In the context of this work, the shower-independent formulation in terms of truncated 
parton showers and an arbitrary jet criterion is employed, which was introduced in~\cite{Hoeche:2009rj}.

The basic idea is to separate the phase space into a matrix-element and a parton-shower 
domain through a cut in the real-emission phase space. The matrix-element domain is then supposed 
to contain hard, well-separated partons only, while the parton-shower domain covers the region 
where resummation effects become important. Throughout the hard domain parton-shower emissions 
are corrected using tree-level matrix elements up to a given maximum multiplicity. 
In the soft domain, the parton shower is applied as is. The separation is achieved in terms 
of a so-called jet criterion, defining the ``hardness'' and/or the separation of a parton 
with respect to others~\cite{Hoeche:2009rj}. This can be thought of as a kind of 
$k_T$-jet measure, cf.\ e.g.~\cite{Catani:1992zp}.
\myfigure{t}{
  \subfloat[][]{\scalebox{0.33}{
  \begin{picture}(328,133) (303,-159)
    \SetWidth{1.0}
    \SetColor{Black}
    \Line[arrow,arrowpos=0.5,arrowlength=5,arrowwidth=2,arrowinset=0.2](352,-91)(416,-43)
    \Line[arrow,arrowpos=0.5,arrowlength=5,arrowwidth=2,arrowinset=0.2](416,-123)(352,-91)
    \Line(629,-116)(581,-116)
    \Line(629,-97)(581,-97)
    \Line[arrow,arrowpos=0.5,arrowlength=5,arrowwidth=2,arrowinset=0.2](630,-107)(582,-107)
    \GOval(582,-107)(22,6)(0){0.882}
    \Line[arrow,arrowpos=0.5,arrowlength=5,arrowwidth=2,arrowinset=0.2](576,-107)(496,-107)
    \Line[arrow,arrowpos=0.5,arrowlength=5,arrowwidth=2,arrowinset=0.2](304,-155)(352,-91)
    \Line[arrow,arrowpos=0.5,arrowlength=5,arrowwidth=2,arrowinset=0.2](352,-91)(304,-27)
    \GOval(352,-91)(16,16)(0){0.6}
    \Line[arrow,arrowpos=0.5,arrowlength=5,arrowwidth=2,arrowinset=0.2](496,-107)(416,-123)
    \Gluon(464,-75)(496,-107){7.5}{4}
    \Gluon(416,-123)(384,-155){7.5}{4}
  \end{picture}}\label{fig:multicore_one}}\hspace{1cm}
  \subfloat[][]{\scalebox{0.33}{
  \begin{picture}(328,163) (303,-142)
    \SetWidth{1.0}
    \SetColor{Black}
    \Line[arrow,arrowpos=0.5,arrowlength=5,arrowwidth=2,arrowinset=0.2](496,-93)(416,-61)
    \Line[arrow,arrowpos=0.5,arrowlength=5,arrowwidth=2,arrowinset=0.2](416,-61)(400,-141)
    \Line[arrow,arrowpos=0.5,arrowlength=5,arrowwidth=2,arrowinset=0.2](576,-93)(496,-93)
    \Line[arrow,arrowpos=0.5,arrowlength=5,arrowwidth=2,arrowinset=0.2](304,-125)(352,-61)
    \Line[arrow,arrowpos=0.5,arrowlength=5,arrowwidth=2,arrowinset=0.2](352,-61)(304,3)
    \GOval(416,-61)(16,16)(0){0.6}
    \Gluon(496,-93)(464,-125){7.5}{4}
    \Photon(352,-61)(400,-61){5}{3}
    \Line(629,-102)(581,-102)
    \Line(629,-83)(581,-83)
    \Line[arrow,arrowpos=0.5,arrowlength=5,arrowwidth=2,arrowinset=0.2](630,-93)(582,-93)
    \GOval(582,-93)(22,6)(0){0.882}
    \Gluon(416,-45)(448,19){7.5}{6}
  \end{picture}}\label{fig:multicore_two}}\hspace{1cm}
  \subfloat[][]{\scalebox{0.33}{
  \begin{picture}(328,163) (303,-159)
    \SetWidth{1.0}
    \SetColor{Black}
    \Line[arrow,arrowpos=0.5,arrowlength=5,arrowwidth=2,arrowinset=0.2](400,-61)(464,-45)
    \Line[arrow,arrowpos=0.5,arrowlength=5,arrowwidth=2,arrowinset=0.2](480,-77)(400,-61)
    \Line[arrow,arrowpos=0.5,arrowlength=5,arrowwidth=2,arrowinset=0.2](304,-125)(352,-61)
    \Line[arrow,arrowpos=0.5,arrowlength=5,arrowwidth=2,arrowinset=0.2](352,-61)(304,3)
    \GOval(496,-77)(16,16)(0){0.6}
    \Line[arrow,arrowpos=0.5,arrowlength=5,arrowwidth=2,arrowinset=0.2](576,-77)(512,-77)
    \Gluon(496,3)(496,-61){7.5}{5}
    \Photon(352,-61)(400,-61){5}{3}
    \Line(629,-86)(581,-86)
    \Line(629,-67)(581,-67)
    \Line[arrow,arrowpos=0.5,arrowlength=5,arrowwidth=2,arrowinset=0.2](630,-77)(582,-77)
    \GOval(582,-77)(22,6)(0){0.882}
    \Gluon(496,-93)(464,-157){7.5}{6}
  \end{picture}}\label{fig:multicore_three}}
}{Schematic view of three possible core process choices in DIS three-jet production. 
  Part~\protect\subref{fig:multicore_one} corresponds to the most probable core process
  being the virtual photon exchange, while additional hard partons are interpreted
  as parton shower emissions. Parts~\protect\subref{fig:multicore_two} 
  and~\protect\subref{fig:multicore_three} depict configurations, where the most probable
  core process is the interaction of the virtual photon with a parton and a pure QCD
  $2\to 2$ process, respectively.
  \label{fig:multicore}}

As pointed out in~\cite{Hoeche:2009rj}, this merging algorithm needs to be refined
if the scale difference between $Q^2$ and the hardness scale $k_T^2$ of additional partons 
is large and negative. In this case, logarithmic corrections are not induced by $Q^2/q^2$, 
but rather by $k_T^2/q^2$, where $q^2$ is the jet resolution scale. In the case of DIS, 
the production of the virtual photon can then be viewed as an electroweak splitting process, 
attached to the core $\gamma^* j\to jj$ interaction, as depicted in Fig.~\ref{fig:multicore_two}.
In the extreme case of very hard jets, the core process does not even include the 
virtual photon. This is visualised in Fig.~\ref{fig:multicore_three}. The correct choice
of the core process is not arbitrary, but is rather fixed by the backwards clustering 
algorithm described in~\cite{Hoeche:2009rj}, cf.\ also~\cite{Krauss:2004bs,*Schalicke:2005nv}.
To allow an inclusive merging procedure, the clustering algorithm must allow to identify the virtual photon 
as a soft particle, which is removed in order to find the core process and reproduced 
later in unfolding the matrix-element branching history. If QED splitting functions 
are included into the parton-shower, the correct method is obtained immediately, 
cf.\ also~\cite{Hoeche:2009xy}.

The above merging algorithm can also be employed to solve the problem outlined in 
Sec.~\ref{sec:shower}. That is, it can be used to fill the complete available real emission
phase space for any given $Q^2$. A similar solution is in fact adopted in Drell-Yan 
lepton-pair production via $\gamma^*/Z$-exchange, where the separation cut $Q_{\rm cut}$ between 
matrix-element and parton-shower domain is set such that $Q_{\rm cut}<m_{ll'}$, with $m_{ll'}$
being the invariant mass of the lepton pair. This situation is particularly simple, since an
experimental cut is usually applied, which enforces $m_{ll'}\approx m_{Z}$.
Therefore $Q_{\rm cut}$ can remain constant at $Q_{\rm cut}=S_{\,\rm DY}\,m_{Z}$,
where $S_{\,\rm DY}$ is an in principle arbitrary constant with $0<S_{\,\rm DY}<1$. 
Of course, $S_{\,\rm DY}$ must be chosen sensibly,
such as not to drive $Q_{\rm cut}$ into the non-perturbative domain. Also, $S_{\,\rm DY}$ should not be 
too close to one, since the proper description of particle spectra in this region largely depends on 
the recoil strategy employed in the shower. In practice, we have $0.1\lesssim S_{\,\rm DY}\lesssim 0.5$.
In deep-inelastic-scattering the situation is slightly different due to the variable value
of $Q^2$. The solution can, however, be identical. We choose
\begin{equation}\label{eq:sliding_qcut}
  Q_{\rm cut}\,=\;\bar{Q}_{\rm cut}\,\sbr{\;1+\frac{\bar{Q}_{\rm cut}^2/S_{\,\rm DIS}^{\,2}}{Q^2}\;}^{-1/2}\;,
\end{equation}
where $\bar{Q}_{\rm cut}$ is a fixed value, much like $Q_{\rm cut}$ in the Drell-Yan pair production case.
It ensures that high-$Q^2$, medium-$E_{T,B}^2$ events are described by matrix elements, rather 
than by the parton shower. At the same time, the factor in the square bracket, including 
$S_{\,\rm DIS}<1$, enforces low-$Q^2$, high-$E_{T,B}^2$ events to be in the matrix-element domain as well,
such that the complete available real-emission phase space can be filled by the Monte-Carlo simulation. Note that, 
contrary to the large freedom in the choice of $\bar{Q}_{\rm cut}$, we are rather limited in the choice
of $S_{DIS}$. Most analyses of deep-inelastic scattering data employ a cut on the photon virtuality which
is of the order of a few GeV$^2$. The Monte-Carlo simulation, however, is bound to have $Q_{\rm cut}$ 
in the perturbative domain with some difference between $Q_{\rm cut}^2$ and $Q^2$, as discussed above.
This introduces rather strict limits on the available range for $S_{\,\rm DIS}$. To be specific,
\begin{equation}\label{eq:sdisrange}
  0.4\lesssim S_{\,\rm DIS}\,\lesssim 0.8\;,
\end{equation}
where the lower bound depends on the experimental setup and the upper bound depends on the 
parton-shower model.

\begin{figure}[t]
  \begin{center}
  \subfloat[][]{\parbox{5.6cm}{
    \includegraphics[width=5.5cm]{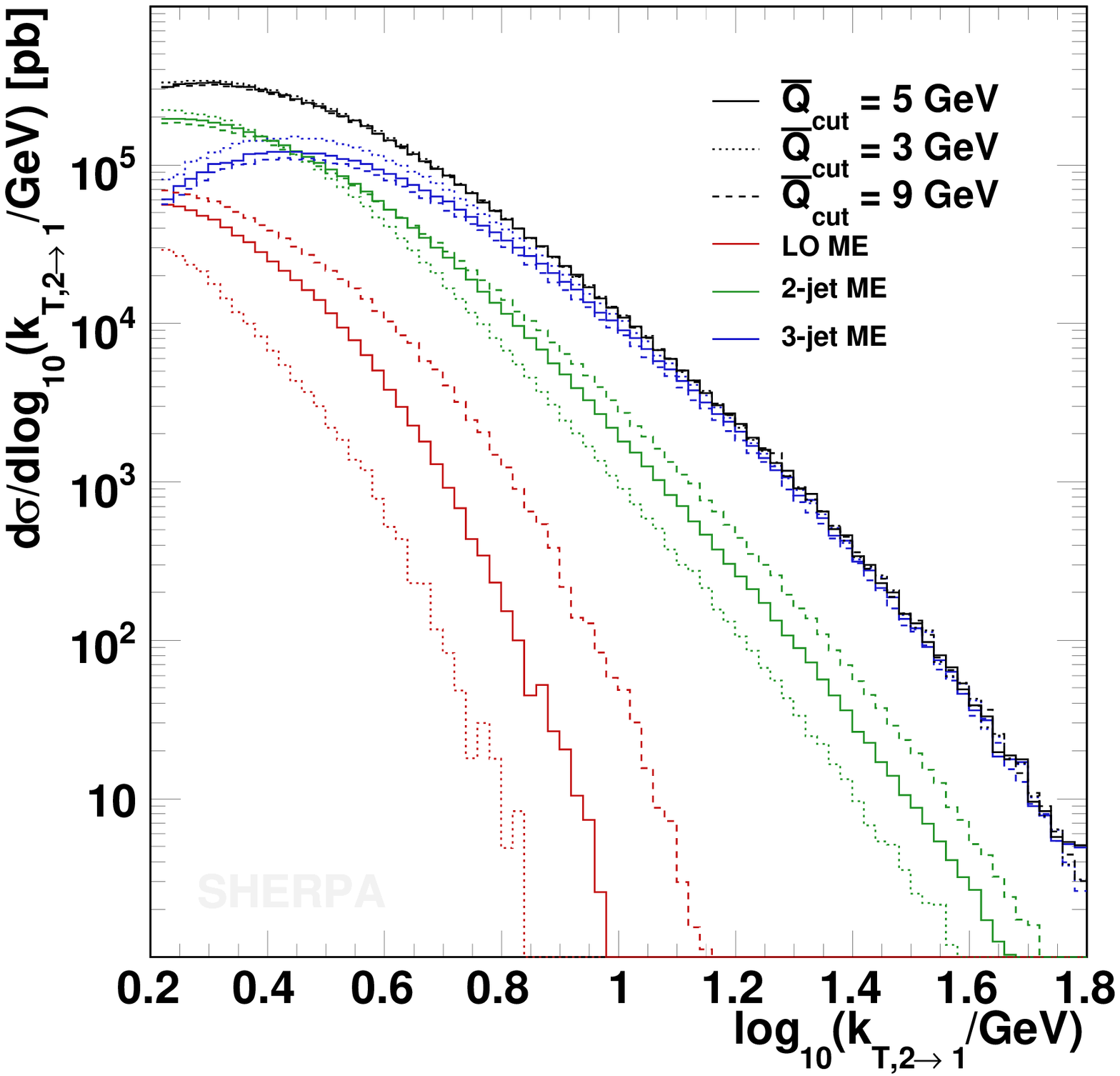}\\
    \includegraphics[width=5.5cm]{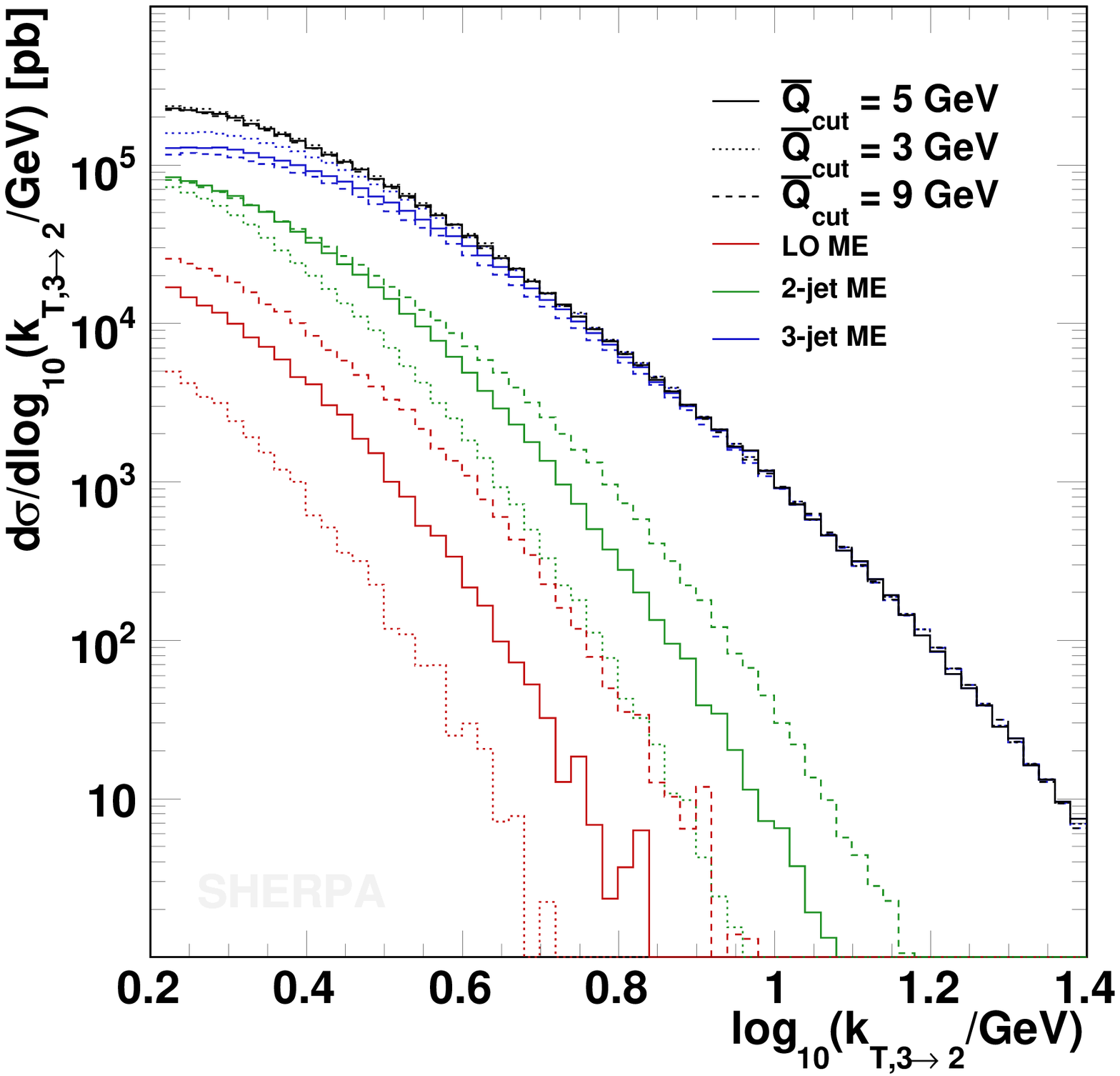}}
    \label{fig:merging_qbarcut}}\hspace*{-5mm}
  \subfloat[][]{\parbox{5.6cm}{
    \includegraphics[width=5.5cm]{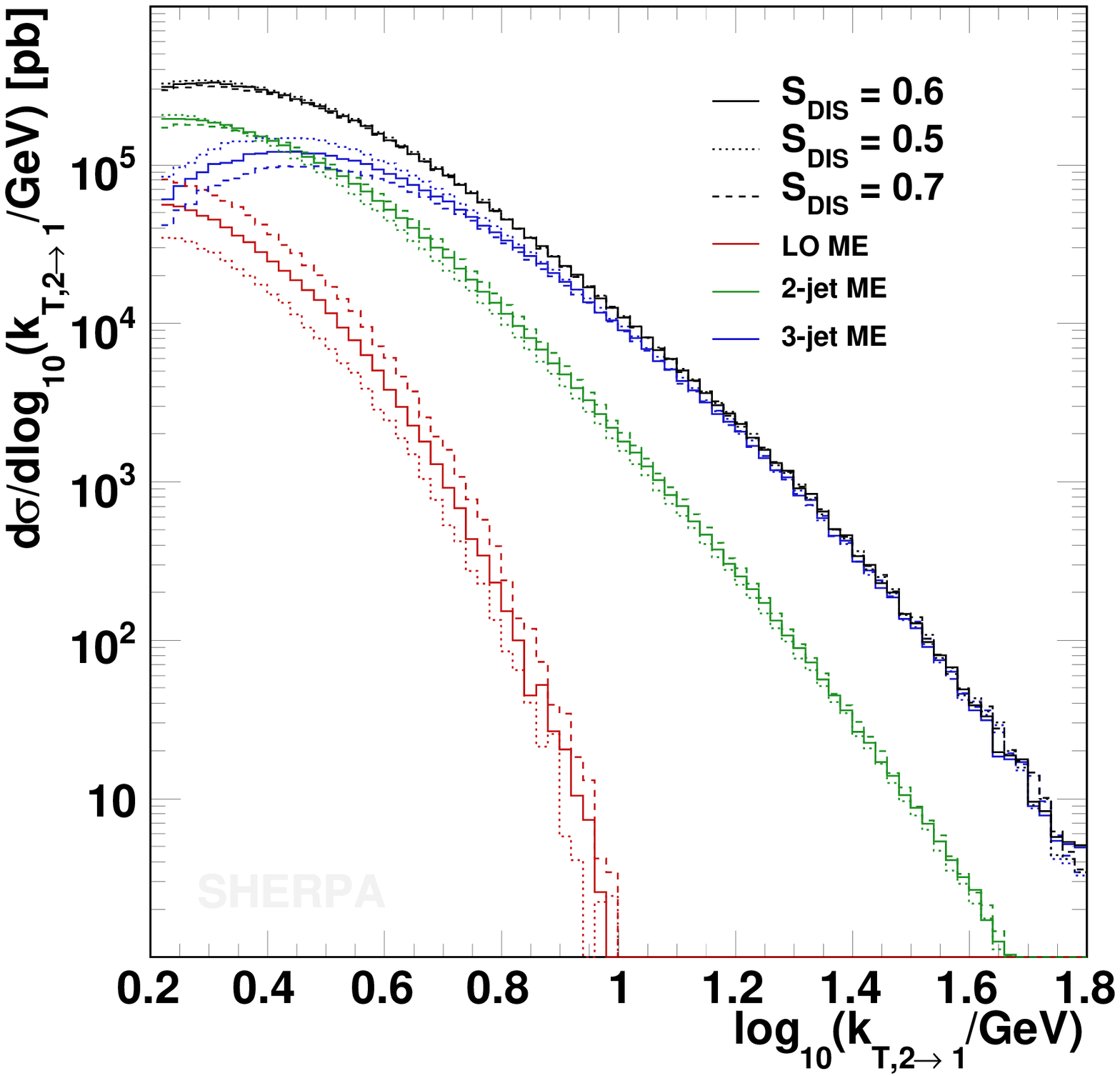}\\
    \includegraphics[width=5.5cm]{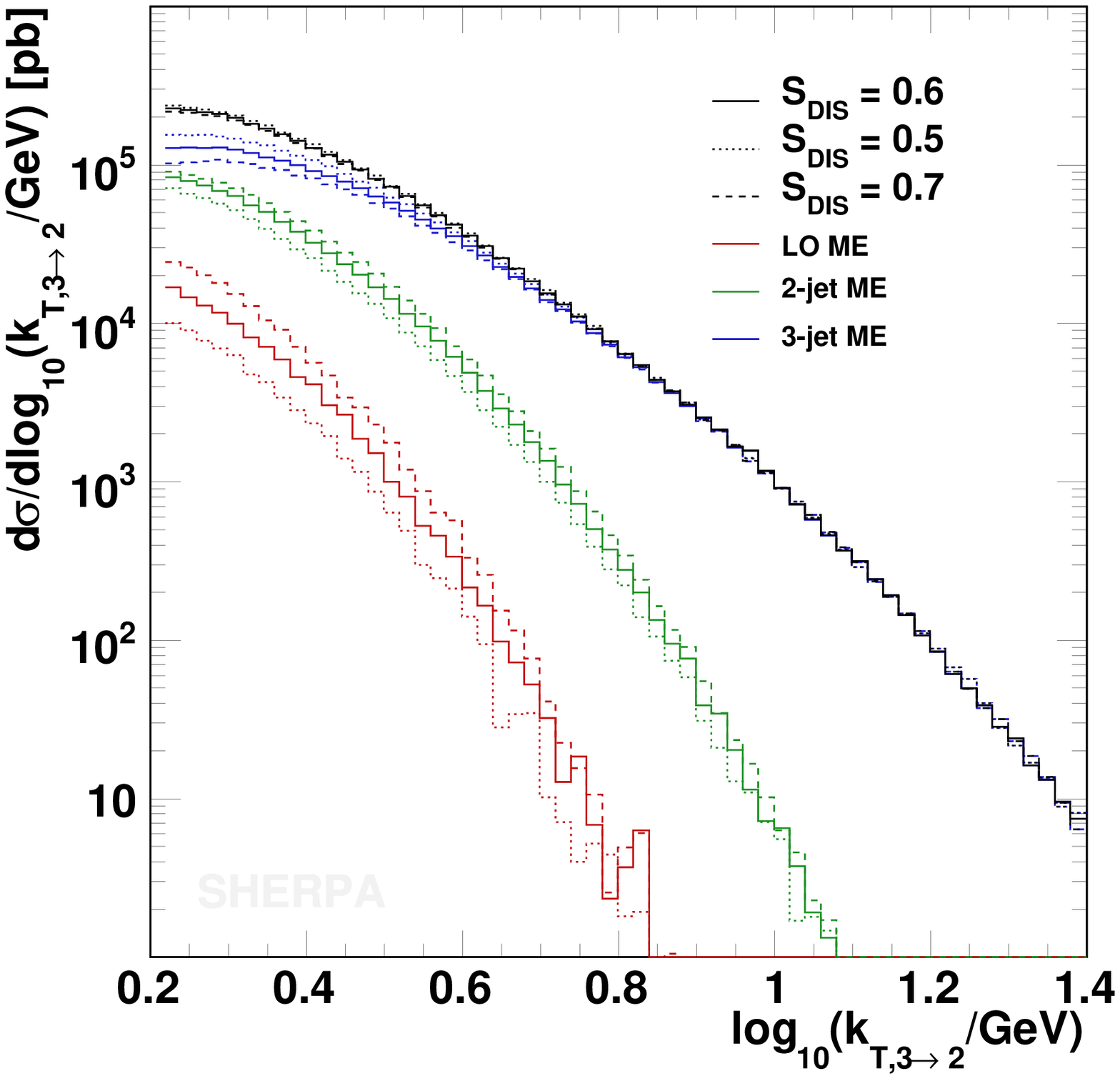}}
    \label{fig:merging_sdis}}\hspace*{-5mm}
  \subfloat[][]{\parbox{5.6cm}{
    \includegraphics[width=5.5cm]{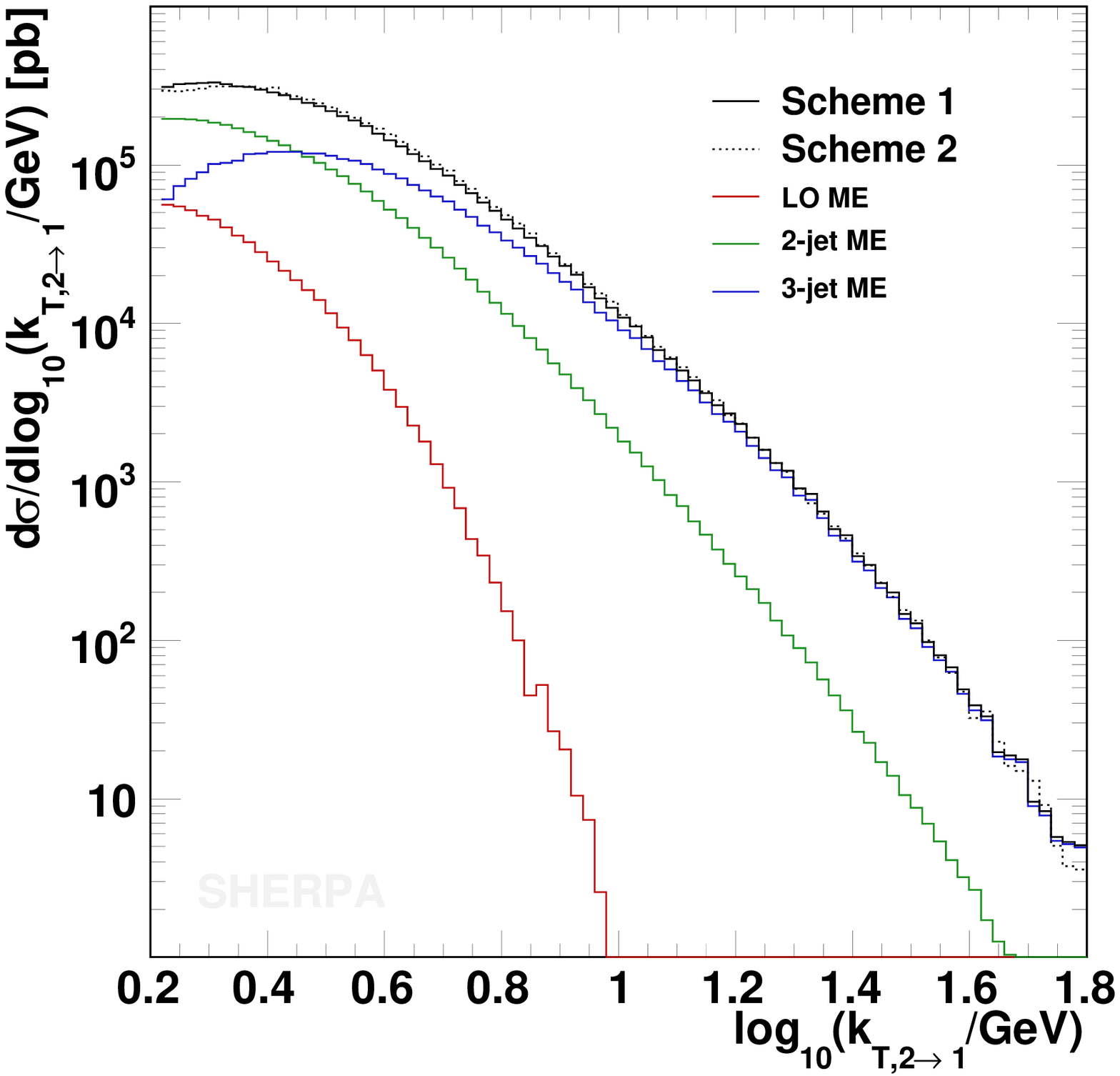}\\
    \includegraphics[width=5.5cm]{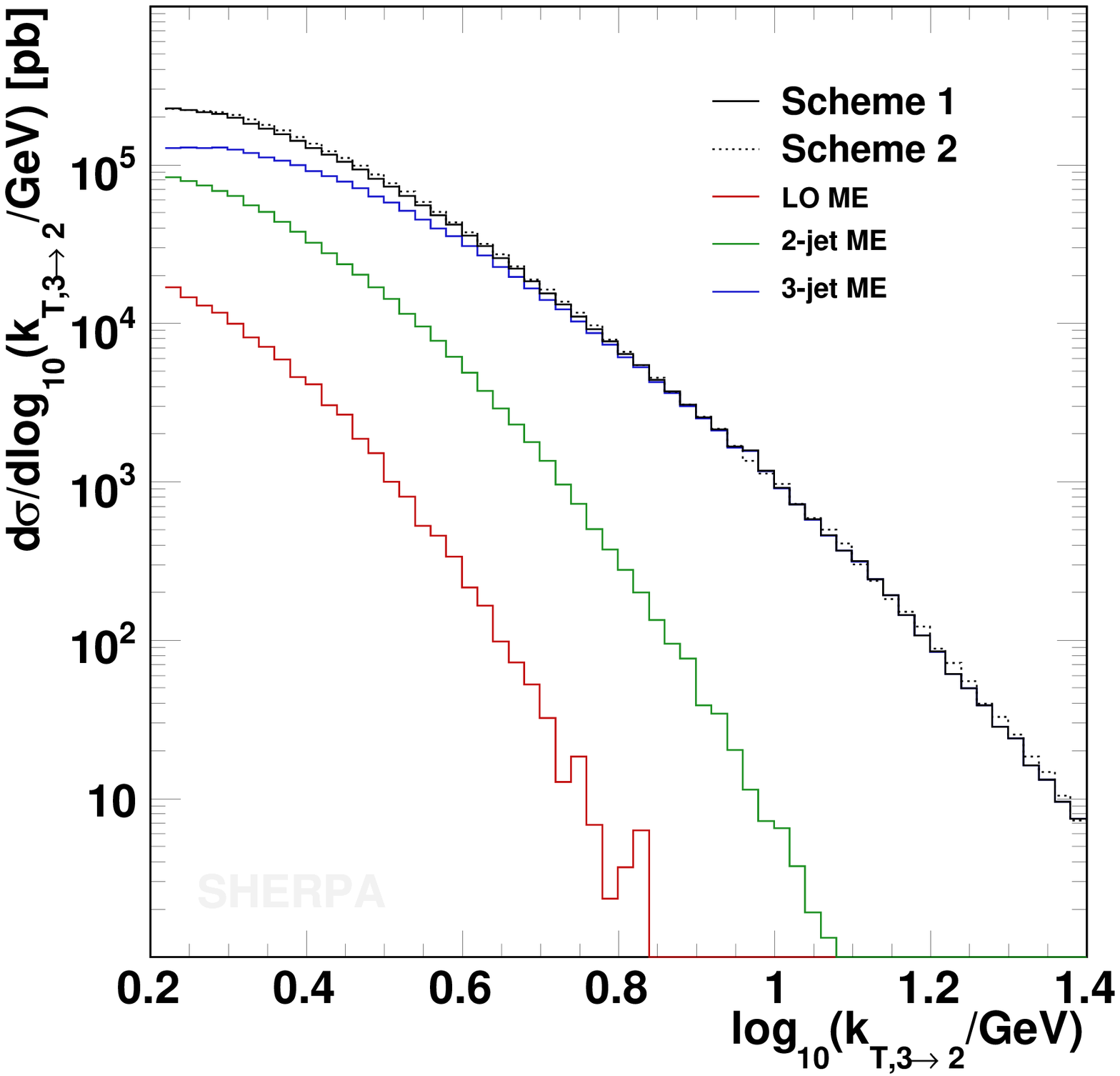}}
    \label{fig:merging_recoil}}
  \end{center}
  \caption{The differential 2- and 3-jet rates in merged event samples of varying 
  $\bar{Q}_{\rm cut}$~\protect\subref{fig:merging_qbarcut}, varying 
  $S_{\,\rm DIS}$~\protect\subref{fig:merging_sdis} and varying shower recoil 
  strategy~\protect\subref{fig:merging_recoil}. See also Fig.~\ref{fig:shower} for notation.
  Coloured lines display the contributions of different final state multiplicity 
  matrix elements. The central parameter value is chosen as $\bar{Q}_{\rm cut}=5$ GeV
  and $S_{\,\rm DIS}=0.6$. The maximum parton multiplicity in hard matrix elements is 
  $N_{\rm max}=3$.\label{fig:merging}}
\end{figure}

Figure~\ref{fig:merging} illustrates the effect of different $\bar{Q}_{\rm cut}$ and different
$S_{\,\rm DIS}$ on the prediction for the differential 2- and 3-jet rates in the Breit frame. 
The Monte-Carlo result remains very stable against corresponding variations. It can also be seen 
that when merging the parton shower with matrix elements, previous differences arising from 
different recoil strategies reduce considerably. This is essentially because the parton-shower
contribution to the observable is largely reduced, such that kinematical effects from shower
branchings have far less influence than in event samples without matrix-element merging.
\section{Comparison with experimental data}
\label{sec:results}

In this section, Monte-Carlo predictions, generated according to Secs.~\ref{sec:shower} 
and~\ref{sec:mets}, are confronted with hadronic final state data taken by the H1 experiment. 
The correct description of the selected measurements is quite challenging for the Monte Carlo 
traditionally used in the analysis of HERA data~\cite{Brook:1995nn,*Brook:1998jd}.
We seek to quantify the effect of varying perturbative input parameters 
and varying intrinsic parameters of the merging approach. 
We are mainly interested in the hard, perturbative domain, and therefore we choose 
to focus particularly on jet analyses. Monte-Carlo predictions stem from the \Sherpa 
program~\cite{Gleisberg:2003xi,*Gleisberg:2008ta}, which, in this context,
employs the matrix-element generator \Comix~\cite{Gleisberg:2008fv} to simulate 
the hard processes. Parton showers are implemented by the dipole-like cascade 
presented in~\cite{Schumann:2007mg}. Hadronisation is simulated either using 
the cluster fragmentation model of \Sherpa~\cite{Krauss:2010xy,*Winter:2003tt}, or the Lund string 
fragmentation model~\cite{Andersson:1983ia,*Andersson:1998tv} in the implementation 
of \Pythia~\cite{Sjostrand:2006za}. Both models were previously tuned to describe LEP data
employing the \Professor program~\cite{Buckley:2009bj}.\footnote{We are indebted to
  Frank Krauss, Hendrik Hoeth and Eike von Seggern for making preliminary sets
  of tuning parameters available.}
Hadron decays are implemented by \Sherpa's internal hadron decay module~\cite{Krauss:2010xx}
or by \Pythia~\cite{Sjostrand:2006za}, depending on the hadronisation model employed.
Photon radiation is simulated by \Sherpa's internal YFS generator~\cite{Schonherr:2008av}.
All analyses are carried out using the HZTool library~\cite{Waugh:2006ip}.

If not stated otherwise, matrix elements with up to five QCD partons in the final state
are employed and the parameters of the matrix-element parton-shower merging according to
Sec.~\ref{sec:mets} are set to $S_{\,\rm DIS}=0.6$ and $\bar{Q}_{\,\rm cut}=5$ GeV.
The default PDF set is NNPDF~1.2~\cite{Ball:2008by} in the implementation with 100 replicas.
The perturbative order of the strong coupling and its value at the reference scale
$m_Z$ is always chosen in accordance with the PDF set.

\subsection{Inclusive jet analysis}
\label{sec:results_incljets}
Following the arguments of Sec.~\ref{sec:shower} and~\ref{sec:mets}, a crucial 
observable is given by the inclusive jet cross section, differential with respect to
$E_{T,B}^2/Q^2$, where $E_{T,B}$ is the jet transverse energy in the Breit frame. 
For $E_{T,B}^2/Q^2>1$ it probes a part of the phase space where 
leading order Monte-Carlo models without the inclusion of low-$x$ effects are bound 
to fail in their description of jet spectra. The question to be answered here is,
whether the incorporation of higher order tree-level matrix elements is sufficient
to improve on this deficiency and yield predictions which are consistent with
experimental data. Inclusive $E_{T,B}^2/Q^2$-spectra were measured for different 
ranges of the jet-pseudorapidity in the laboratory frame, $\eta_{lab}$, 
in the low-$Q^2$ domain $5<Q^2<100\,\rm GeV^2$ 
by the H1 collaboration~\cite{Adloff:2002ew}. In this analysis jets were defined using 
the inclusive $k_T$-algorithm~\cite{Ellis:1993tq,*Catani:1993hr} and were constrained 
to $E_{T,B}>5\,\rm GeV$ and the pseudorapidity range $-1<\eta_{lab}<2.8$. It was 
found that next-to-leading order QCD calculations can describe the data reasonably well,
while large differences between leading-order and next-to-leading order results 
have been observed, especially in the forward region $1.5<\eta_{lab}<2.8$.

Results from our analysis are compared to the H1 data in Figs.~\ref{fig:etq2} to~\ref{fig:etq2_3}.
Figure~\ref{fig:etq2_njet} shows that the Monte-Carlo prediction gradually 
improves with a growing number of final-state partons in the hard matrix elements.
This behaviour is expected. Including matrix elements of larger final-state
multiplicity corresponds to opening the full phase space for high-$E_{T,B}$ jet production.
If all possible channels are to be incorporated, matrix elements with at least three final state 
partons must be available (3-parton sample). Indeed we observe that if only matrix elements 
with up to two final-state partons are considered (2-parton sample), the Monte-Carlo
prediction is far off the data. While the 3-parton sample gives an improved description,
the data are described satisfactorily only by a 4-parton sample. Including one additional
emission in the matrix elements does not alter the results too much. This can be 
understood as an effect of the definition of the observable. Since {\rm inclusive}
jet spectra are investigated, the proper description of the kinematics of sub-leading 
jets can be as important as the simulation of the leading jet.

Theoretical uncertainties of the Monte-Carlo prediction are shown in 
Figs.~\ref{fig:etq2_qcut} to~\ref*{fig:etq2_pdferr}. The variation of
renormalisation and factorisation scales has the largest impact on our results. This is 
expected, since an improved leading order approach is employed, which does not allow 
to predict the total cross section correctly. Higher-order virtual corrections
are missing in this algorithm. However, the systematic inclusion of higher-order
real corrections allows the prediction of the jet spectra in arbitrary multi-jet 
topologies at leading order, a feature which is inherently not present in any 
fixed-order calculation. The uncertainties associated with a variation of the 
intrinsic parameters of the merging algorithm are small compared to the variation 
found when altering the renormalisation and factorisation scale. This is exemplified
in Fig.~\ref{fig:etq2_qcut}.

Figure~\ref{fig:etq2_pskin} shows that the uncertainty related to the parton shower 
recoil strategy is negligible. This is a direct consequence of the merging approach 
and has been discussed in Sec.~\ref{sec:mets}, cf.\ Fig.~\ref{fig:merging}. 
We also compare the influence of different PDF sets on the results of our analysis.
All those PDFs are based on next-to-leading order fits. 
Figure~\ref{fig:etq2_pdf} shows that the corresponding results are essentially 
compatible, while a slight preference can currently be given to the NNPDF~1.2 set,
cf.\ also Fig.~\ref{fig:q2_pdf}.

\subsection{Inclusive jet and di-jet analysis}
\label{sec:results_incltwojets}
It is interesting to investigate jet properties in some more detail.
The analysis presented in~\cite{Adloff:2000tq} covers a wider range of $Q^2$ 
and presents jet-$E_{T,B}$ spectra doubly differential in $Q^2$ and $E_{T,B}$.
Jet pseudorapidities ($\eta_{lab}$) and pseudorapidity differences 
($\eta'=\abs{\eta_{B,1}-\eta_{B,2}}/2$) were also analysed.
Next-to-leading order calculations turn out to be particularly important 
for the latter, with deviations from leading-order predictions being most pronounced 
in the region of large $\eta_{lab}$ of the forward jet.
The acceptance region of this measurement is $5<Q^2<15000\,\rm GeV^2$ and 
$-1<\eta_{lab}<2.5$. Jet transverse energies are subject to the cuts 
$E_{T,B\,1,2}>5\,\rm GeV$ and $E_{T,B\,1}+E_{T,B\,2}>17\,\rm GeV$.
The latter requirement is introduced to avoid $E_{T,B\,1}\approx E_{T,B\,2}$, which is
the region of the phase space where next-to-leading order corrections are unstable 
due to implicit restrictions on soft emissions~\cite{Frixione:1997ks}.

A good probe of the proper Monte-Carlo simulation of such events is the di-jet cross section
shown in Fig.~6 of~\cite{Adloff:2000tq}. While still a relatively inclusive quantity
compared to the double differential spectra in the rest of this analysis, it tests the 
proper behaviour of jet production with decreasing $Q^2$ and is therefore as important
as the jet-$E_{T,B}^2/Q^2$ spectra shown previously. It can be seen in Fig.~\ref{fig:q2_njet}
that a large number of partons simulated with hard matrix elements is needed to describe 
this observable properly. Once a good description is obtained, however, we are also capable of
predicting the double differential jet spectra in $E_{T,B}$ and $\eta'$, 
cf.\ Figs.~\ref{fig:etb_twojet} and~\ref{fig:eta_twojet}. The agreement is excellent over the complete
$Q^2$-range of the measurement, which implies that the merging approach is well capable 
to describe the dynamics of multi-jet final states, if the maximum multiplicity in
hard matrix elements is large enough. We show theoretical uncertainties associated 
with our predictions of the $Q^2$-spectrum in Figs.~\ref{fig:q2_qcut} to~\ref*{fig:q2_pdferr}.
The same comments as for the previous section apply. It can also be seen in 
Fig.~\ref{fig:q2_scale}, that a variation of renormalisation and factorisation scales
does not only result in a global $K$-factor, i.e.\ a redefinition of the total cross section.
Varying these scales can instead induce a distortion of jet- and particle spectra and
it is therefore important to assess the related uncertainties.

\subsection{Low-$x$ di-jet analysis}
\label{sec:results_lowxtwojets}
In DIS di-jet events, next-to-leading order corrections are especially large
when the two jets in the Breit frame have similar transverse energy~\cite{Frixione:1997ks}.
It is thus interesting to study an observable, which singles out the corresponding 
region of the phase space. A dedicated measurement which defines such an observable
was carried out for the low-$Q^2$, low-$x$ domain by the H1 collaboration~\cite{Aktas:2003ja}.
Jets were defined using the inclusive $k_T$-algorithm~\cite{Ellis:1993tq,*Catani:1993hr} 
and were constrained to $E_{T,B}>5\,\rm GeV$ and the pseudorapidity range $-1<\eta_{lab}<2.5$. 
Deep-inelastic scattering events were selected in the kinematic range $5\,\rm GeV^2\,<Q^2<100\,\rm GeV^2$ 
and $10^{-4}<x<10^{-2}$. A variable $\Delta$ was defined by the requirement 
$E^*_{T,\rm max}>E^*_{T\,\rm cut}+\Delta$, where $E^*_{T\,\rm cut}$ is the minimum 
jet transverse energy and $E^*_{T\,\rm max}$ is the transverse energy of the hardest jet.
All quantities marked with an asterisk are given in the hadronic centre-of-mass frame,
which is related to the Breit frame simply by a longitudinal boost~\cite{Aktas:2003ja}.
An observable $\Delta\eta^*$ was defined as the pseudorapidity difference between the two hardest jets 
in the event for a fixed value of $\Delta=2$ GeV. 

The results of our Monte-Carlo analysis are compared to the H1 data in 
Figs.~\ref{fig:delta_twojet} and~\ref{fig:deltaeta_twojet}. For the $\Delta$ spectra we present 
parton level predictions and hadron level results. It can be seen that the effect of hadronisation 
on this observable is rather large in the region of very low $x$. The fact that hadronisation corrections 
are not uniform over the phase space indicates the importance of multi-purpose Monte-Carlo
event generators. Full hadron-level events can easily be simulated by such programs. Also, the effect 
of different hadronisation models can be studied. We observe that the overall description 
of the data improves once hadronisation corrections are included. The fragmentation model
employed here is the cluster fragmentation of~\cite{Krauss:2010xy,*Winter:2003tt}.

\subsection{Three-jet analysis}
\label{sec:results_threejets}
The three-jet analysis presented in~\cite{Adloff:2001kg} allows to test perturbative 
QCD predictions for events including one additional hard jet. Two angles, $\theta_3$ 
and $\psi_3$, were introduced in~\cite{Abachi:1995zv}, which, together with the scaled 
energies of the jets, i.e.\ the jet energies w.r.t.\ the invariant mass of the three-jet system, 
can be used to parametrise the phase space 
of three-jet events. At the same time they exhibit some sensitivity to the 
correct simulation of the QCD dynamics, cf.~\cite{Abachi:1995zv,Adloff:2001kg}.
They are defined in the three-jet centre-of-mass frame, with $\theta_3$ being the
angle of the most energetic jet w.r.t.\ the proton beam direction and $\psi_3$
the angle between the plane containing the most energetic jet and the proton beam
and the plane containing all three jets.
The inclusive $k_T$-algorithm~\cite{Ellis:1993tq,*Catani:1993hr} was employed 
in~\cite{Adloff:2001kg} to define jets, which were then constrained to the region 
$E_{T,B}>5\,\rm GeV$ and $-1<\eta_{lab}<2.5$. Due to the construction of the H1
detector, the phase-space of the measurement is slightly different in the low-$Q^2$ 
($5\,\rm GeV^2\,<Q^2<100\,\rm GeV^2$) and in the high-$Q^2$ ($150\,\rm GeV^2\,<Q^2<5000\,\rm GeV^2$) analysis.
Details can be found in the original publication.

Figure~\ref{fig:threejet_obs} compares the results of our Monte-Carlo analysis with data.
The distribution of the angles $\theta_3$ and $\psi_3$ is relatively well described 
by the simulation. We also show the $Q^2$-distribution of the three-jet events,
where the comments of Sec.~\ref{sec:results_incltwojets} apply. 
We observe that the $Q^2$-spectrum is matched very well by the Monte-Carlo prediction, 
which again indicates the relevance of including high-multiplicity matrix elements
into the simulation.
A particularly useful observable to test the correct description of multi-jet rates,
which is not available in di-jet events, is the ratio of the three- over the two-jet 
cross section, $R_{32}$. This quantity is independent of the overall normalisation 
of the event sample and is therefore especially suited to validate the Monte-Carlo
models employed by leading-order event generators. We find satisfactory agreement 
with the corresponding data over the complete observed $Q^2$-range.

\subsection{Jet-shape analysis}
\label{sec:jet_shapes}
The analysis presented by the H1 collaboration in~\cite{Adloff:1998ni} investigates
shapes and sub-jet rates of jets defined using either the inclusive 
$k_T$-algorithm~\cite{Ellis:1993tq,*Catani:1993hr} or a 
cone algorithm~\cite{DelPozo:1993th,*Akers:1994wj}. 
While the jet shape $\Psi(r)$ receives sizeable contributions
from non-perturbative effects over the whole radial range of the jet, the sub-jet rate
becomes fairly independent of non-perturbative dynamics at large values of the resolution 
parameter~\cite{Adloff:1998ni}. It is therefore a useful observable for measuring 
the perturbative dynamics of jet final states and can be used in particular to validate 
our Monte-Carlo simulation of parton evolution.

We exemplify in Figs.~\ref{fig:jetshapes} and~\ref{fig:subjetrates} that these observables
are described satisfactorily by our Monte-Carlo approach. In fact, this can be expected 
once jet rates and event shapes are fitted in $e^+e^-$ experiments. In this respect, 
we present a simple but necessary cross-check on the universality of the parton shower and 
hadronisation algorithms, which tests the nontrivial extension from pure final-state 
parton evolution to combined final- and initial-state evolution.

\subsection{Energy-flow analysis}
\label{sec:energy_flow}
Energy flows are crucial observables to determine the properties of QCD final states
in the region where perturbative and non-perturbative effects are equally important.
It has been pointed out~\cite{Kuhlen:1995yv,*Edin:1995gi,*Edin:1996mw} that fragmentation might have as big an impact
on these observables as has the perturbative input from hard processes and parton showering.
This statement is supported by the observation that a large part of the spectra can be tuned
to described the data by varying fragmentation parameters only. However, the influence of 
the perturbative input, i.e.\ the distribution of partons in the phase space and their colour 
correlations after the termination of the parton cascade, cannot be neglected.
Transverse energy flows thus constitute an ideal observable to test the interplay between
the hard, perturbative event phase and the hadronisation phase in Monte Carlo programs.
The analysis presented by the H1 collaboration in~\cite{Adloff:1999ws} extended previous 
measurements to a larger $\eta$-range and higher $Q^2$, where the usage of a
forward calorimeter (PLUG) allowed the determination of data points at very low $\eta^*$,
the particle rapidity in the hadronic centre-of-mass frame. As pointed out in~\cite{Adloff:1999ws},
the analysis of the transverse energy flow in this frame of reference isolates
the physically interesting part of the distribution.

Figures~\ref{fig:eflow_lowq} and~\ref{fig:eflow_highq} compare our Monte Carlo predictions
to the H1 Data. We find very good agreement when employing the cluster fragmentation model
of~\cite{Krauss:2010xy,*Winter:2003tt}, while the Lund string fragmentation~\cite{Andersson:1983ia,*Andersson:1998tv} 
gives predictions, which are slightly off the data. It should be noted, that
a set of parameters can be found, with which the string fragmentation model 
gives better results for this particular observable.
The fact that these parameters do not match those for which the model has been tuned to LEP data
indicates the importance of a combined analysis when tuning
intrinsic parameters of Monte-Carlo event generators.

\subsection{Charged particle spectra analysis}
\label{sec:charged_spectra}
Due to the large dependence of the transverse energy flows on hadronisation effects, 
transverse momentum spectra and pseudorapidity spectra of charged particles have been 
measured additionally by the H1 collaboration in~\cite{Adloff:1996dy}. 
It was argued in~\cite{Kuhlen:1996et,*Kuhlen:1996yf} that these observables provide 
a more direct measure of parton dynamics through a strong correlation between
partons and final-state particles and might therefore be crucial to distinguish between 
DGLAP- and BFKL-driven parton evolution. The influence of hadronisation 
should be more pronounced in the low-$p_T$ region, while the high-$p_T$ tail of 
the distributions is more sensitive to perturbative effects.
Significant discrepancies have been observed in the high-$p_T$ domain between the data 
and predictions from DGLAP-based Monte-Carlo models. Deviations also occur in the 
particle flow for $p_T>1$ GeV tracks.

We observe similar effects when comparing our Monte-Carlo results with the H1 data.
Even though up to five-parton final states are included in our Monte-Carlo simulation,
Fig.~\ref{fig:charged_pt} shows discrepancies especially in the high-$p_T$ region.
The particle flow for tracks with transverse momentum larger than 1 GeV, shown in
Fig.~\ref{fig:charged_multi}, projects onto the critical part of the phase space.
We show results from two different Monte-Carlo setups, labeled ``Set~1'' and ``Set~2''.
While ``Set~1'' was produced using the cluster hadronisation model in combination with 
the NNPDF~1.2 PDF set ``Set~2'' displays predictions from the Lund string hadronisation
in combination with the CTEQ~6L1 PDF set. We observed that the particle flow can be 
described satisfactorily by ``Set~2''. The corresponding results for other
observables are, however, not matching the data. Using this parameterisation, for instance, 
transverse energy flows can not be described satisfactorily. Hence, at present, there is 
no agreement with data for these observables. This finding highlights the importance 
of including HERA data into the global tuning of hadronisation parameters. 

\subsection{Charged multiplicity analysis}
\label{sec:multiplicities}
Multiplicity distributions are one of the basic observables in hadronic final states.
Much like the transverse energy flows, they allow a validation of the interplay between
perturbative and nonperturbative parts of a Monte-Carlo simulation of detector events. 
The evolution of charged particle multiplicities with the hadronic centre-of-mass energy,
$W$, and their dependence on the allowed pseudorapidity range was studied by the 
H1 collaboration in~\cite{Aid:1996cb}.

We present a comparison between our Monte-Carlo results and the data 
in Fig.~\ref{fig:multi}. Good agreement over the complete $W$ and $\eta$ range is observed.

\section{Conclusions}
\label{sec:conclusions}

In this publication, we have extended the \Sherpa event-generation framework 
to describe hadronic final states in 
deep-inelastic lepton-nucleon scattering processes. 
\Sherpa is a modern event generator, which implements the 
merging of matrix-element based  event generation with a parton shower. 

The merging procedure relies on a backward clustering algorithm according to
an inverted parton shower, which determines a hard core process 
at the origin of the matrix element or the parton shower. For a fully 
inclusive matrix-element parton-shower merging, the 
clustering must be performed on all outgoing particles.
Depending on the final state kinematics, characterised by the photon virtuality 
$Q^2$ and the transverse momenta of final state jets, 
the core process in deep-inelastic scattering is then
either found as electron-quark scattering, photon-quark scattering or a partonic 
$2\to 2$ scattering process.  

To account for the kinematical situation in deep-inelastic scattering, the merging procedure 
in \Sherpa was refined,  taking into account that most hadronic final states 
are characterised by several hard scales, not only by $Q^2$. By choosing 
appropriate merging scales, a successful
description of processes in all kinematical regions (including the low-$x$ 
region, and including high-$E_{T,B}^2$, low-$Q^2$ processes) could be obtained
in a theoretically consistent manner, consistent with factorisation. To reduce 
the merging uncertainty, modifications to the parton shower kernels were made.
Finite terms were added to the previously used dipole kernel to ensure 
that the kernels amount to the full matrix elements associated to the 
splitting process (like in 
antenna-based showers). With these refinements, \Sherpa is the first 
multi-purpose event generator program for deep-inelastic processes which 
incorporates a full merging of leading-order matrix elements with parton showers. 

We validated our results on a multitude of HERA data on hadronic final 
states in deep-inelastic scattering, including jet cross sections, 
jet-transition rates and hadronic particle spectra. All observables 
considered are described in a very satisfactory manner.
We quantified the uncertainties due to 
scale choices, merging parameters, parton-shower schemes, parton distribution 
functions and hadronisation models.

The comparison with HERA data provides an important validation of the 
\Sherpa initial- and final-state parton-shower schemes in
non-trivial kinematical situations hardly accessible in other collider 
environments. It has important consequences for LHC studies in similar 
kinematical situations (like low-mass Drell-Yan or vector-boson plus multi-jet
production). 

Using the HERA data set on different aspects of hadronic final states will 
allow for a validation and tuning of hadronisation models, 
which was based up to now purely on data from $e^+e^-$ annihilation. Inclusion 
of DIS data probes different flavour combinations, and will help to 
resolve parameter degeneracies, thereby leading to important improvements 
of the hadronisation models.

\section*{Acknowledgements}
We thank Hendrik Hoeth for providing a dedicated tune of the Lund string fragmentation model
in the implementation of Pyhtia 6.4.18 and Eike von Seggern for providing a tune of the cluster
fragmentation model in \Sherpa. We gratefully acknowledge discussions with Alberto Guffanti,
Frank Krauss, Steffen Schumann and Frank Siegert.

This work was funded in part by the Swiss National Science Foundation (SNF, contract 
number 200020-126691) and by the University of Z{\"u}rich (Forschungskredit number 57183003). 

\appendix
\section{Matrix element correction of the splitting kernels}
\label{sec:mecorrection}

The original parton shower algorithm~\cite{Schumann:2007mg}, based on Catani-Seymour 
dipoles, can be modified easily to improve the radiation pattern in deep-inelastic 
scattering. The key idea is to add some nonsingular bits to the original
spin-averaged dipole functions, such that combining the radiation functions 
of emitter and spectator yields the exact NLO
real radiation matrix element. This correction 
does not spoil the logarithmic accuracy of the parton shower. However,
merging the shower with higher-order tree-level matrix elements
along the lines of Sec.~\ref{sec:mets} is then alleviated for the first
emission, because the radiation patterns are formally identical.\footnote{
  In fact the identity of radiation patterns largely depends on the
  recoil scheme of the parton shower, cf.~\cite{Kleiss:1986re,*Seymour:1994we}.
  In this context, we only refer to the processes $\gamma^*g\to q\bar{q}$
  and $\gamma^*q\to qg$, averaged over the virtual photon spin.}
We focus on massless partons.
\myfigure{t}{
  \subfloat[][]{\scalebox{0.9}{\begin{picture}(200,100)(0,0)
    \Line(100, 50)( 10, 20)
    \Line(100, 50)(160, 70)
    \LongArrow( 20,16)( 46, 25)
    \LongArrow( 54,46)( 34, 66)
    \LongArrow(125,51)(152, 60)
    \Line( 55, 35)( 20, 70)
    \Vertex( 55,35){2.5}
    \GCirc(100,50){10}{1}
    \put( 73, 25){$\widetilde{ai}$}
    \put(  5, 10){$a$}
    \put( 20, 75){$i$}
    \put(165, 60){$k$}
    \put( 83, 67){$\abr{{\bf V}^{ai,k}}$}
    \put(140, 45){$p_k$}
    \put( 32, 10){$p_a$}
    \put( 48, 60){$p_i$}
  \end{picture}}\label{fig:split_if}}
  \subfloat[][]{\scalebox{0.9}{\begin{picture}(200,100)(0,10)
    \Line(190, 80)(100, 50)
    \Line( 40, 30)(100, 50)
    \LongArrow( 53,27)( 79, 36)
    \LongArrow(138,74)(112,83)
    \LongArrow(158,62)(184, 71)
    \Line(145, 65)(100, 80)
    \Vertex(145,65){2.5}
    \GCirc(100,50){10}{1}
    \put(125, 45){$\widetilde{ij}$}
    \put( 35, 20){$a$}
    \put( 90, 80){$j$}
    \put(195, 70){$i$}
    \put( 53, 57){$\abr{{\bf V}^{ij,a}}$}
    \put(172, 55){$p_i$}
    \put( 66, 20){$p_a$}
    \put(130, 87){$p_j$}
  \end{picture}}\label{fig:split_fi}}
}{Schematic view of the splittings of an initial-state parton with a 
  final-state spectator and the splitting of a final-state parton with 
  an initial-state spectator.  The blob denotes the hard matrix element. 
  Incoming and outgoing lines label initial- and final-state partons, 
  respectively.\label{fig:split_if_fi}}

The situation in final-state parton splittings with initial-state spectator
is sketched in Fig.~\ref{fig:split_fi}. We employ the variables
\begin{align}
  \tilde{z}_i\,&=\;\frac{p_ip_a}{\rbr{p_i+p_j}p_a}
  &\text{and}&
  &x_{ij,a}\,=\;1-\frac{p_ip_j}{\rbr{p_i+p_j}p_a}\;.
\end{align}
The corresponding spin-averaged splitting functions are given 
in~\cite{Schumann:2007mg}. While $\abr{{\bf V}^a_{g^i g^j}}$
is left unchanged, we redefine, in the context of this work
\begin{equation}
  \begin{split}
  \abr{{\bf V}^a_{q_i g_j}}(\tilde{z}_i,x_{ij,a})&=
    C_F\,\cbr{\frac{2}{1-\tilde{z}_i+\rbr{1-x_{ij,a}}}-\rbr{1+\tilde{z}_i}
      +C^{\rm DIS}_{q_ig_j}(\tilde{z}_i,x_{ij,a})}\;,\\
  \abr{{\bf V}^a_{q_i q_j}}(\tilde{z}_i,x_{ij,a})&=
    T_R\,\cbr{\rbr{\vphantom{\sum}1-2\,\tilde{z}_i\rbr{1-\tilde{z}_i}}
      \rbr{1-\frac{C^{\rm DIS}_{q_iq_j}(x_{ij,a},\tilde{z}_i)}{
        2\,\rbr{1-\tilde{z}_{i}}}}
      +C^{\rm DIS}_{q_iq_j}(\tilde{z}_i,x_{ij,a})}\;.
  \end{split}
\end{equation}
These functions differ from the original evolution kernels
by the additional nonsingular factors
\begin{equation}\label{eq:def_cdis}
  \begin{split}
  C^{\rm DIS}_{q_ig_j}(\tilde{z}_i,x_{ij,a})\,&=\;
    \rbr{1-x_{ij,a}}\sbr{\vphantom{K^i_j}\,1+3\,x_{ij,a}\,\tilde{z}_i\,}\;,\\
  C^{\rm DIS}_{q_iq_j}(\tilde{z}_i,x_{ij,a})\,&=\;
    4\,\rbr{1-x_{ij,a}}\,\tilde{z}_i\rbr{1-\tilde{z}_i}\;.
  \end{split}
\end{equation}
As required, the corrections vanish in the soft and collinear limits
and the original evolution kernels remain.

An initial-state parton splitting with final-state spectator
is sketched in Fig.~\ref{fig:split_if}. We employ the variables
\begin{align}
  u_i\,&=\;\frac{p_ip_a}{\rbr{p_i+p_k}p_a}
  &\text{and}&
  &x_{ik,a}\,=\;1-\frac{p_ip_k}{\rbr{p_i+p_k}p_a}\;.
\end{align}
The corresponding spin-averaged splitting functions $\abr{\bf V}$ 
are presented in~\cite{Schumann:2007mg}. While
$\abr{{\bf V}_k^{q_a q_i}}$ and $\abr{{\bf V}_k^{g_a g_i}}$
are left unchanged, we redefine, in the context of this work
\begin{equation}
  \begin{split}
  \abr{{\bf V}_k^{q_a g_i}}(x_{ik,a},u_i)&=
    C_F\,\cbr{\frac{2}{1-x_{ik,a}+u_i}-\rbr{1+x_{ik,a}}
      +C^{\rm DIS}_{q_ag_i}(x_{ik,a},u_k)}\;,\\
  \abr{{\bf V}_k^{g_a q_i}}(x_{ik,a},u_i)&=
    T_R\,\cbr{\rbr{\vphantom{\sum}1-2\,x_{ik,a}\rbr{1-x_{ik,a}}}
      \rbr{1-\frac{C^{\rm DIS}_{q_{ai}q_i}(u_k,x_{ik,a})}{2\,x_{ik,a}}}
      +C^{\rm DIS}_{q_{ai}q_i}(x_{ik,a},u_k)}\;.
  \end{split}
\end{equation}

Using the above modifications, it can be shown that the combination
of appropriate splitting kernels indeed reproduces the complete
real-emission matrix elements for the processes $\gamma^* g\to q\bar{q}$
and $\gamma^* q\to qg$.

\bibliographystyle{bib/amsunsrt_mod}  
\bibliography{bib/journal}

\begin{figure}[p]
  \begin{center}
  \subfloat[][]{\includegraphics[width=10cm]{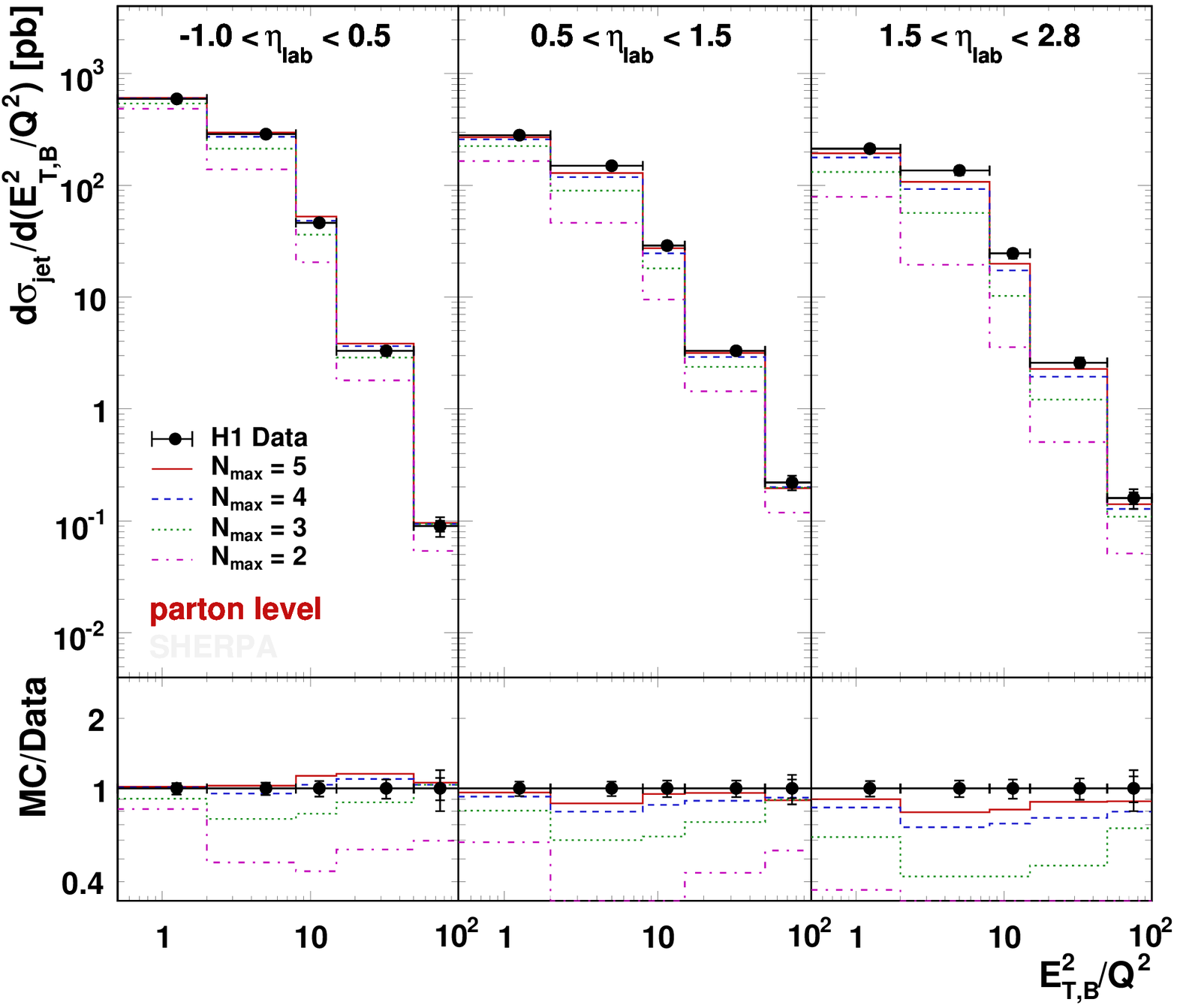}
    \label{fig:etq2_njet}}\\[-3mm]
  \subfloat[][]{\includegraphics[width=10cm]{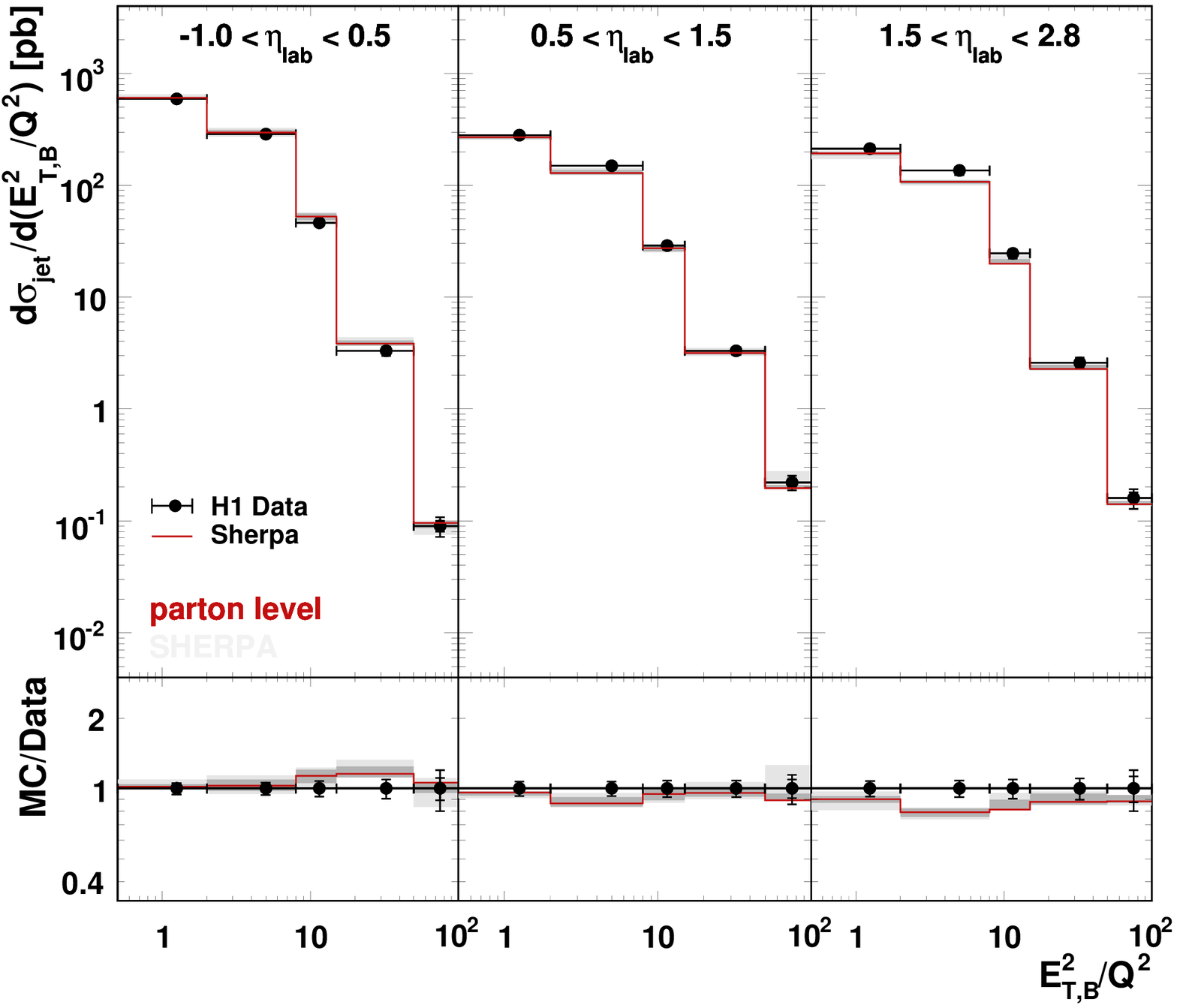}
    \label{fig:etq2_qcut}}
  \end{center}
  \caption{The inclusive jet cross section as a function of $E_{T,B}^2/Q^2$ in bins
  of $\eta_{lab}$, measured by the H1 Collaboration~\protect\cite{Adloff:2002ew}.
  $E_{T,B}^2$ is the jet transverse energy in the Breit frame, while $\eta_{lab}$
  denotes the jet rapidity in the laboratory frame. 
  Part~\protect\subref{fig:etq2_njet} displays the influence of the maximum parton multiplicity,
  $N_{\,\rm max}$, from hard matrix elements. We show the uncertainty originating from varying
  $S_{\,\rm DIS}$ between 0.5 and 0.7 (light grey band) and from varying $\bar{Q}_{\rm cut}$ 
  between 3 GeV and 9 GeV (dark grey band) in part~\protect\subref{fig:etq2_qcut}.}
  \label{fig:etq2}
\end{figure}
\begin{figure}[p]
  \begin{center}
  \subfloat[][]{\includegraphics[width=10cm]{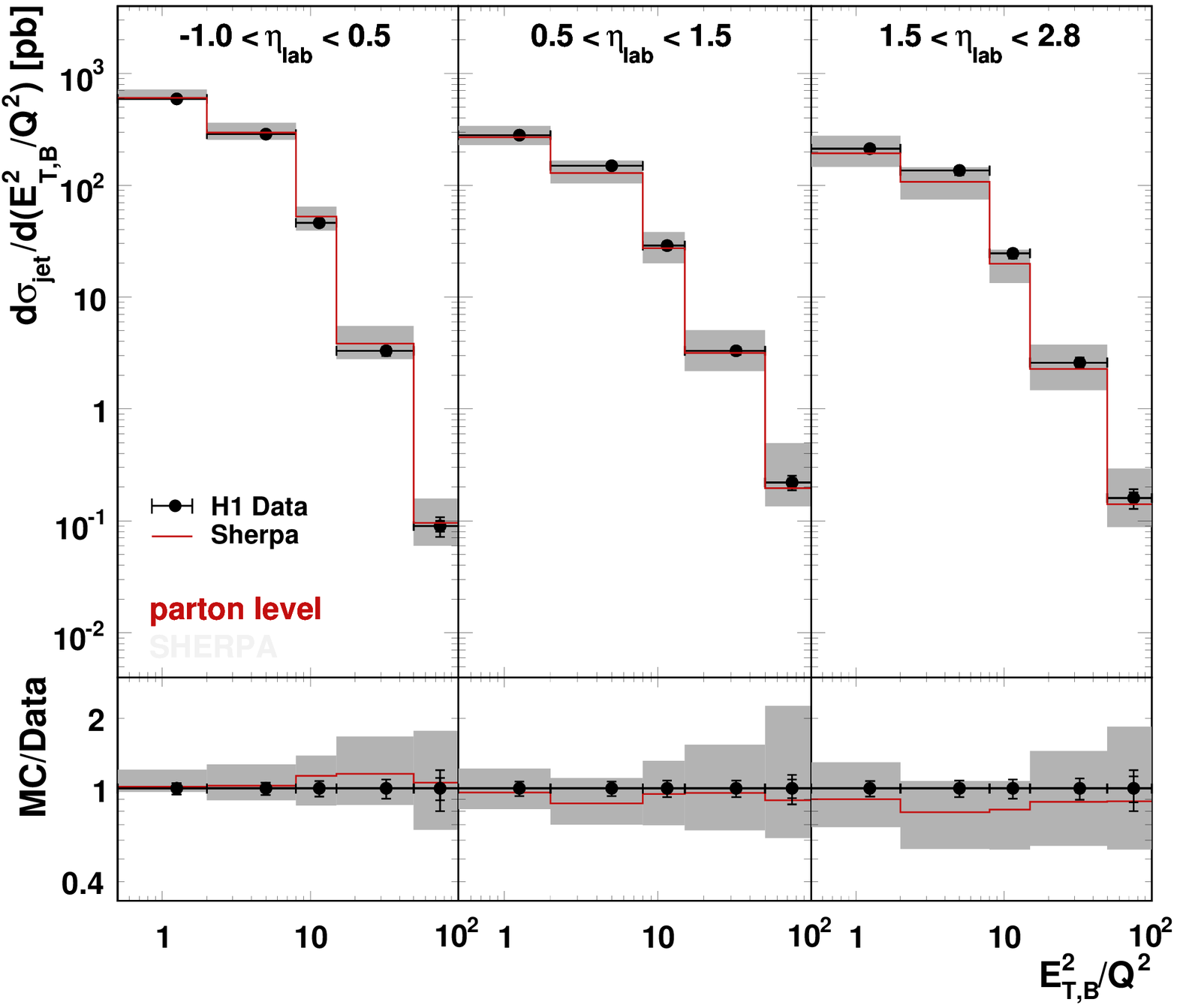}
    \label{fig:etq2_scale}}\\[-3mm]
  \subfloat[][]{\includegraphics[width=10cm]{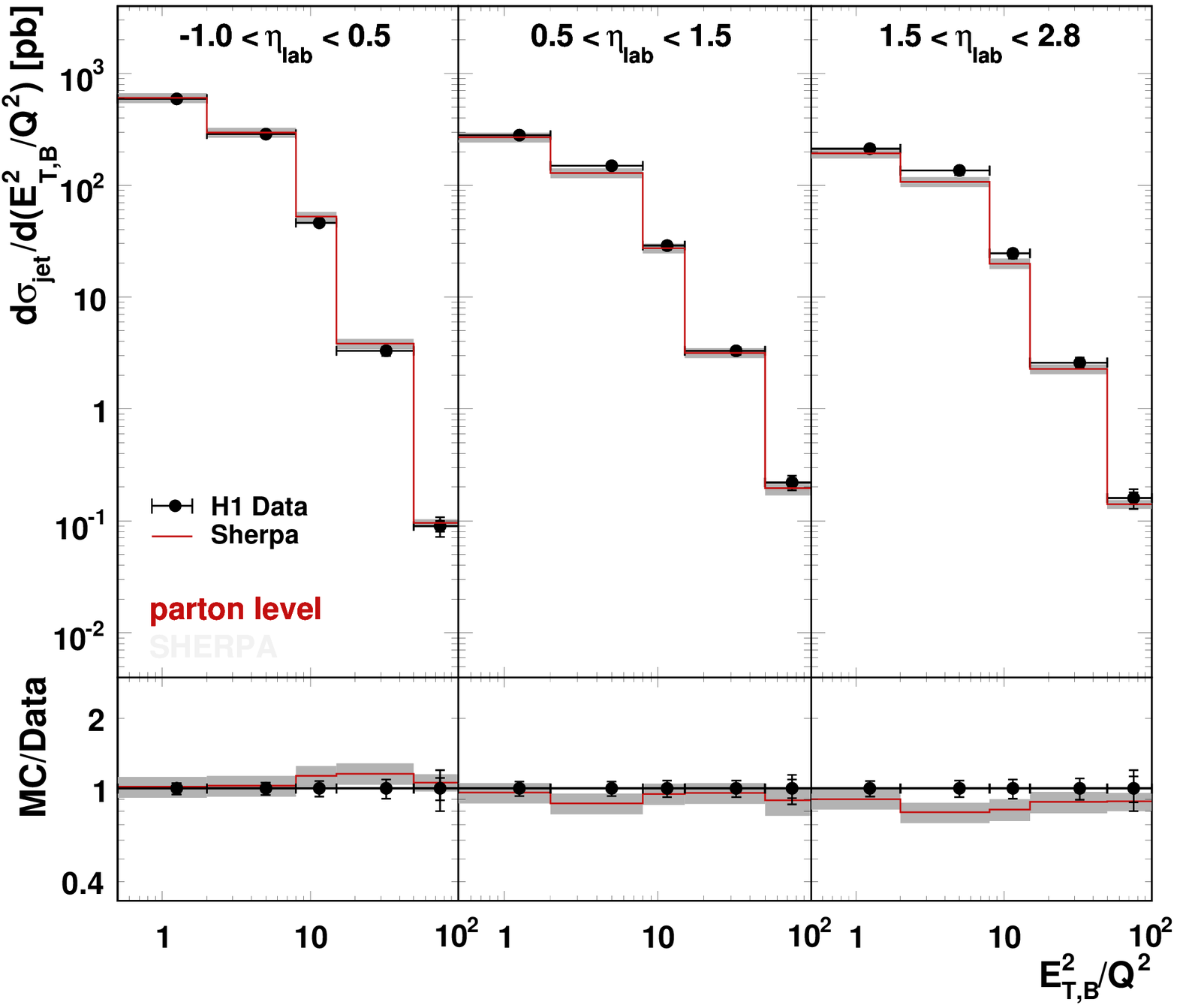}
    \label{fig:etq2_pdferr}}
  \end{center}
  \caption{The inclusive jet cross section as a function of $E_{T,B}^2/Q^2$ in bins
  of $\eta_{lab}$, measured by the H1 Collaboration~\protect\cite{Adloff:2002ew}, 
  cf.\ Fig.~\ref{fig:etq2}. Part~\protect\subref{fig:etq2_scale}
  presents the uncertainty associated with the variation of factorisation and renormalisation
  scales by factors of $1/2$ and $2$. Part~\protect\subref{fig:etq2_pdferr} shows the PDF uncertainty
  for the NNPDF~1.2 PDF set with 100 replicas~\protect\cite{Ball:2008by}.}
  \label{fig:etq2_2}
\end{figure}
\begin{figure}[p]
  \begin{center}
  \subfloat[][]{\includegraphics[width=10cm]{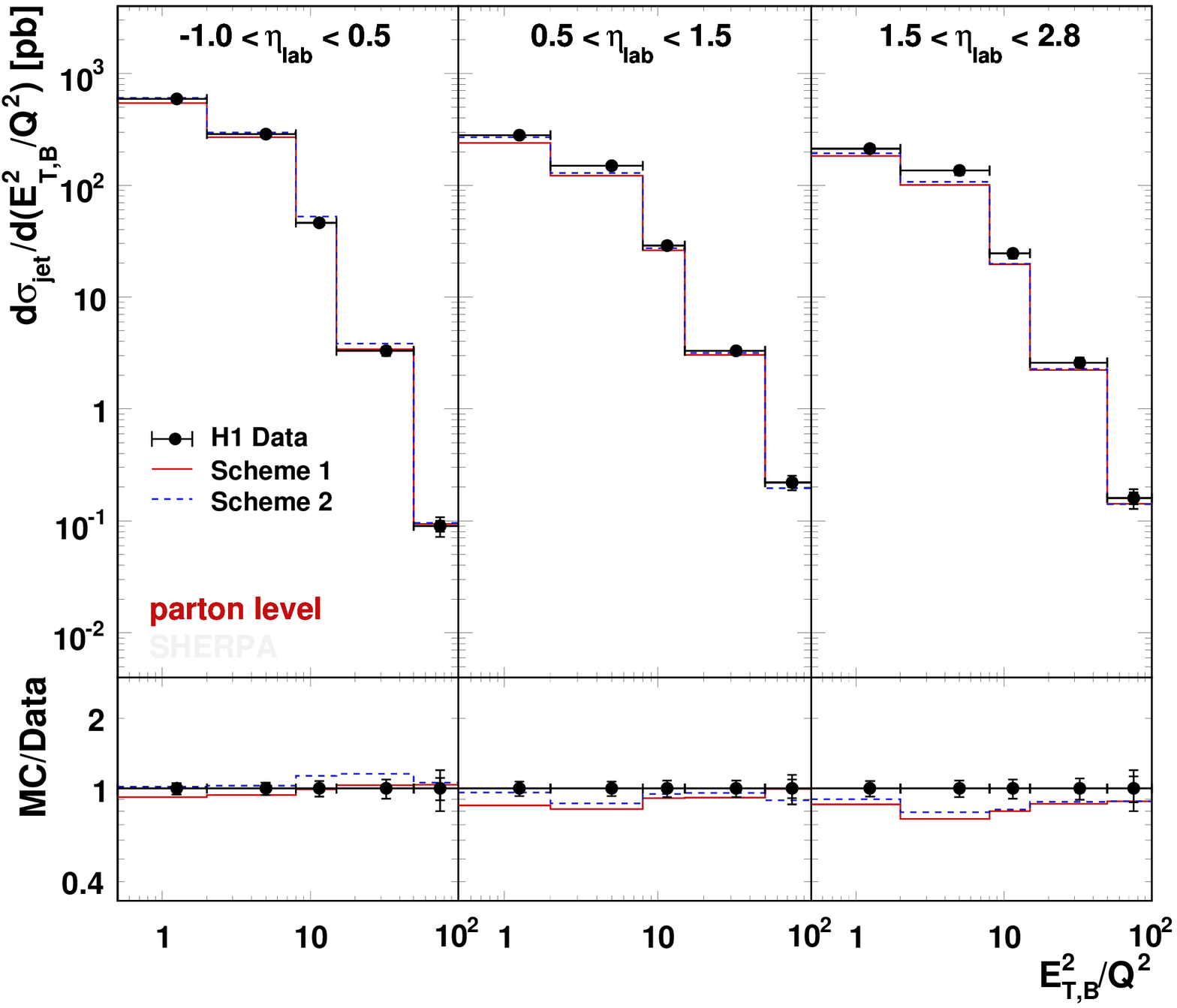}
    \label{fig:etq2_pskin}}\\[-3mm]
  \subfloat[][]{\includegraphics[width=10cm]{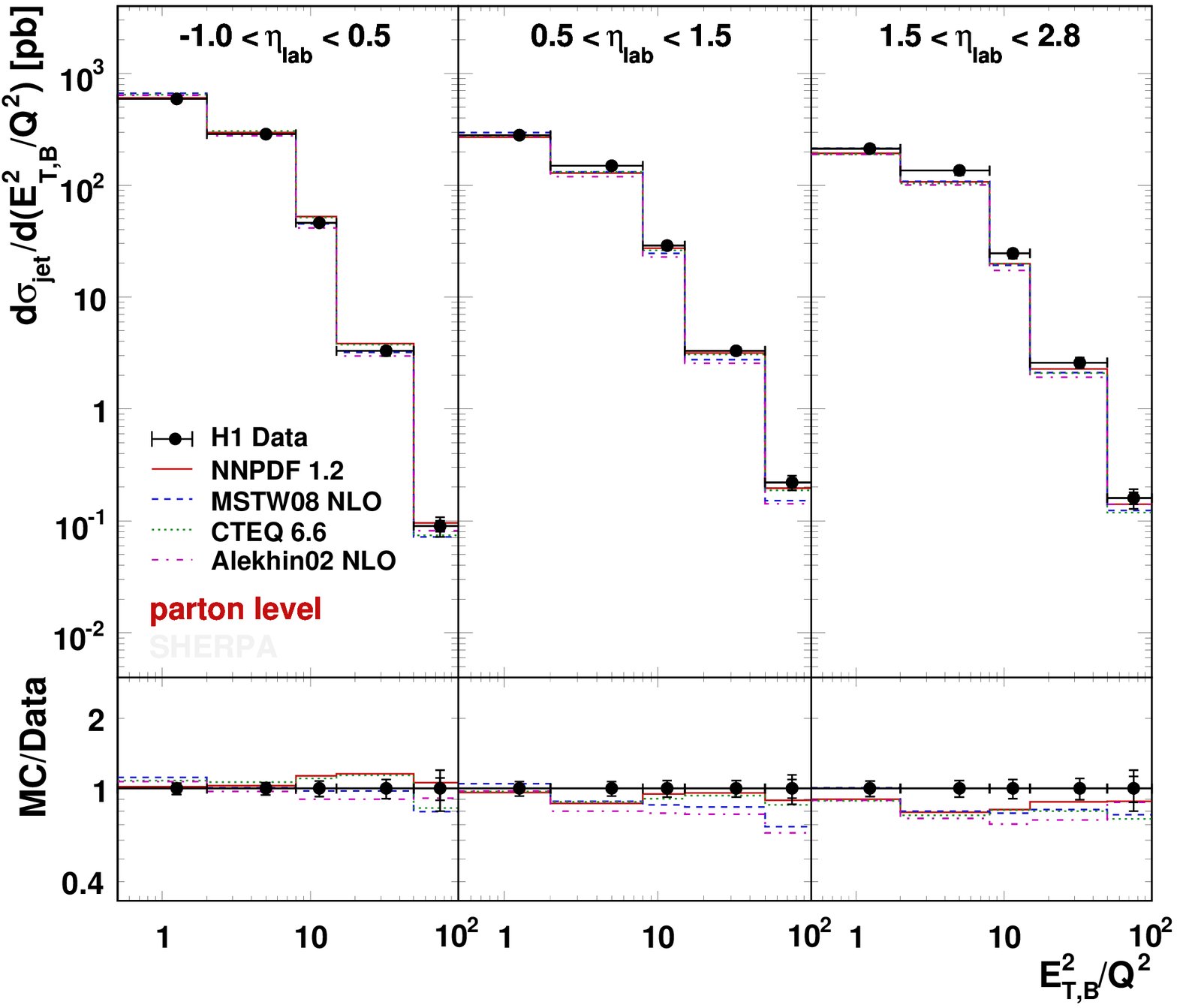}
    \label{fig:etq2_pdf}}
  \end{center}
  \caption{The inclusive jet cross section as a function of $E_{T,B}^2/Q^2$ in bins
  of $\eta_{lab}$, measured by the H1 Collaboration~\protect\cite{Adloff:2002ew}, cf.\ Fig.~\ref{fig:etq2}.
  Part~\protect\subref{fig:etq2_pskin} displays the results for the two different parton-shower
  recoil strategies discussed in Sec.~\ref{sec:shower}, cf.\ Fig.~\ref{fig:shower}.
  Part~\protect\subref{fig:etq2_pdf} shows results obtained with the NLO PDF sets
  NNPDF~1.2~\protect\cite{Ball:2008by}, MSTW~2008~\protect\cite{Martin:2009iq},  
  CTEQ~6.6~\protect\cite{Nadolsky:2008zw} and Alekhin~2002~\protect\cite{Alekhin:2002fv}.}
  \label{fig:etq2_3}
\end{figure}

\begin{figure}[p]
  \begin{center}
  \subfloat[][]{\includegraphics[width=7.75cm]{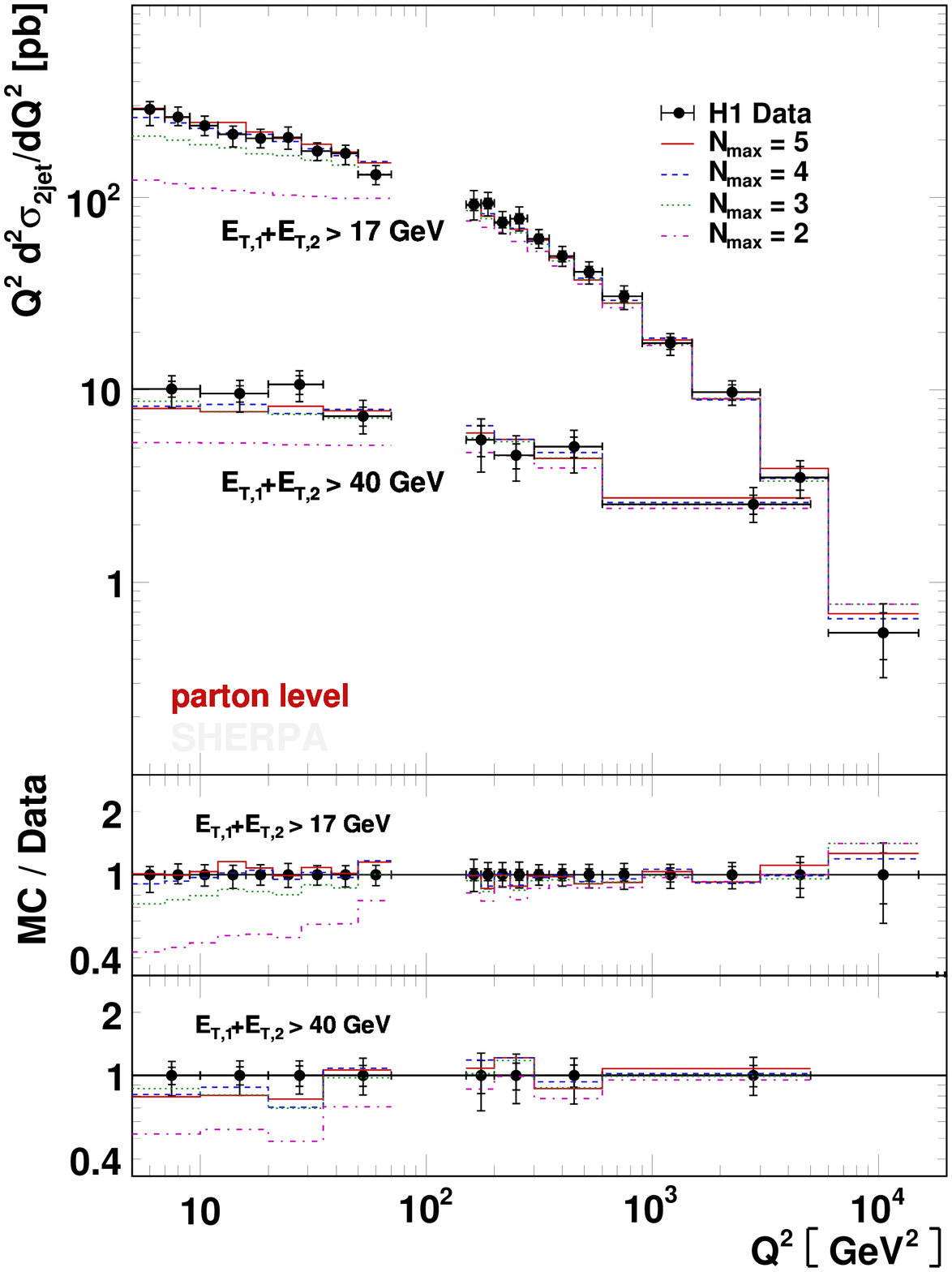}
    \label{fig:q2_njet}}\hspace*{2mm}
  \subfloat[][]{\includegraphics[width=7.75cm]{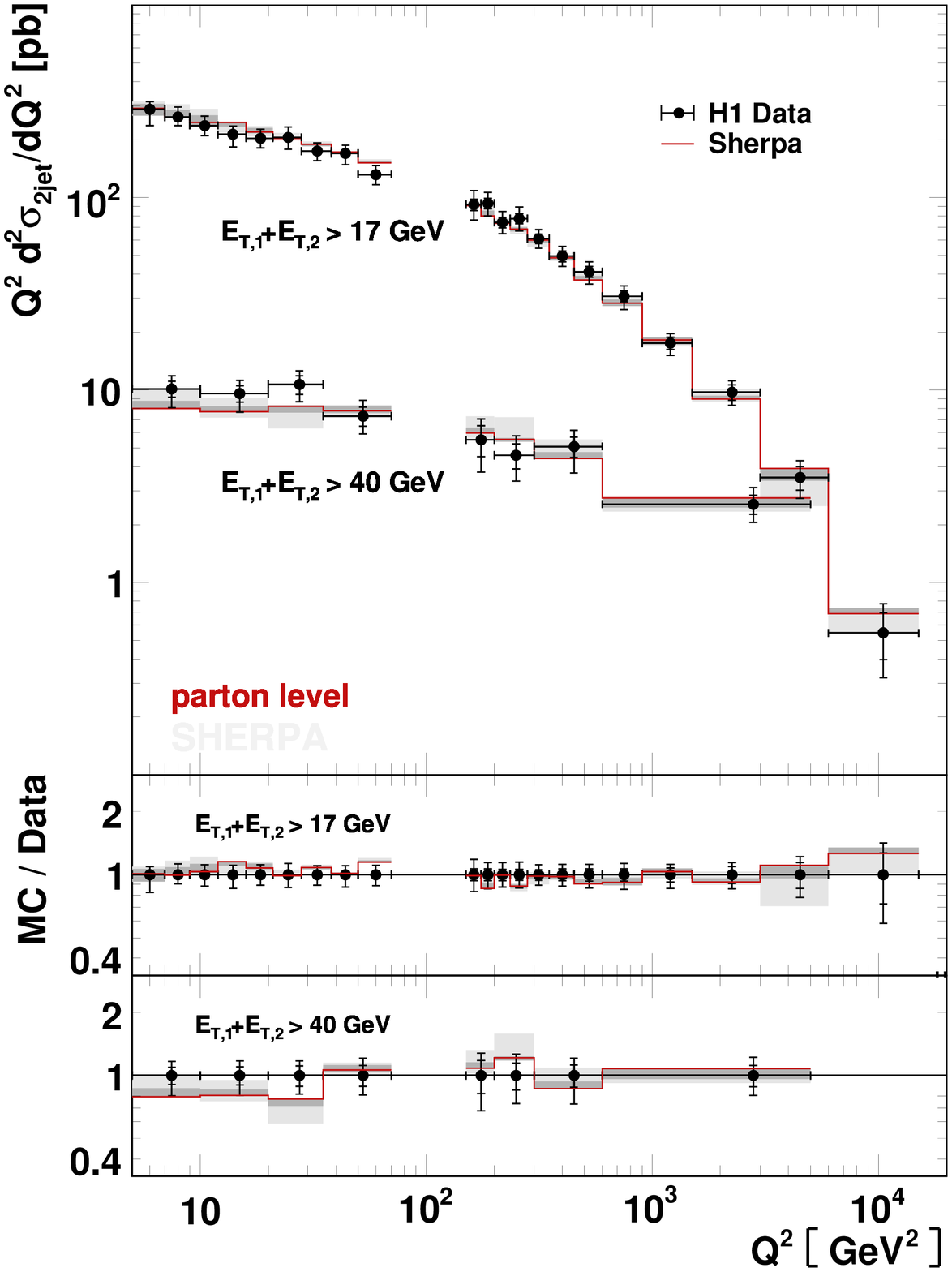}
    \label{fig:q2_qcut}}\\
  \subfloat[][]{\includegraphics[width=7.75cm]{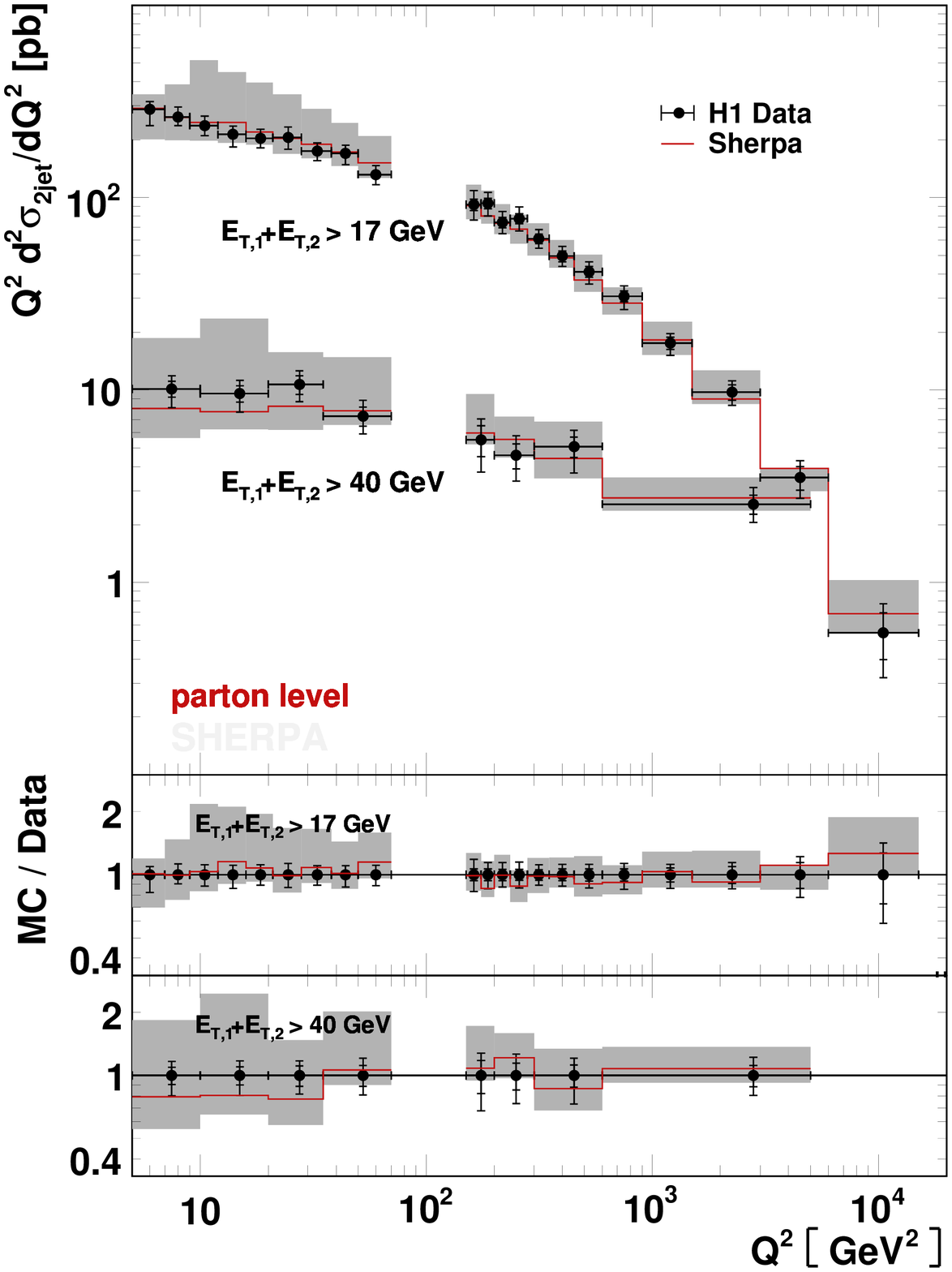}
    \label{fig:q2_scale}}\hspace*{2mm}
  \subfloat[][]{\includegraphics[width=7.75cm]{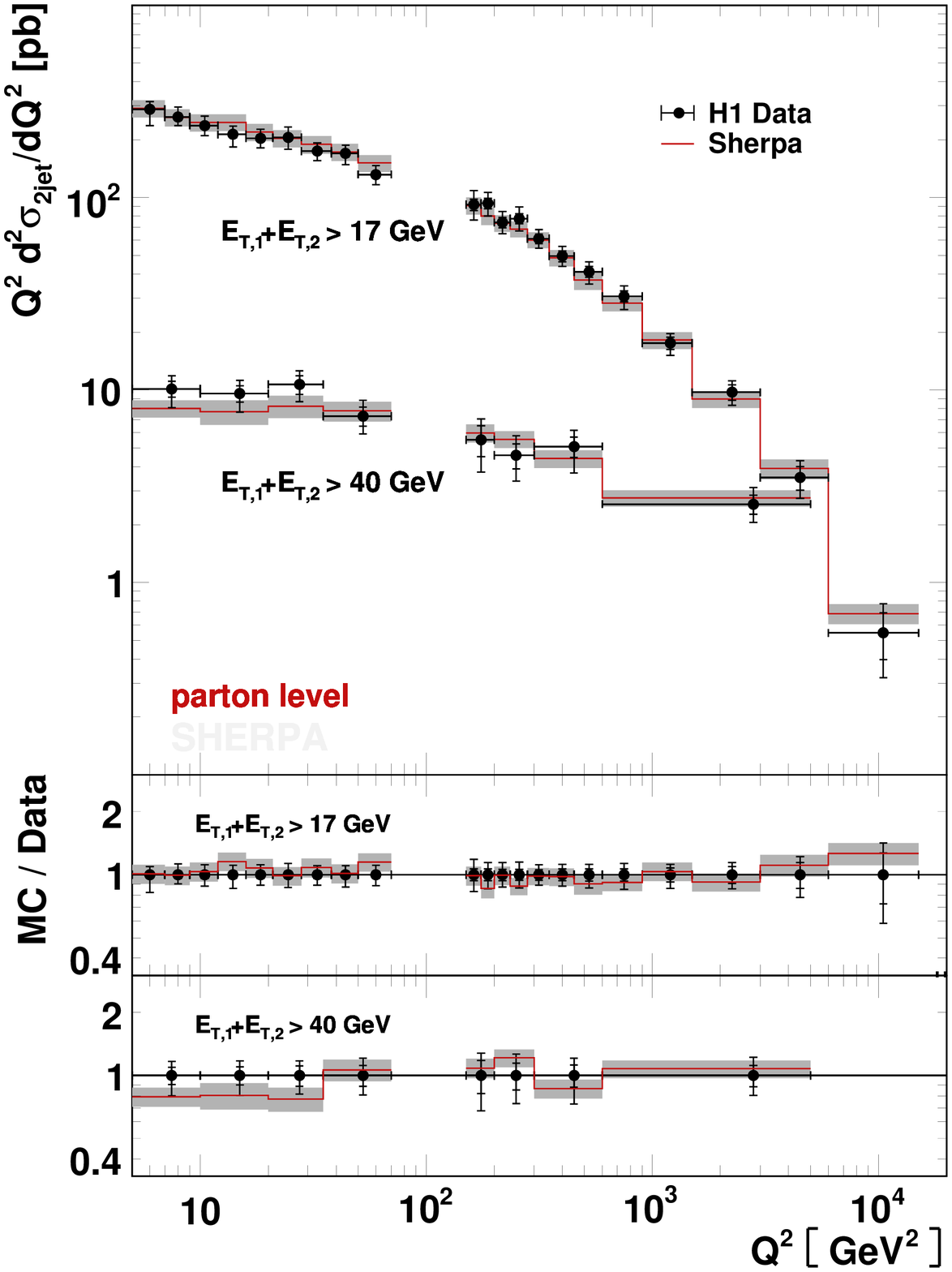}
    \label{fig:q2_pdferr}}
  \end{center}
  \caption{The di-jet cross section as a function of $Q^2$ in bins of $E_{T,1}+E_{T,2}$,
  measured by the H1 Collaboration~\protect\cite{Adloff:2000tq}.
  See Figs.~\ref{fig:etq2} and~\ref{fig:etq2_2} for an explanation of 
  parts~\protect\subref{fig:q2_njet} through~\protect\subref{fig:q2_pdferr}.}
  \label{fig:q2}
\end{figure}
\begin{figure}[p]
  \begin{center}
  \subfloat[][]{\includegraphics[width=7.75cm]{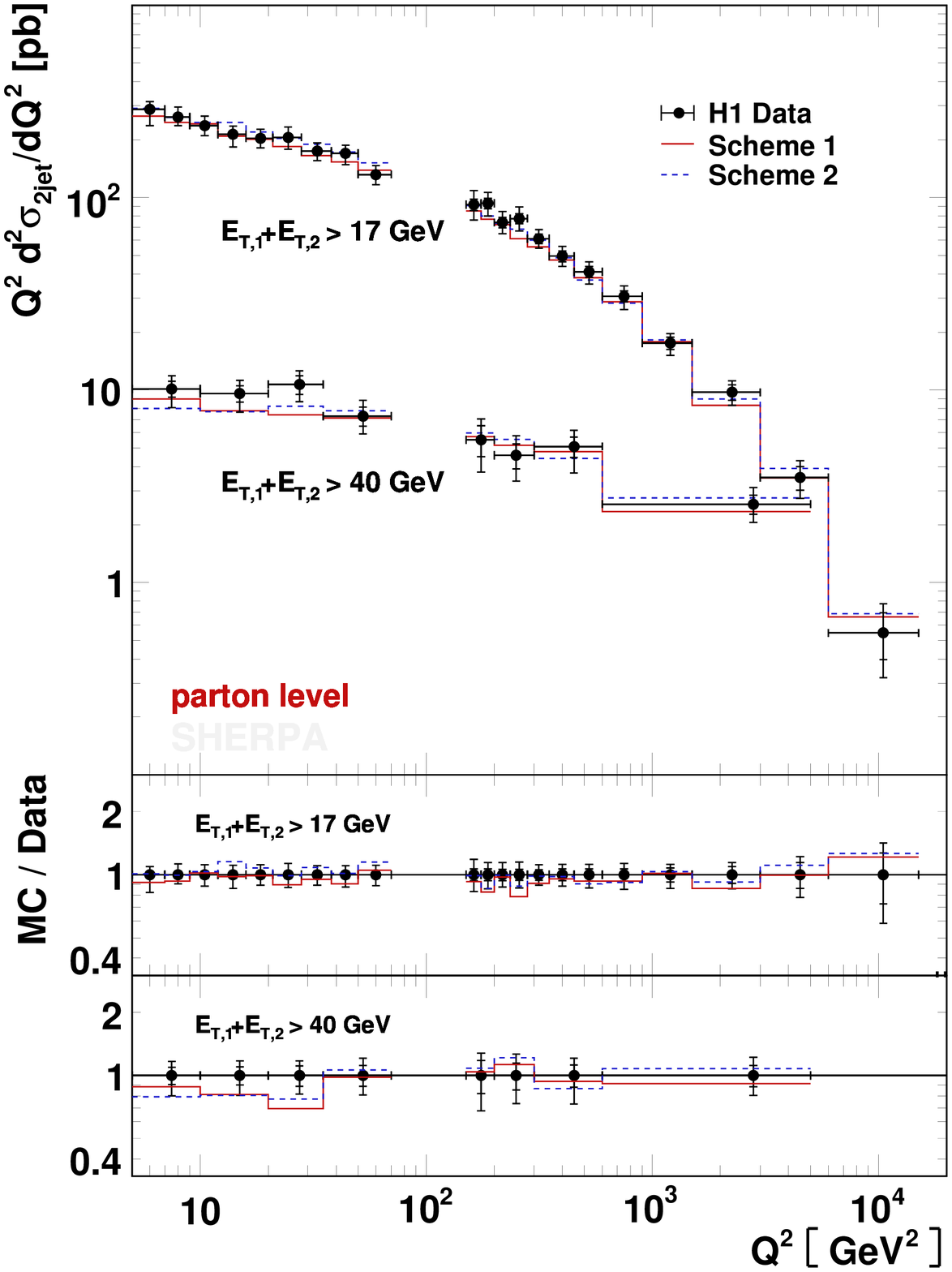}
    \label{fig:q2_pskin}}\hspace*{2mm}
  \subfloat[][]{\includegraphics[width=7.75cm]{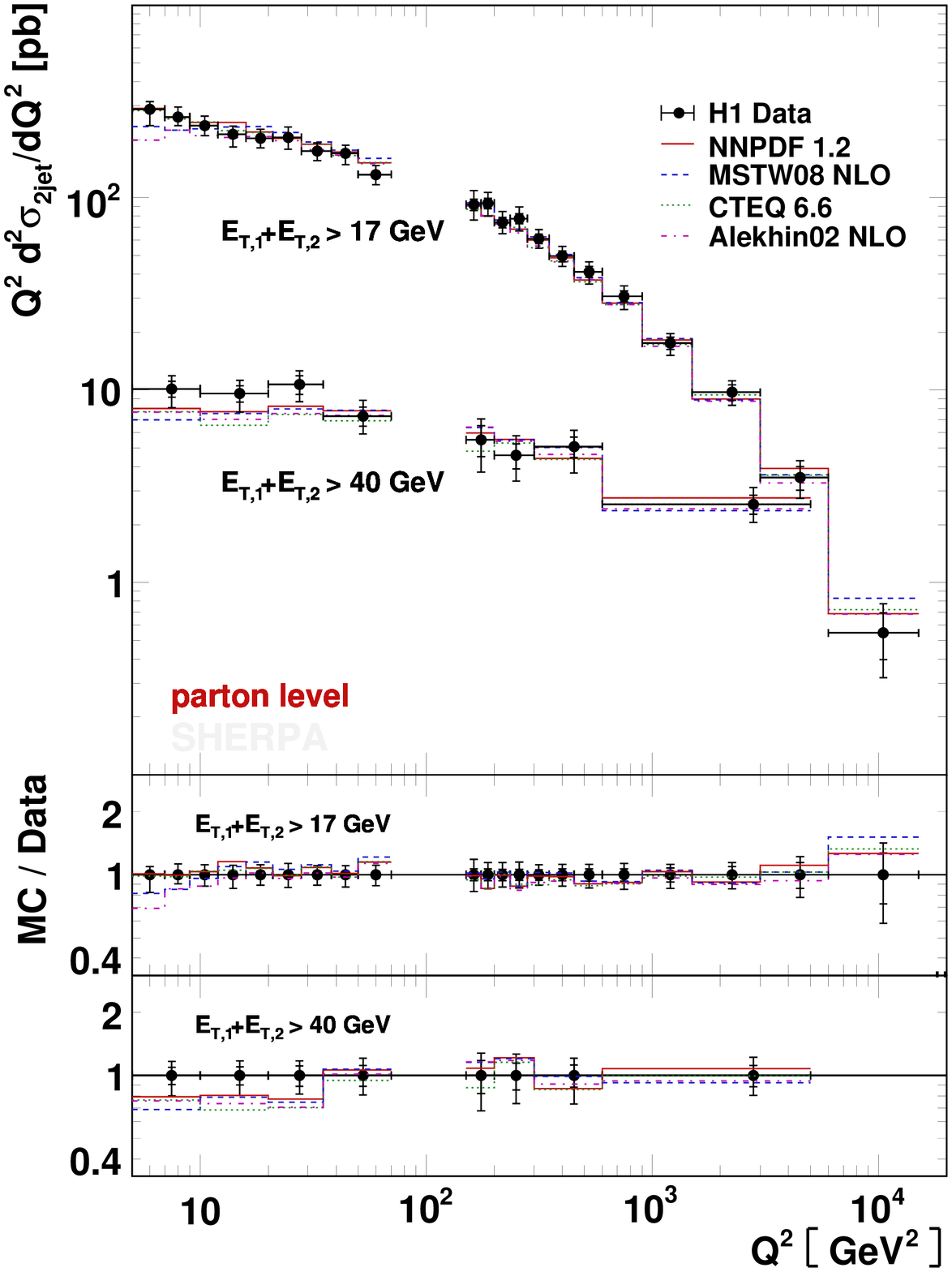}
    \label{fig:q2_pdf}}\\
  \end{center}
  \caption{The di-jet cross section as a function of $Q^2$ in bins of $E_{T,1}+E_{T,2}$,
  measured by the H1 Collaboration~\protect\cite{Adloff:2000tq}.
  See Fig.~\ref{fig:etq2_3} for an explanation of parts~\protect\subref{fig:q2_pskin} 
  and~\protect\subref{fig:q2_pdf}.}
  \label{fig:q2_2}
\end{figure}

\begin{figure}[p]
  \begin{center}
    \includegraphics[width=8cm]{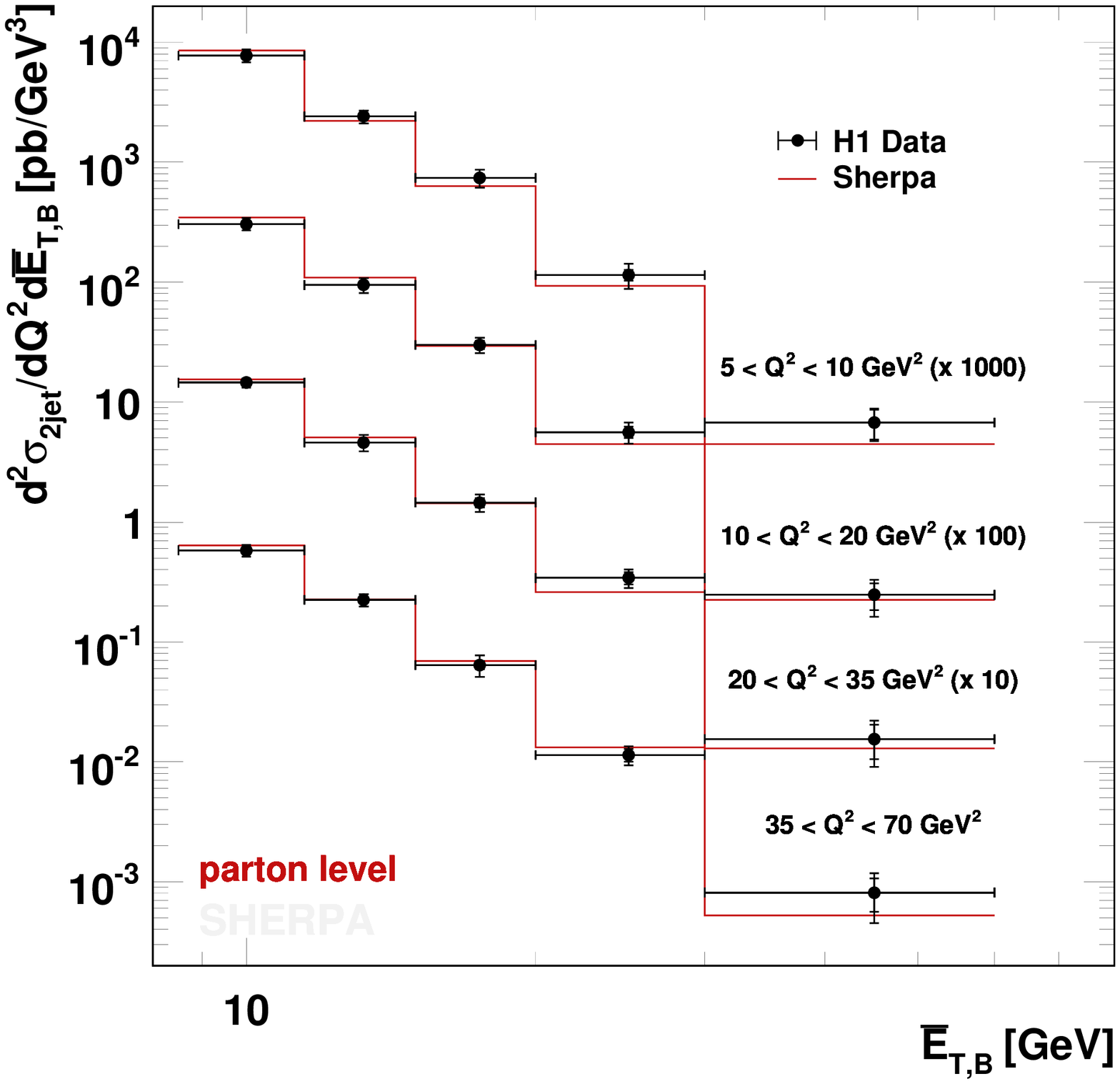}\hspace*{5mm}
    \includegraphics[width=8cm]{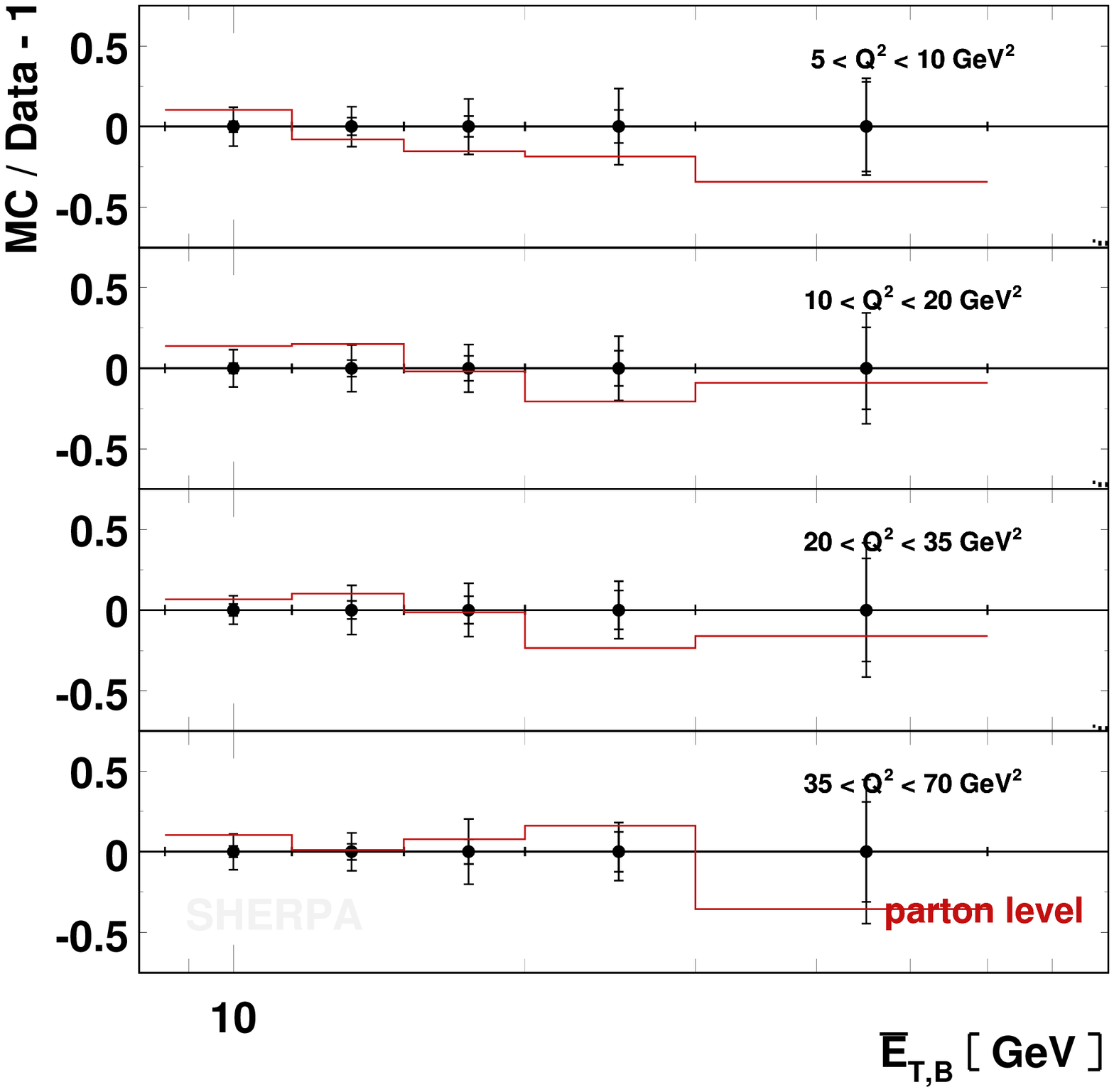}\\
    \includegraphics[width=8cm]{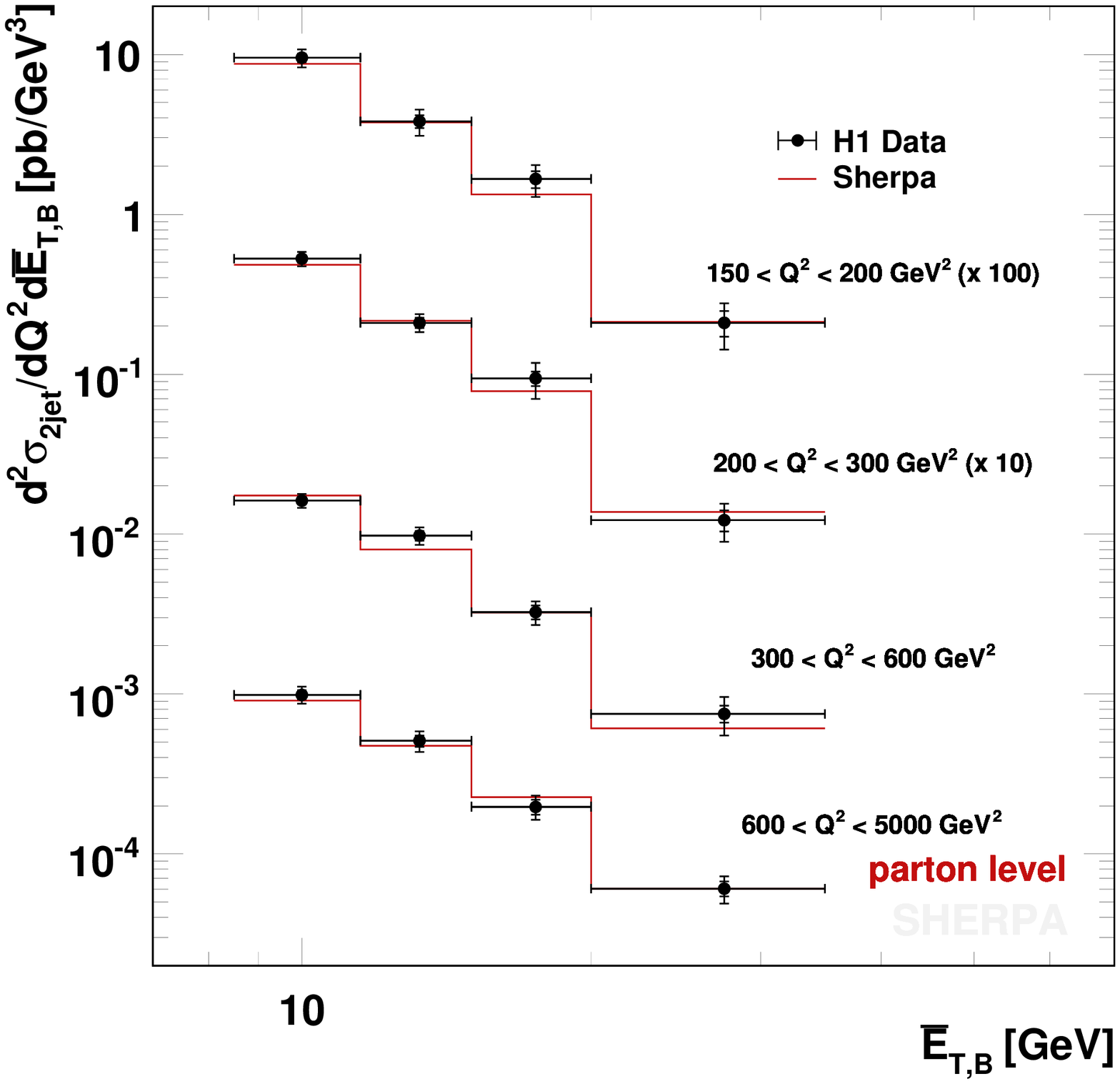}\hspace*{5mm}
    \includegraphics[width=8cm]{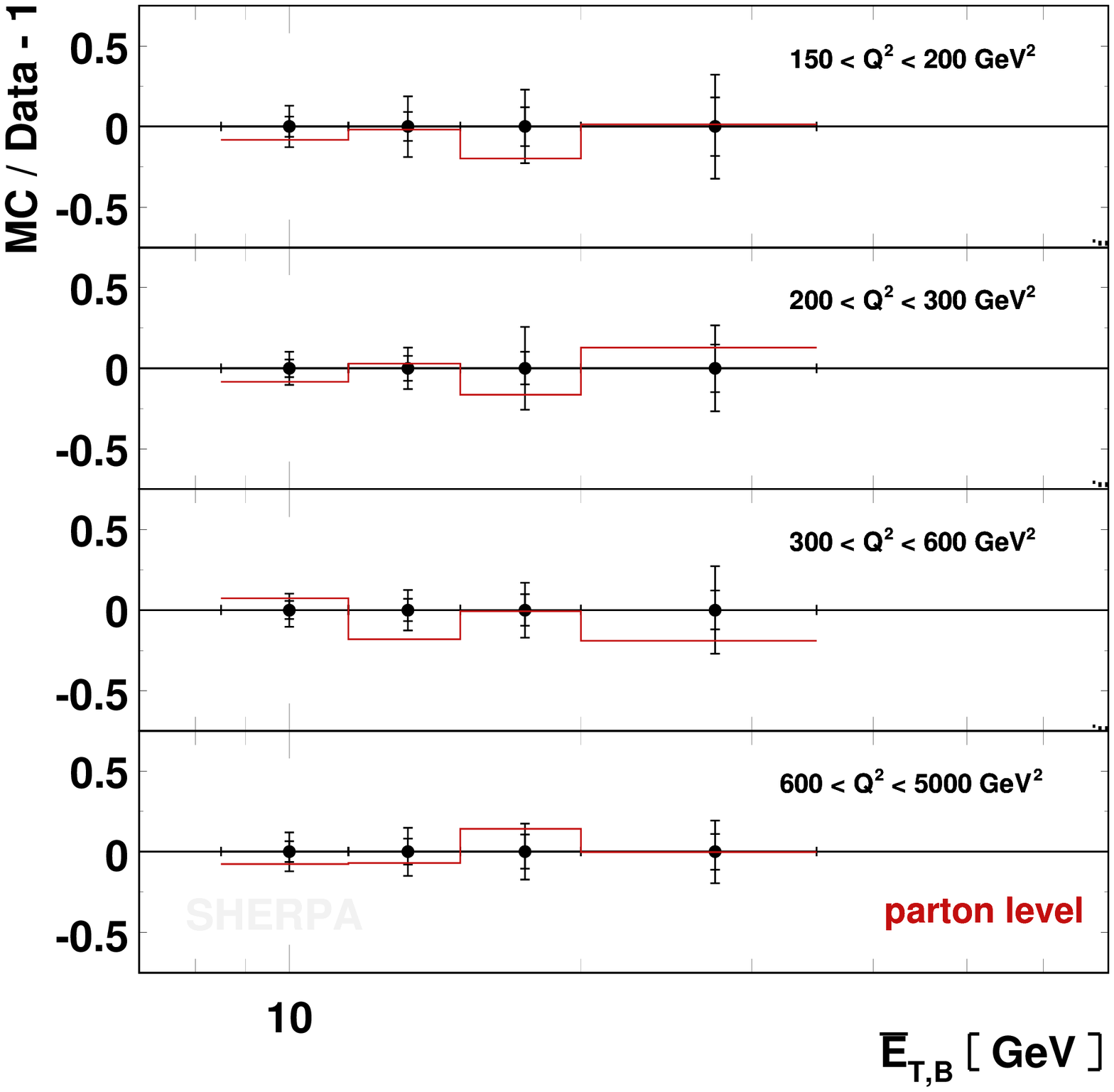}
  \end{center}
  \caption{The di-jet cross section as a function of $\bar{E}_{T,B}$, the mean jet transverse 
    energy in the Breit frame, measured by the H1 Collaboration~\protect\cite{Adloff:2000tq}.
    \label{fig:etb_twojet}}
\end{figure}
\begin{figure}[p]
  \begin{center}
    \includegraphics[width=8cm]{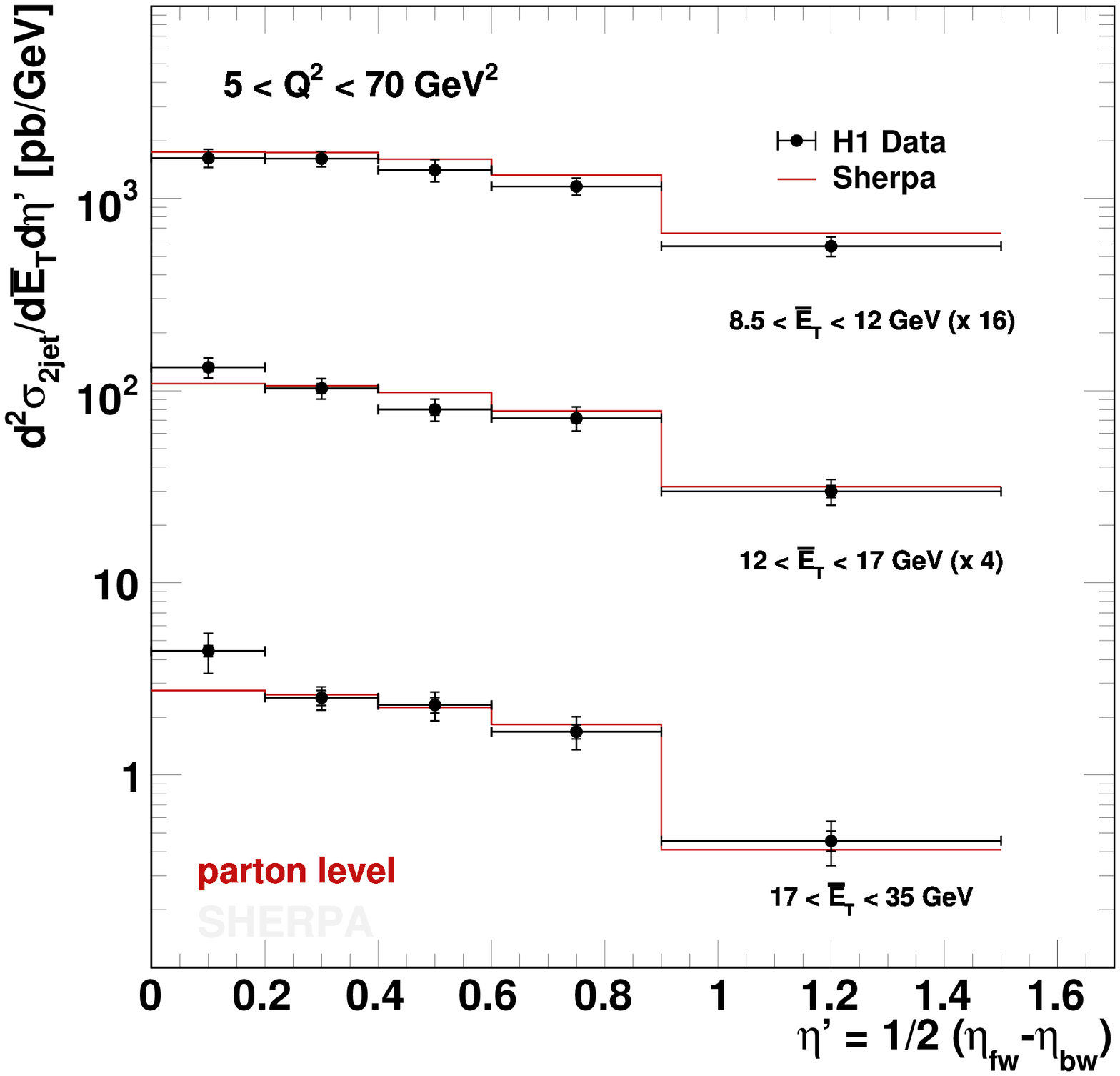}\hspace*{5mm}
    \includegraphics[width=8cm]{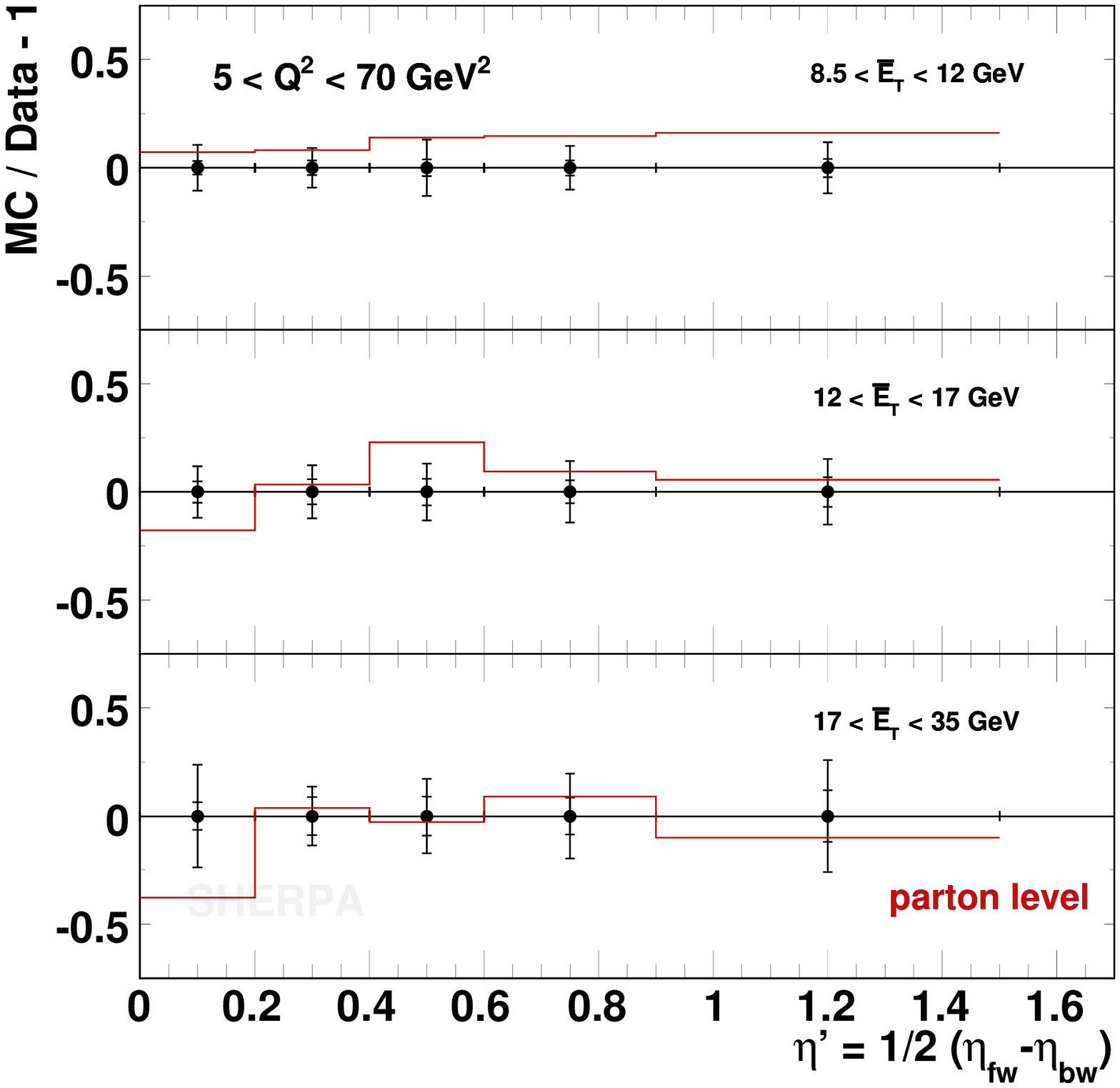}\\
    \includegraphics[width=8cm]{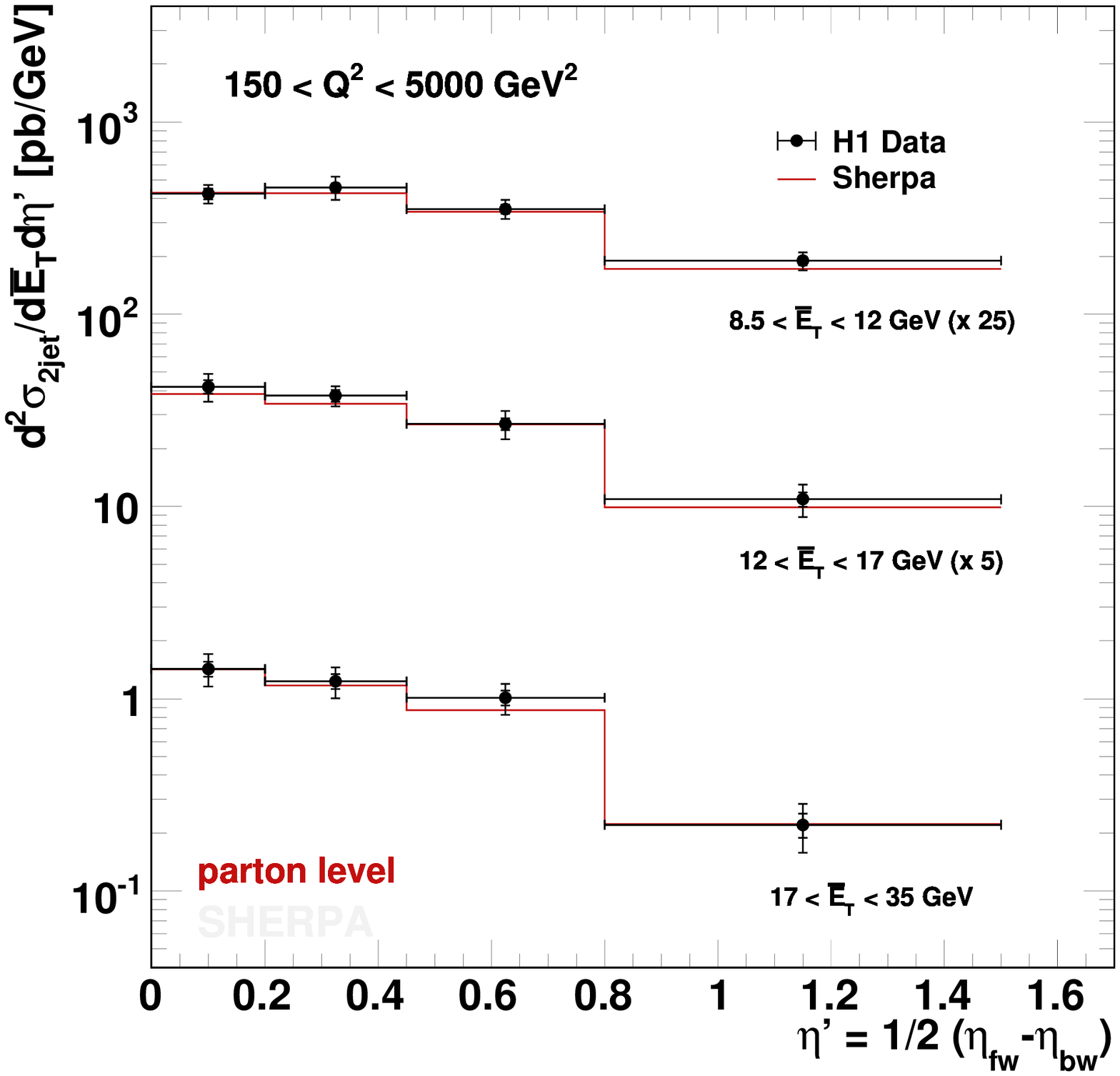}\hspace*{5mm}
    \includegraphics[width=8cm]{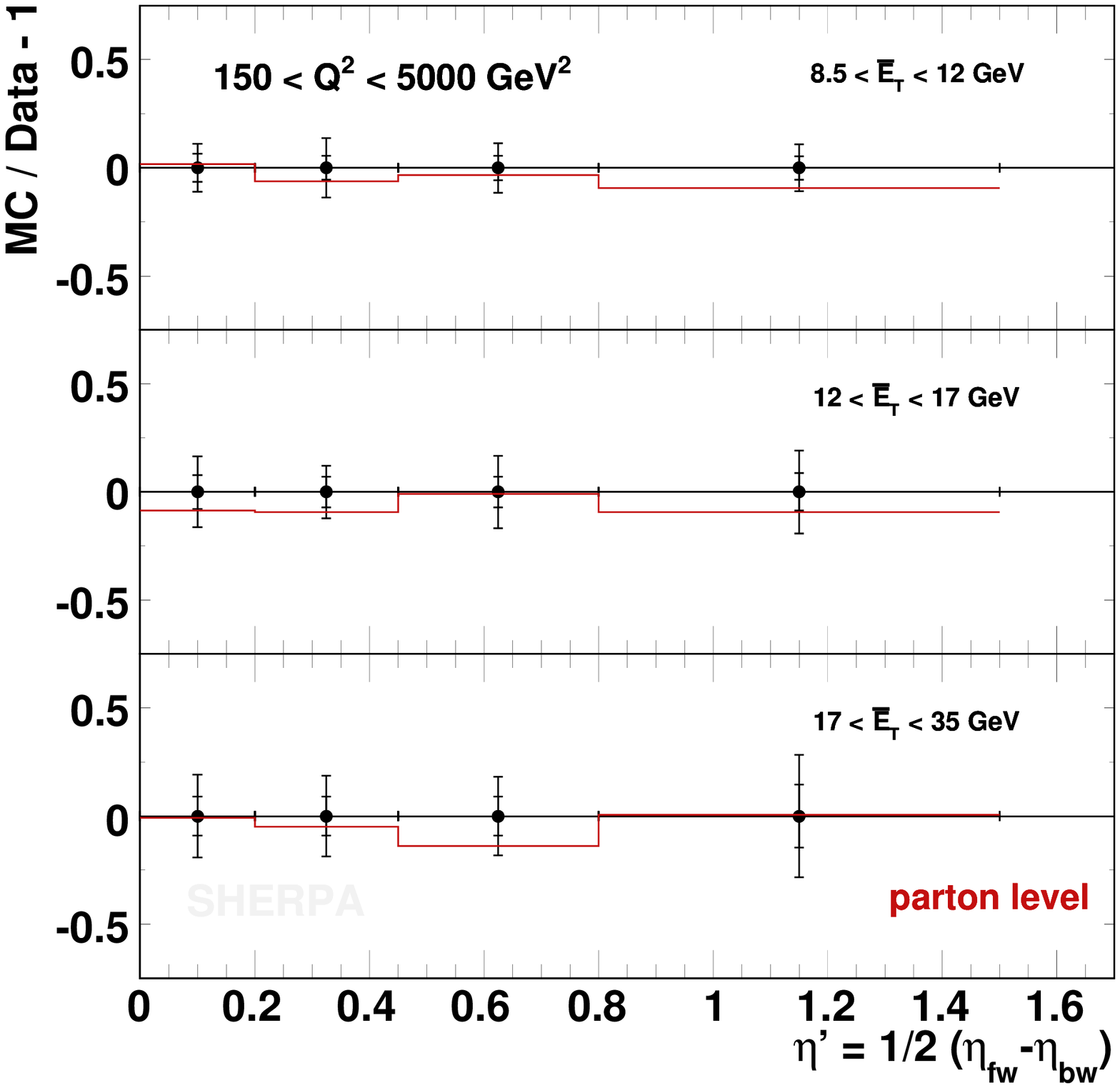}
  \end{center}
  \caption{The di-jet cross section as a function of $\eta'$, measured by the 
    H1 Collaboration~\protect\cite{Adloff:2000tq}. $\eta'$ denotes half the rapidity difference 
    of the two leading jets in the Breit frame.
    \label{fig:eta_twojet}}
\end{figure}

\begin{figure}[p]
  \begin{center}
    \includegraphics[width=\textwidth]{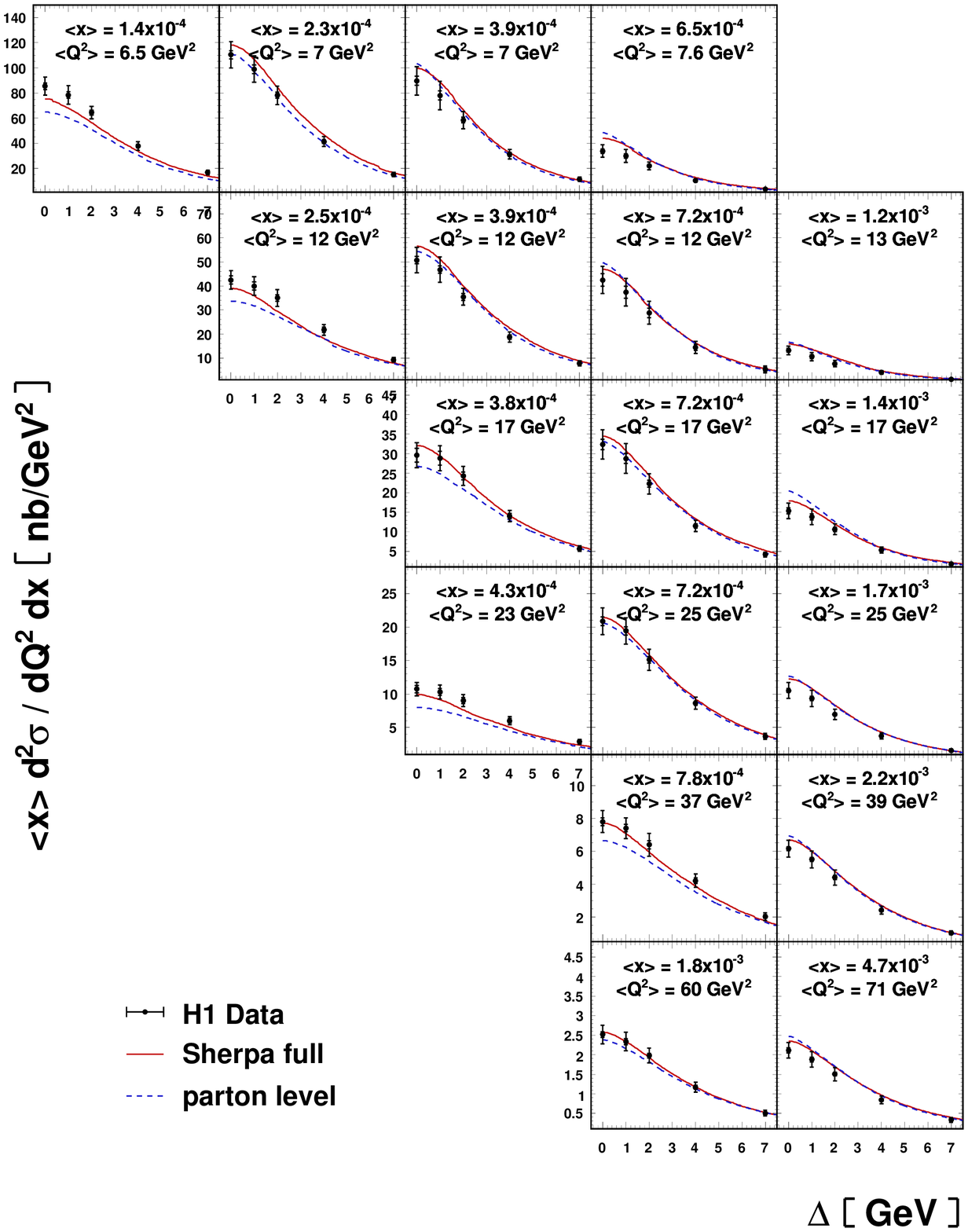}
  \end{center}
  \caption{The differential di-jet cross section as a function of $\Delta$ in
    bins of mean $x$ and $Q^2$, measured by the H1 Collaboration~\protect\cite{Aktas:2003ja}.
    $\Delta$ is defined as $E^*_{T,\rm max}>E^*_{T\,\rm cut}+\Delta$, where $E^*_{T\,\rm cut}$ 
    is the minimum jet transverse energy and $E^*_{T\,\rm max}$ is the transverse energy 
    of the hardest jet.
    \label{fig:delta_twojet}}
\end{figure}
\begin{figure}[p]
  \begin{center}
    \includegraphics[width=7.25cm]{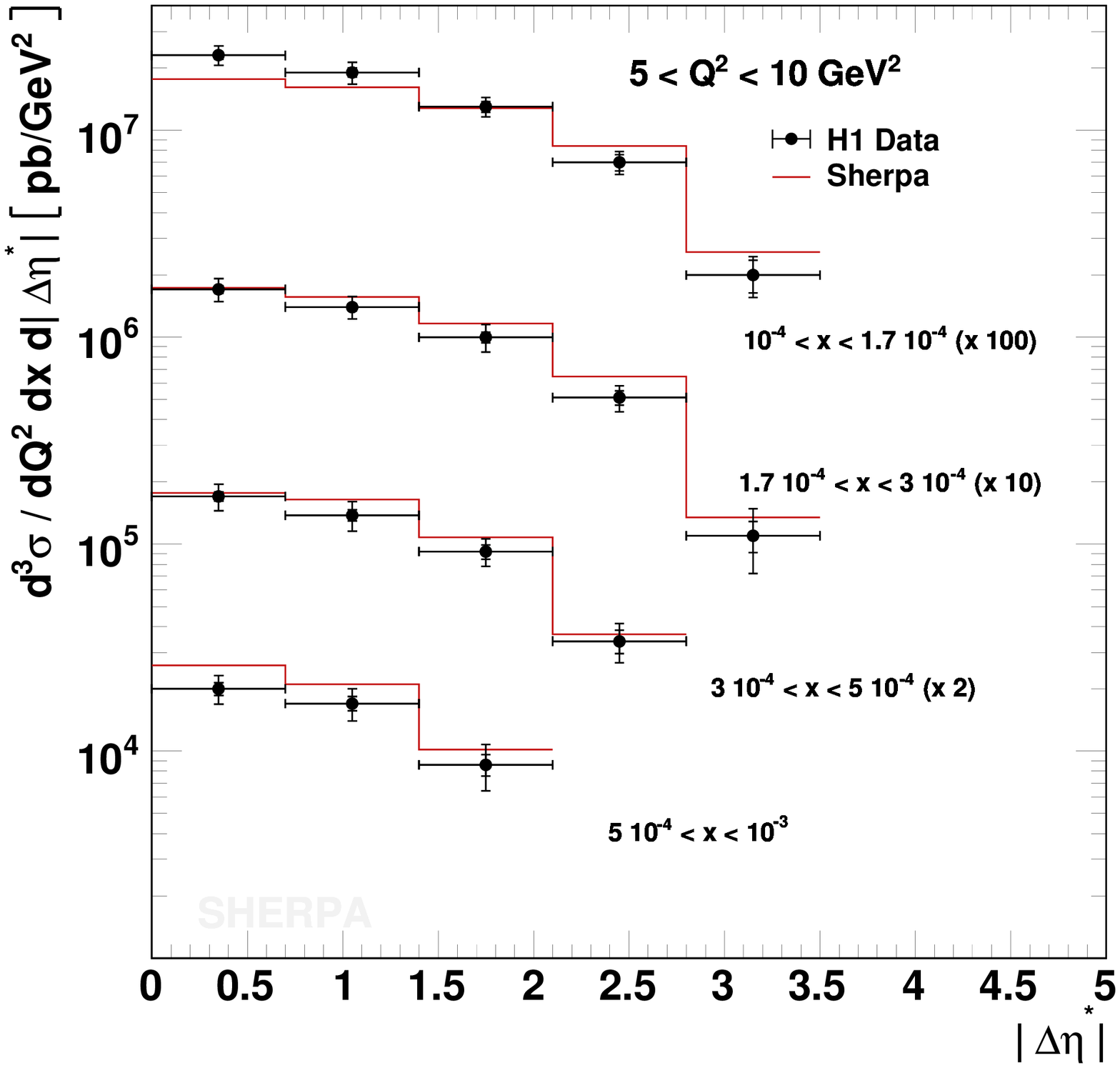}
    \includegraphics[width=7.25cm]{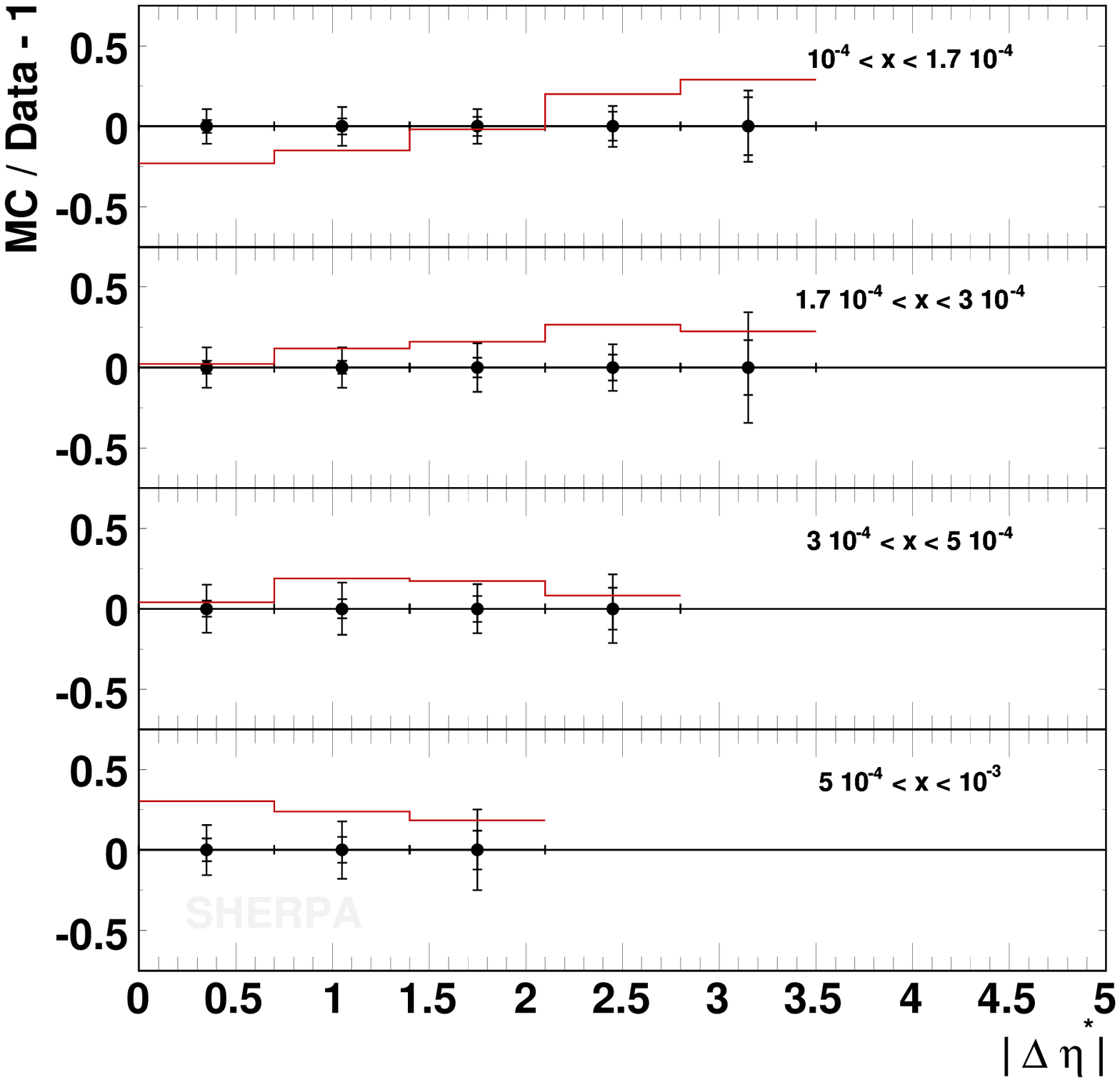}\\
    \includegraphics[width=7.25cm]{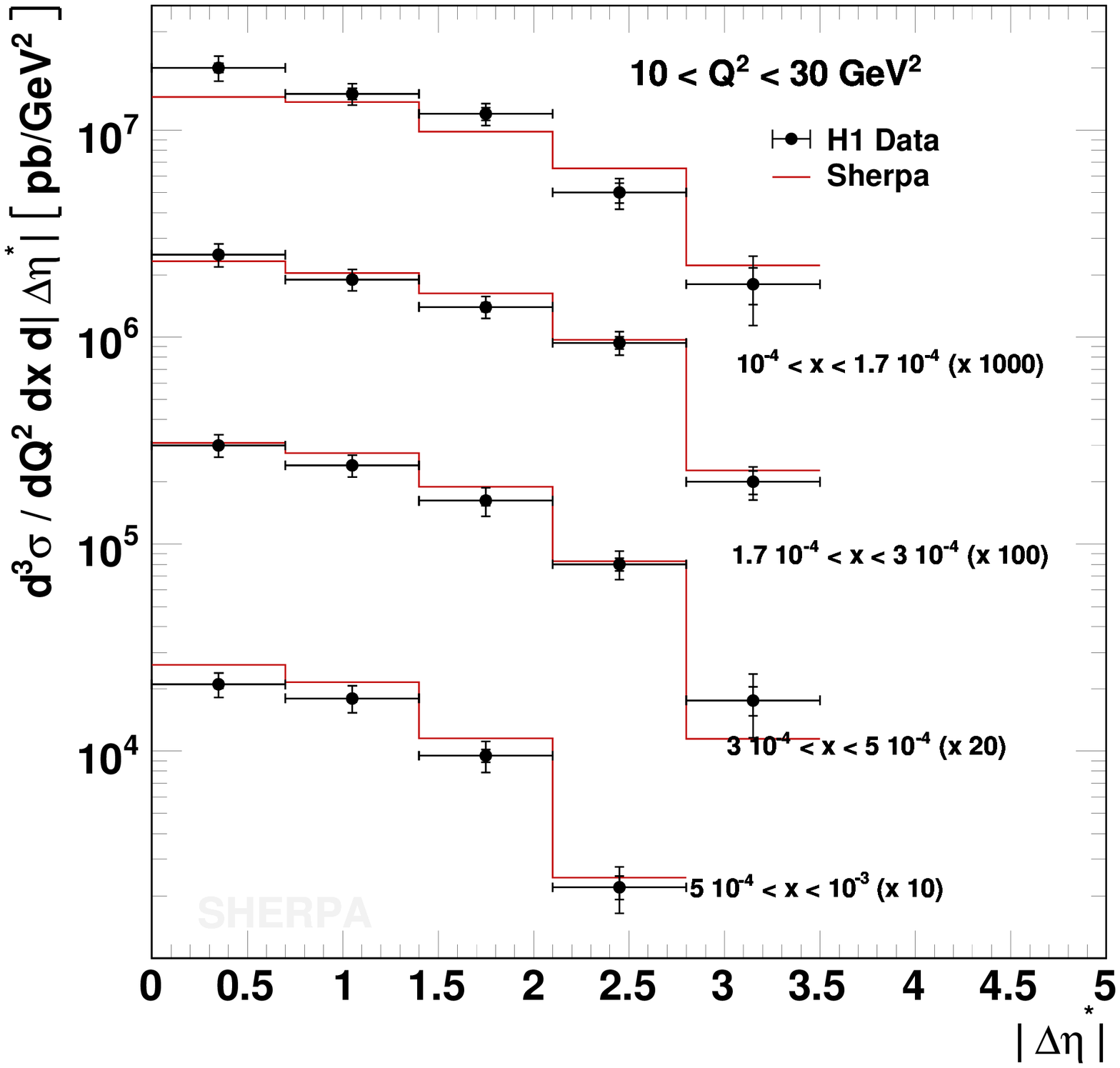}
    \includegraphics[width=7.25cm]{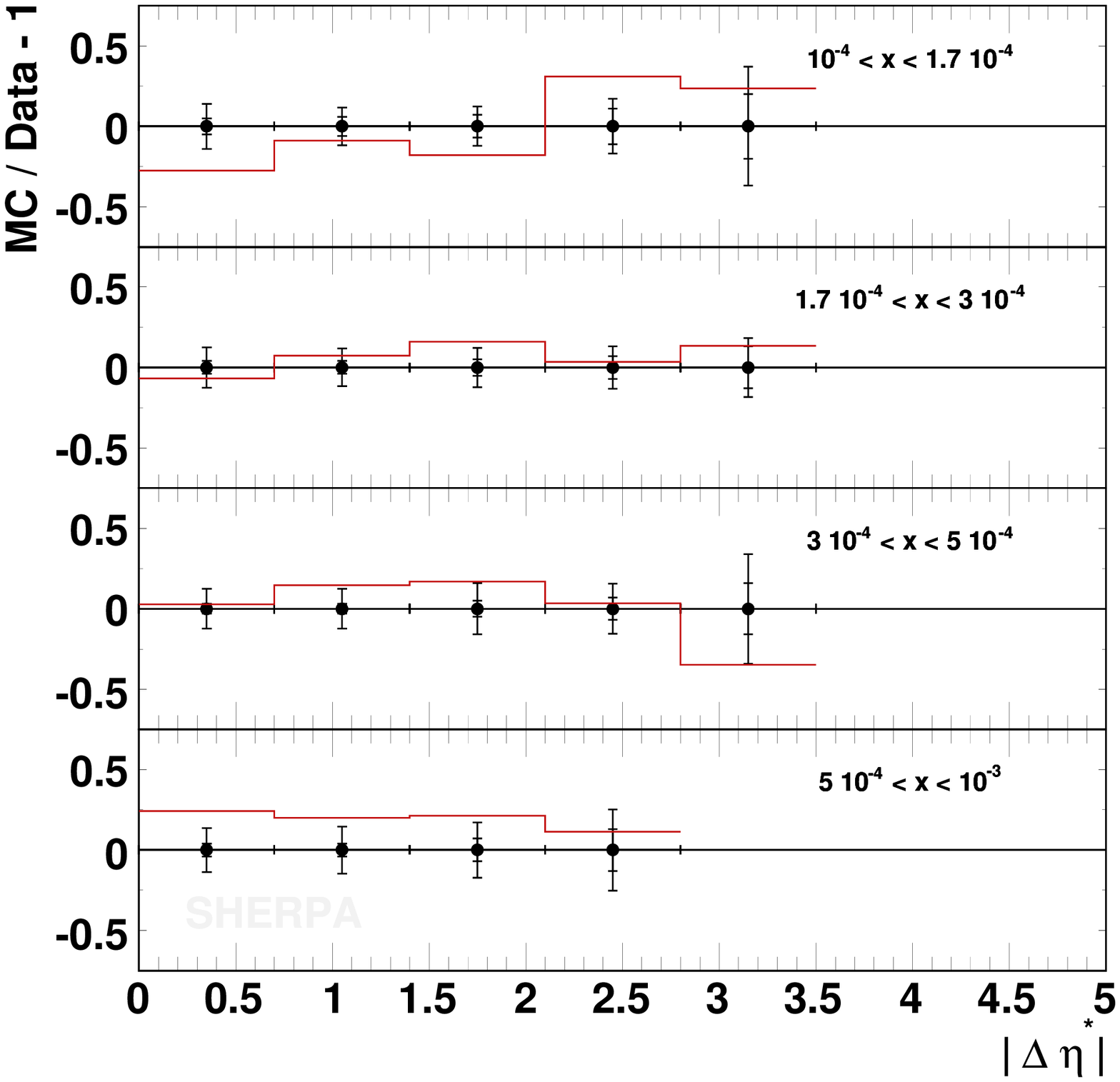}\\
    \includegraphics[width=7.25cm]{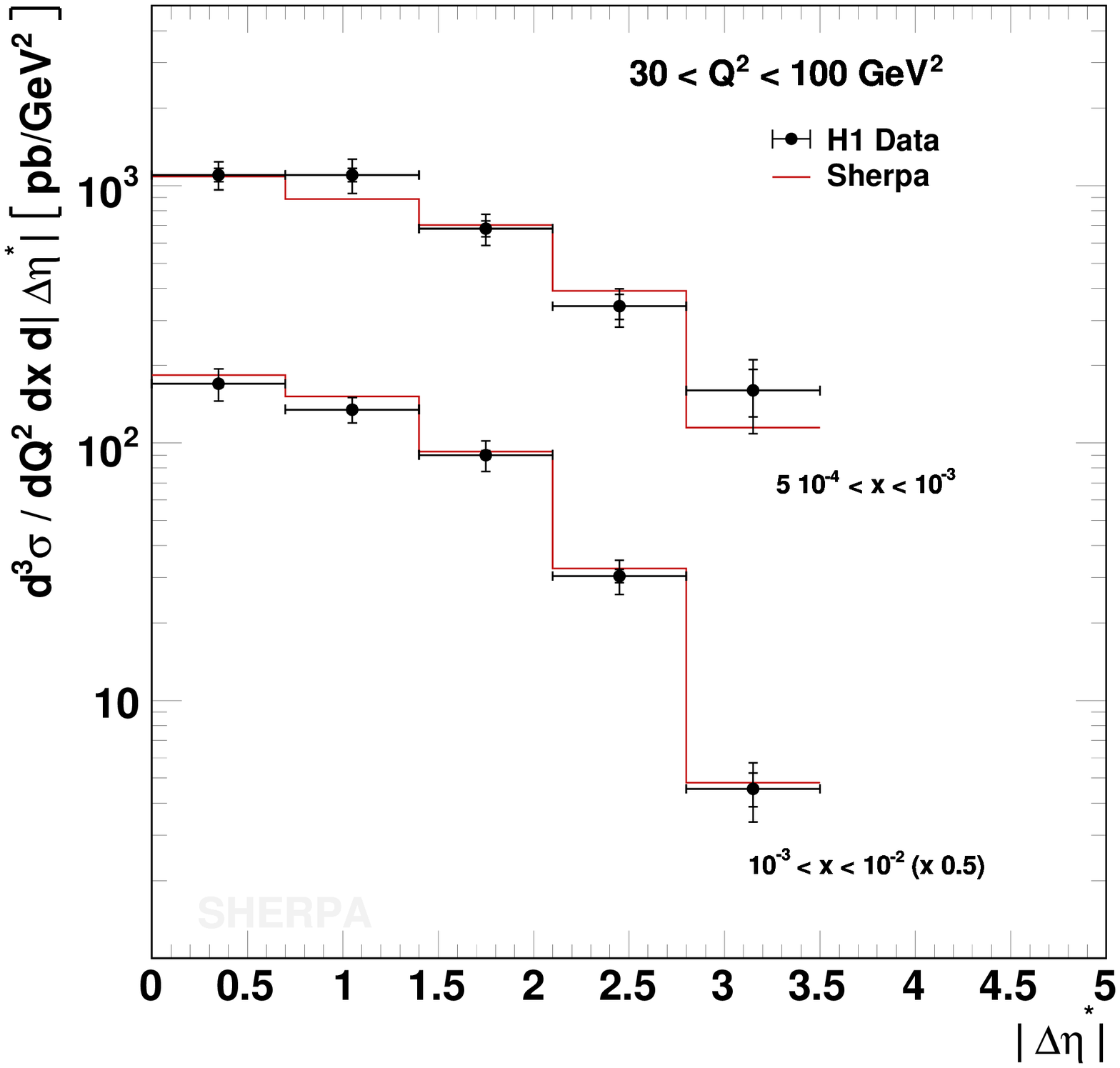}
    \includegraphics[width=7.25cm]{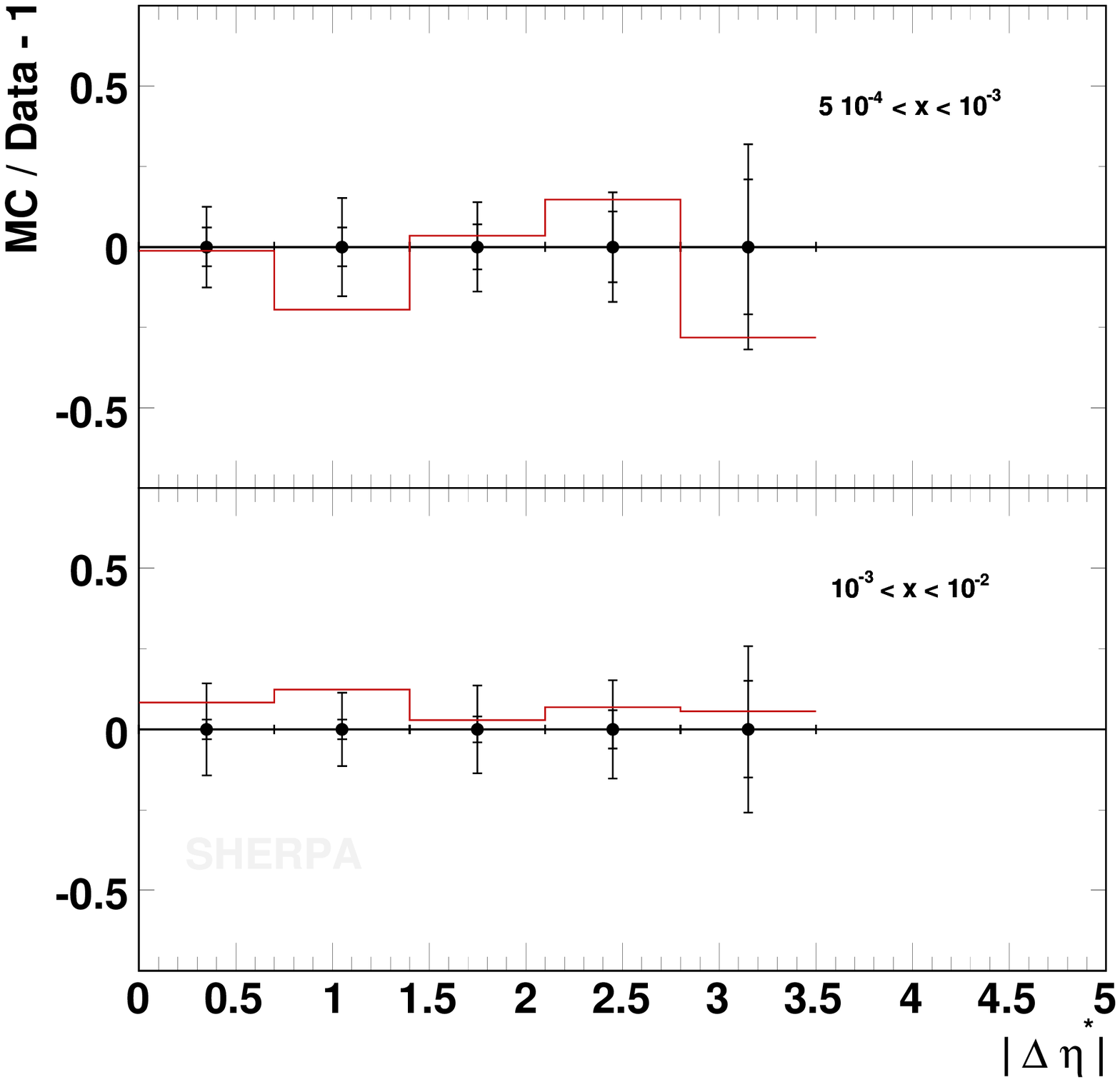}
  \end{center}
  \caption{The di-jet differential cross section for $\Delta=2$ GeV as a function of
    $|\Delta\eta^*|$ in bins of $x$ and $Q^2$, measured by the H1 Collaboration~\protect\cite{Aktas:2003ja}.
    \label{fig:deltaeta_twojet}}
\end{figure}

\begin{figure}[p]
  \begin{center}
  \subfloat[][]{\includegraphics[width=8.5cm]{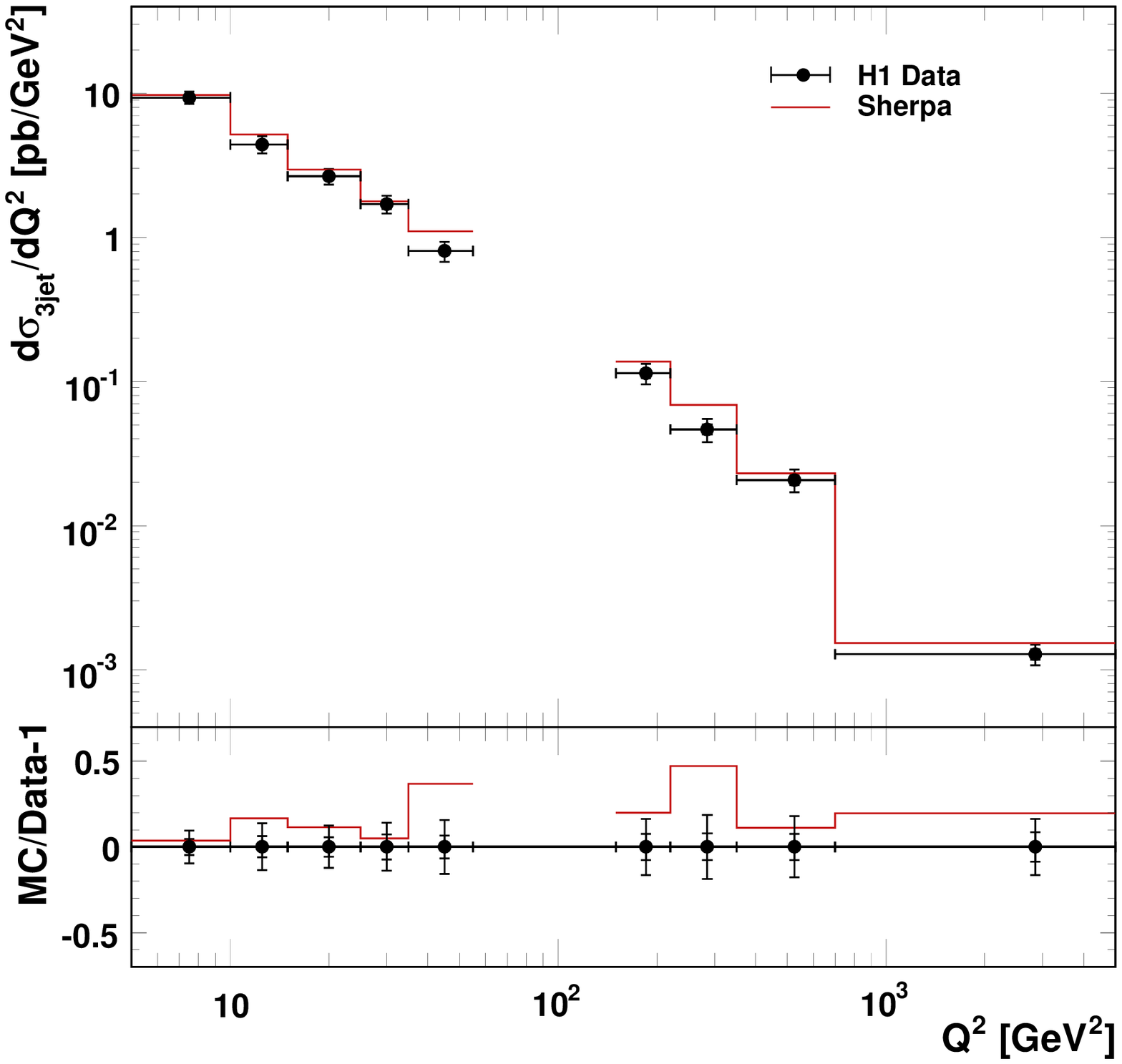}
    \label{fig:q2_3jet}}
  \subfloat[][]{\includegraphics[width=8.5cm]{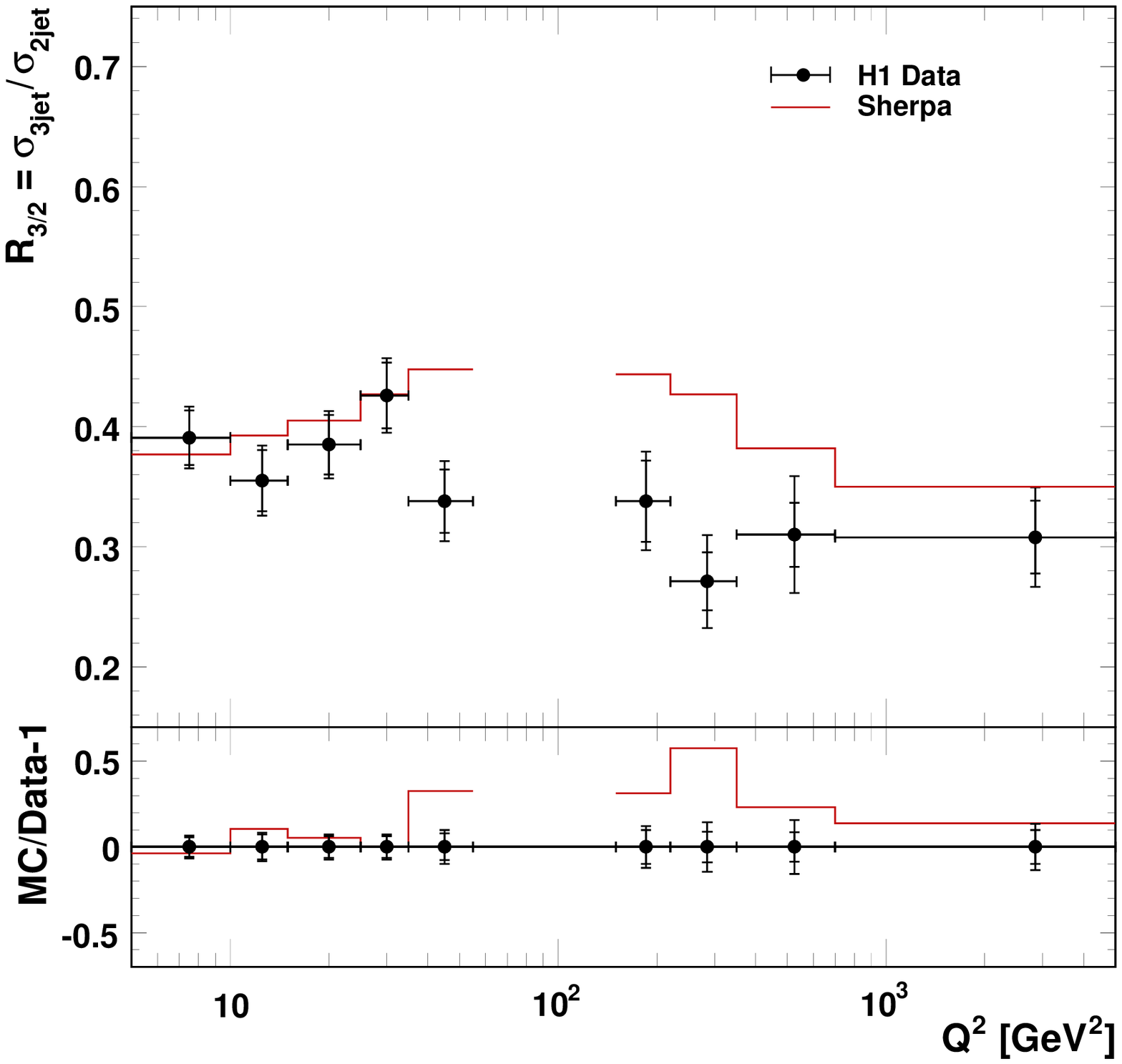}
    \label{fig:R32}}\\
  \subfloat[][]{\includegraphics[width=8.5cm]{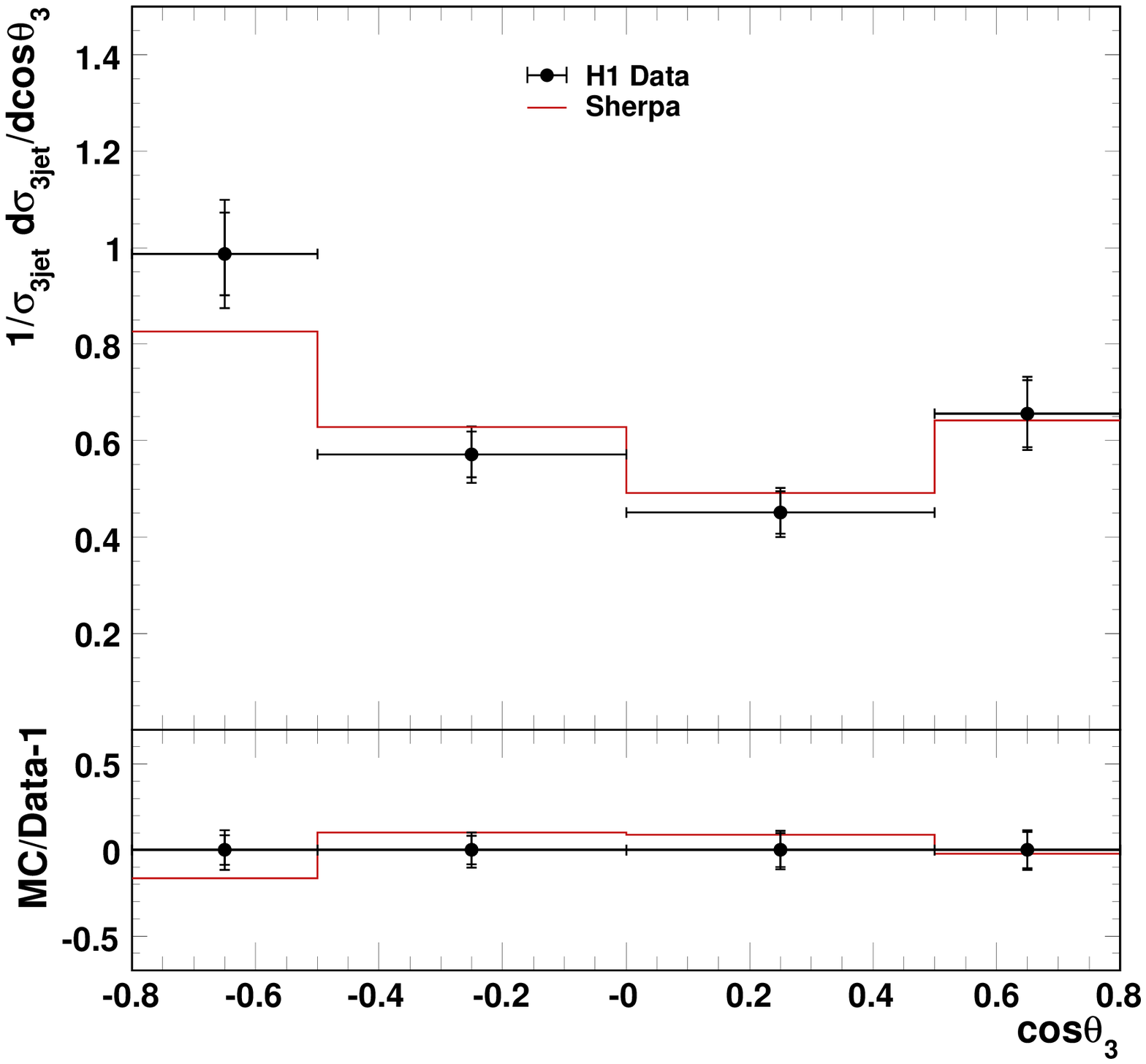}
    \label{fig:theta3}}
  \subfloat[][]{\includegraphics[width=8.5cm]{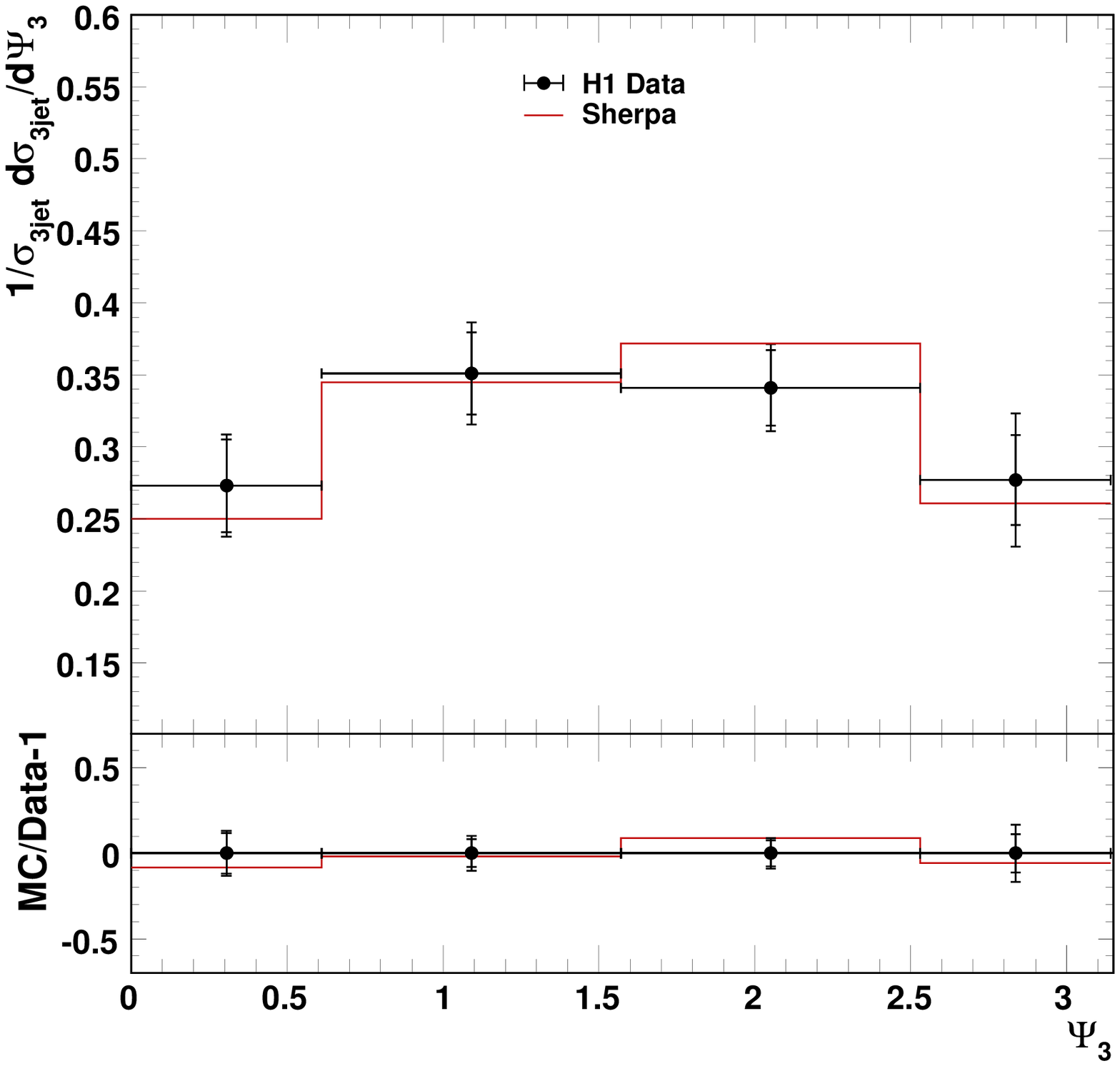}
    \label{fig:psi3}}
  \end{center}
  \caption{The three-jet cross section as a function of $Q^2$~\protect\subref{fig:q2_3jet}, 
    $\cos\theta_3$~\protect\subref{fig:theta3} and $\psi_3$~\protect\subref{fig:psi3} and the
    ratio of the three- over the two-jet rate as a function of $Q^2$~\protect\subref{fig:R32}, measured by the 
    H1 Collaboration~\protect\cite{Adloff:2001kg}.}\label{fig:threejet_obs}
\end{figure}
 
\begin{figure}[p]
  \begin{center}
    \includegraphics[width=0.9\textwidth]{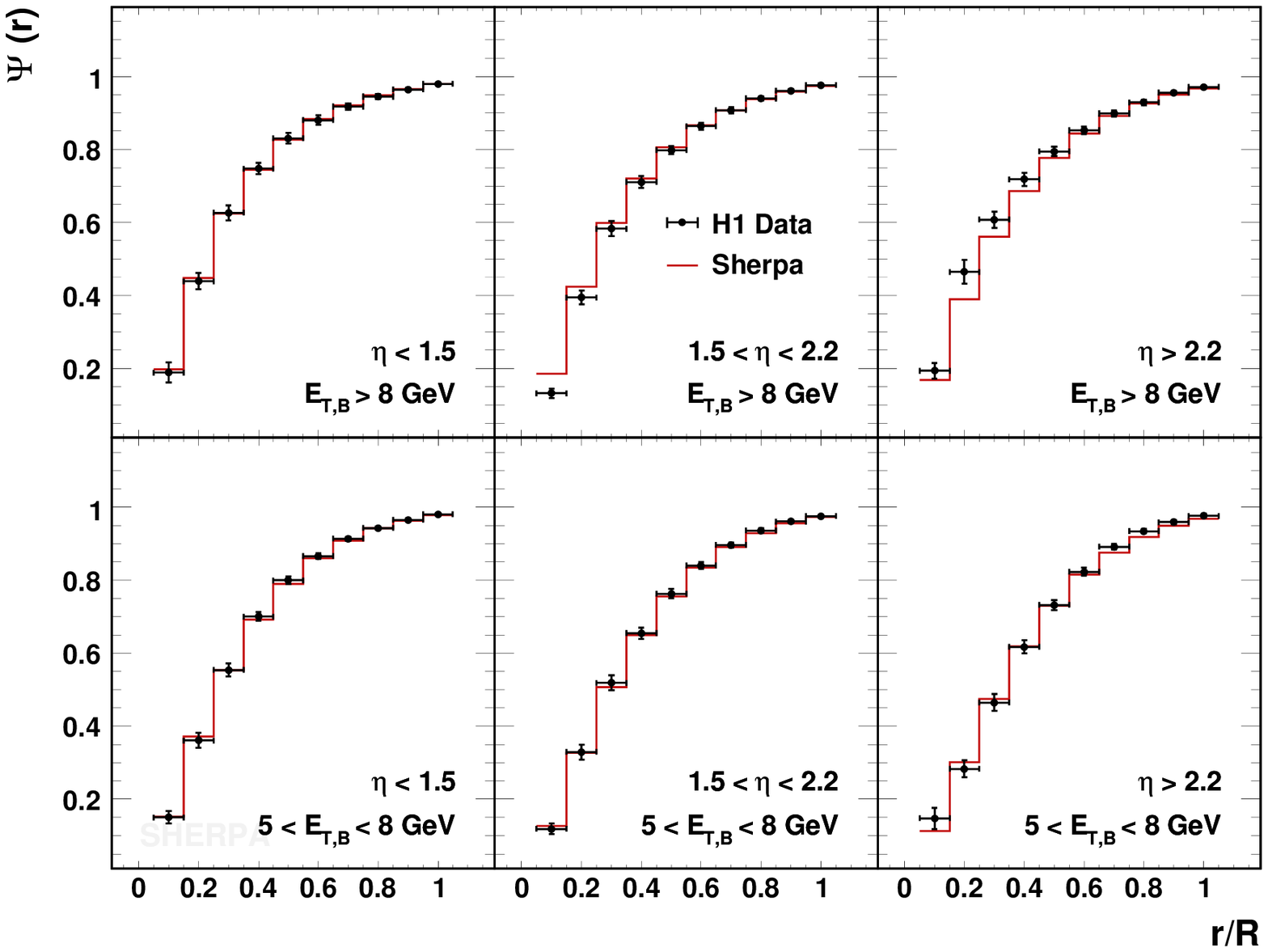}
  \end{center}
  \caption{Jet shapes 
    in bins of jet transverse energy and jet pseudorapidity in the Breit frame,
    measured by the H1 Collaboration~\protect\cite{Adloff:1998ni}.
    \label{fig:jetshapes}}
\end{figure}
\begin{figure}[p]
  \begin{center}
    \includegraphics[width=0.9\textwidth]{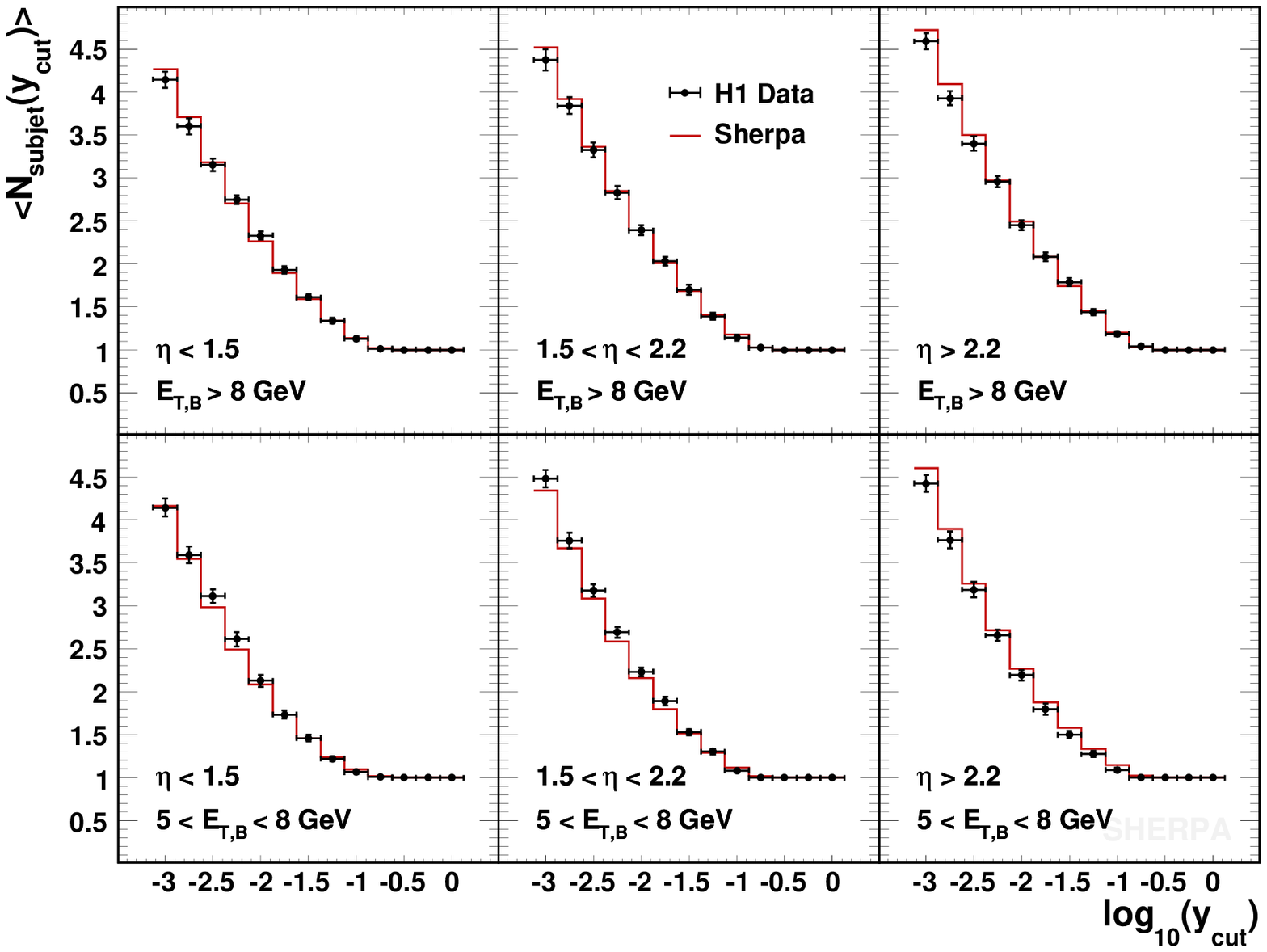}
  \end{center}
  \caption{Sub-jet rates as a function of the resolution parameter $y_{\rm cut}$
    in bins of jet transverse energy and jet pseudorapidity in the Breit frame,
    measured by the H1 Collaboration~\protect\cite{Adloff:1998ni}.
    \label{fig:subjetrates}}
\end{figure}

\begin{figure}[p]
  \begin{center}
    \includegraphics[width=\textwidth]{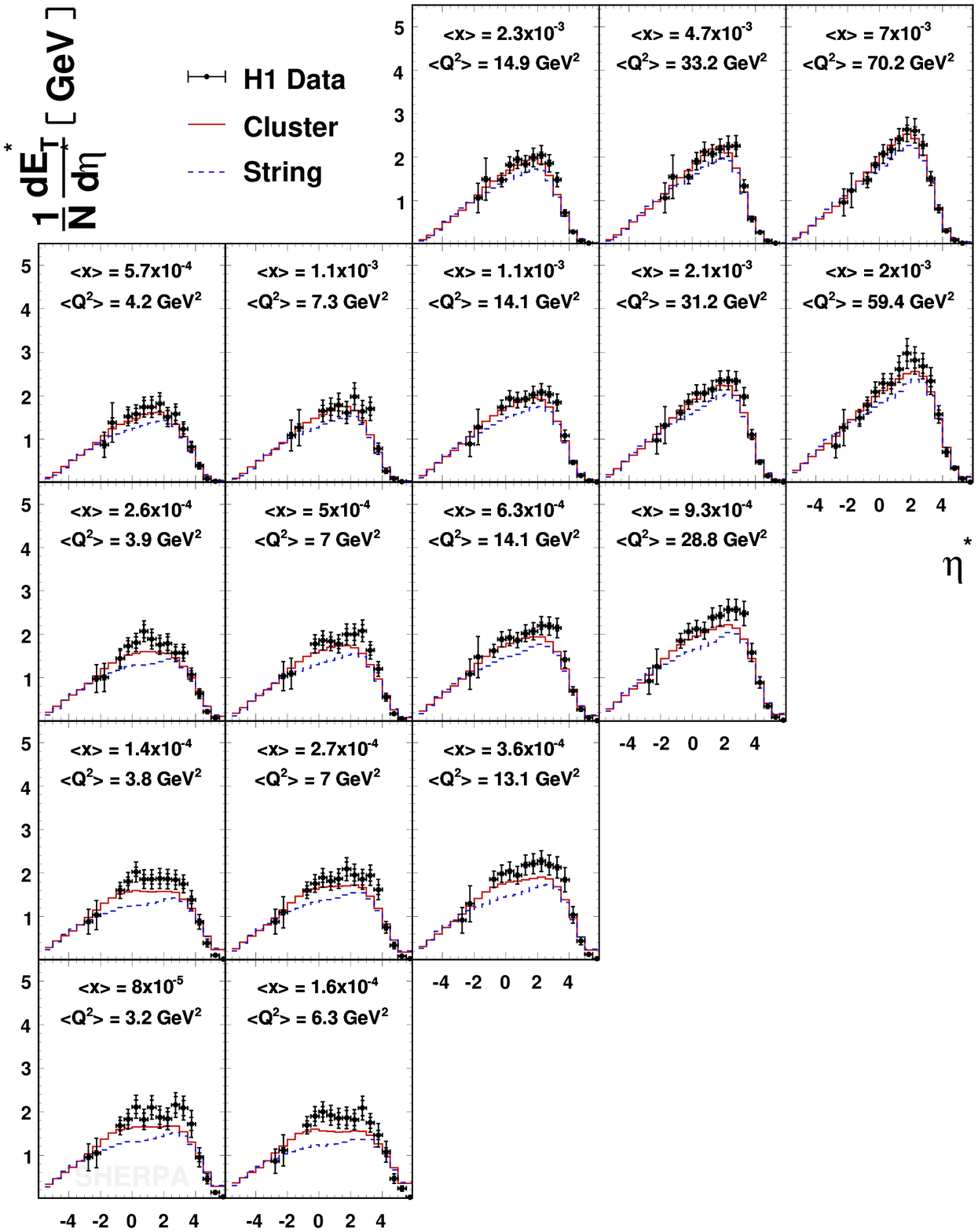}
  \end{center}
  \caption{Transverse energy flows measured by the
    H1 Collaboration~\protect\cite{Adloff:1999ws}.
    The histogram labeled ``Cluster'' displays results obtained with the
    cluster hadronisation model of~\protect\cite{Krauss:2010xy},
    while ``String'' shows predictions of the Lund string 
    hadronisation~\protect\cite{Andersson:1983ia,*Andersson:1998tv}.}
    \label{fig:eflow_lowq}
\end{figure}
\begin{figure}[p]
  \begin{center}
    \includegraphics[width=\textwidth]{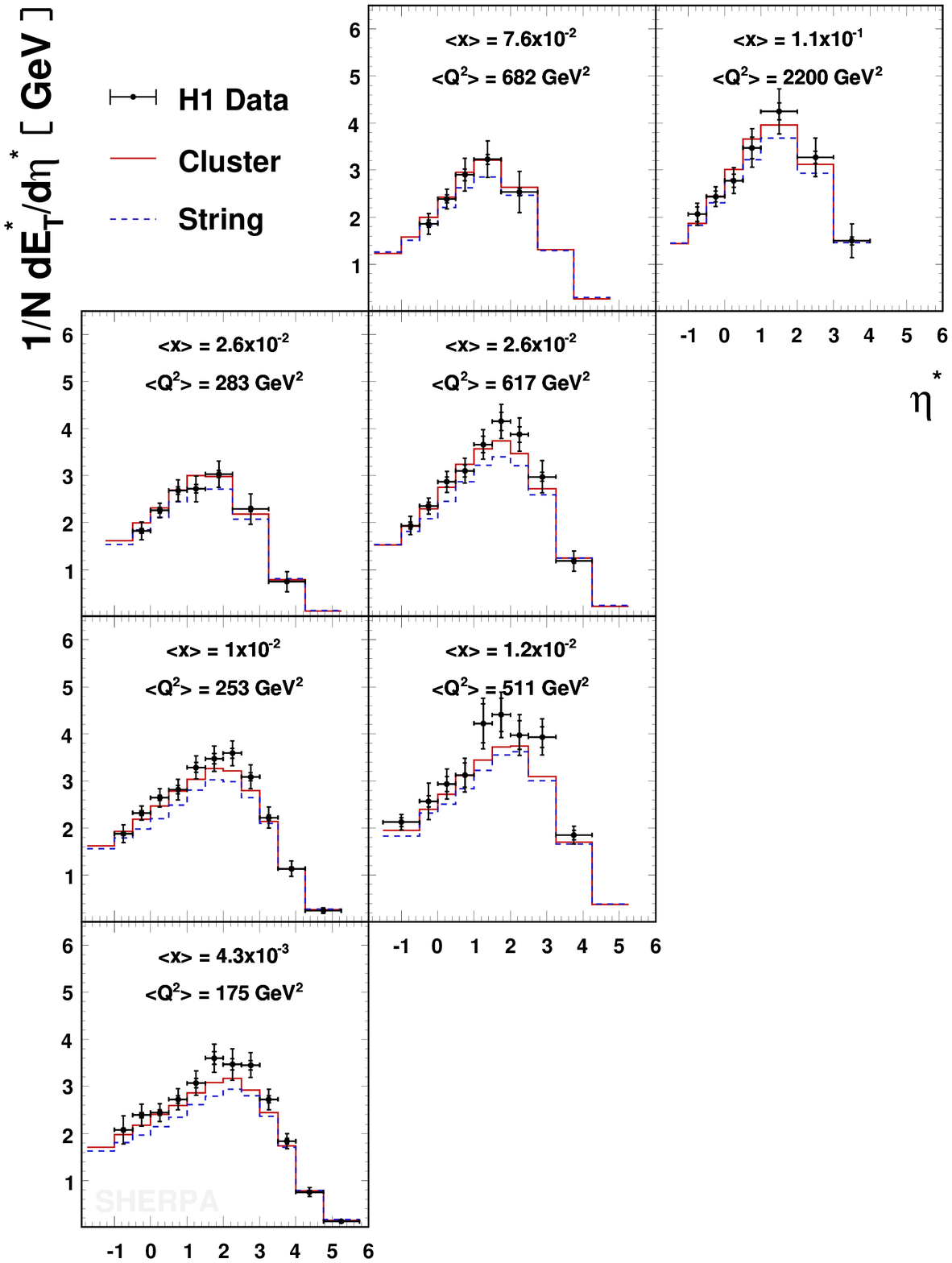}
  \end{center}
  \caption{Transverse energy flows measured by the
    H1 Collaboration~\protect\cite{Adloff:1999ws}.
    See Fig.~\ref{fig:eflow_lowq} for notation.}
    \label{fig:eflow_highq}
\end{figure}

\begin{figure}[p]
  \begin{center}
    \includegraphics[width=\textwidth]{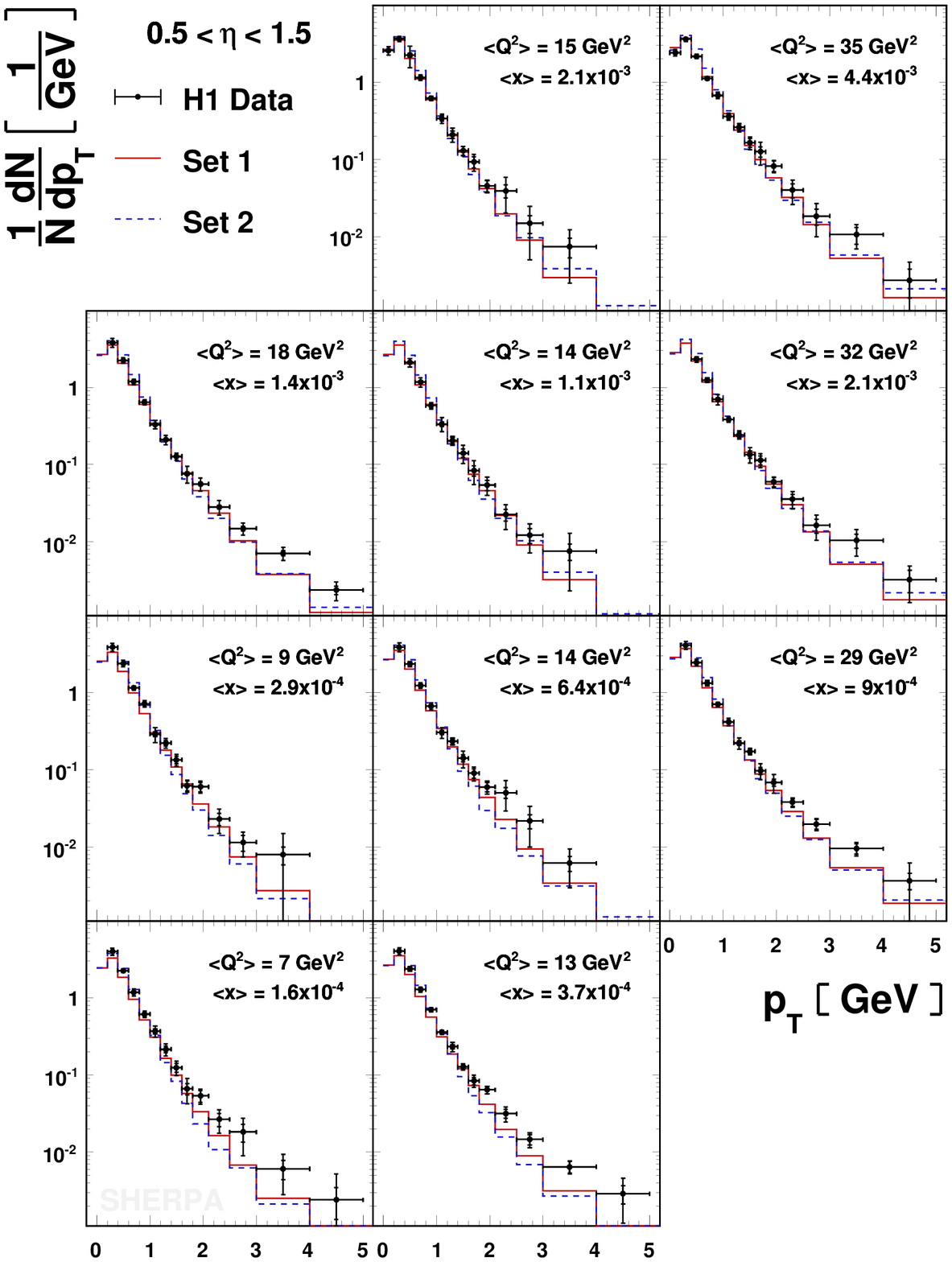}
  \end{center}
  \caption{Charged particle transverse momentum spectra measured by the
    H1 Collaboration~\protect\cite{Adloff:1996dy}.
    The histogram labeled ``Set~1'' shows predictions from the cluster 
    hadronisation model of~\protect\cite{Krauss:2010xy} in combination with 
    the NNPDF~1.2 PDF set~\protect\cite{Ball:2008by}, while ``Set~2'' displays 
    predictions from the Lund string 
    hadronisation~\protect\cite{Andersson:1983ia,*Andersson:1998tv}
    in combination with the CTEQ~6L1 PDF set~\protect\cite{Pumplin:2002vw}.}
    \label{fig:charged_pt}
\end{figure}

\begin{figure}[p]
  \begin{center}
    \includegraphics[width=\textwidth]{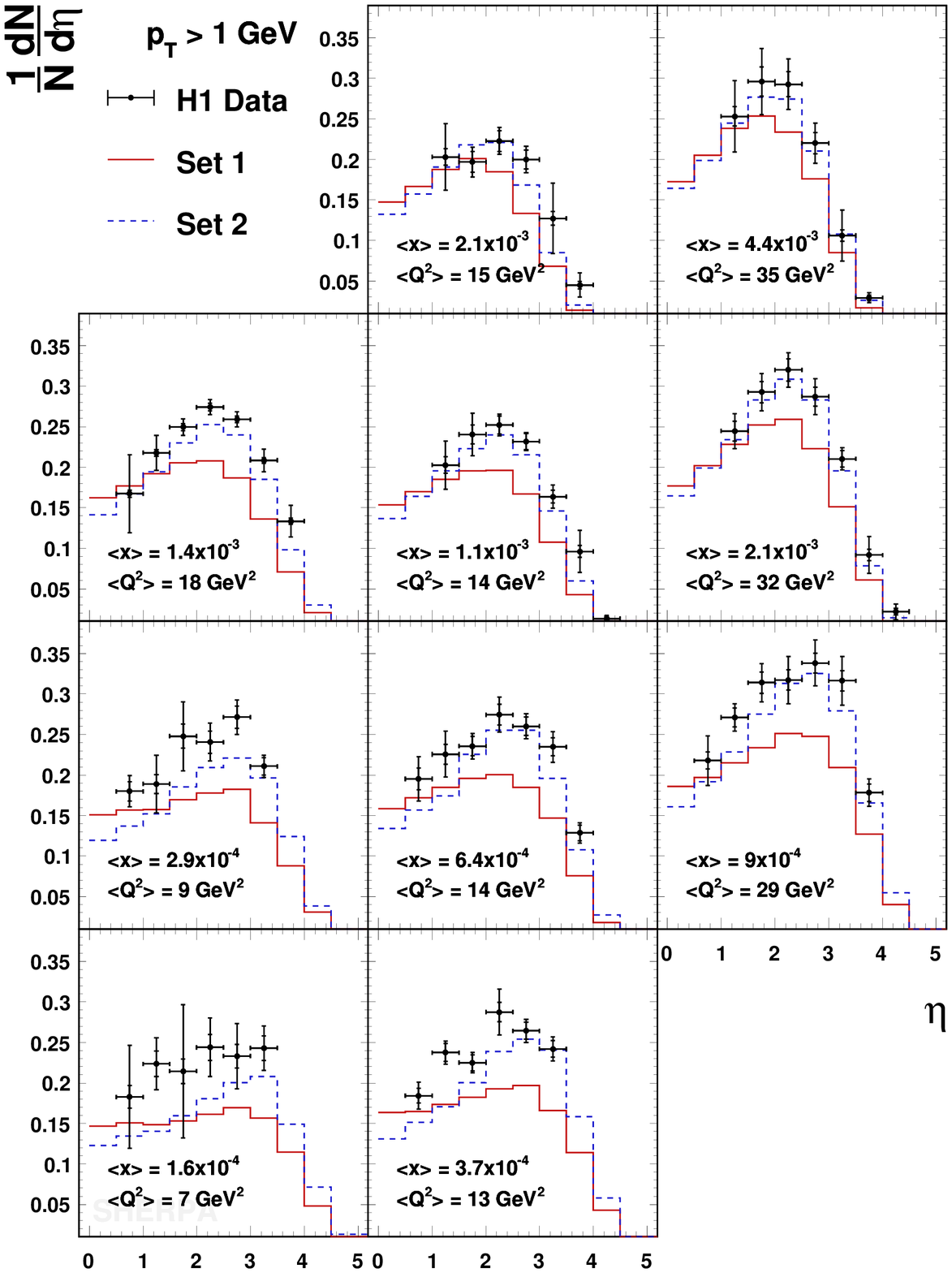}
  \end{center}
  \caption{Charged multiplicity flow for single-particle transverse momenta
    larger than 1 GeV, measured by the
    H1 Collaboration~\protect\cite{Adloff:1996dy}.
    See Fig.~\ref{fig:charged_pt} for notation.
    \label{fig:charged_multi}}
\end{figure}

\begin{figure}[p]
  \begin{center}
    \includegraphics[width=\textwidth]{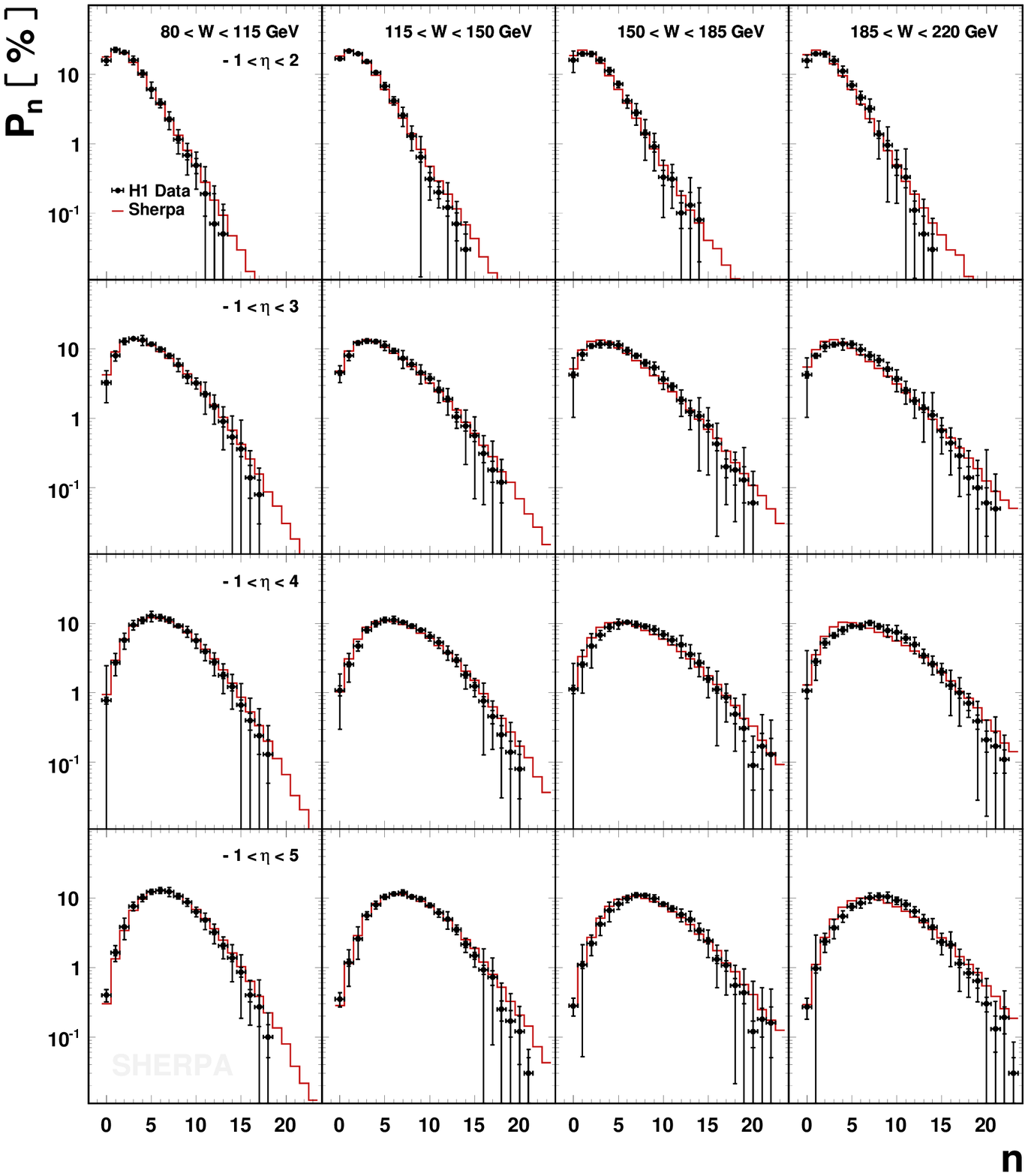}
  \end{center}
  \caption{Charged multiplicity distributions in bins of $\eta$ and $W$,
    measured by the H1 Collaboration~\protect\cite{Aid:1996cb}.
    \label{fig:multi}}
\end{figure}


\end{document}